\documentclass[structabstract]{aa}  
\usepackage{graphicx}
\usepackage{subcaption}
\usepackage{color}
\usepackage{url}
\usepackage{amssymb}
\usepackage{amsmath}
\usepackage{bm}
\usepackage{txfonts}
\usepackage{multirow}
\usepackage{lscape}
\usepackage[flushleft]{threeparttable}
\usepackage{float}
\usepackage{natbib}
\usepackage{stmaryrd}
\usepackage{supertabular}
\usepackage{rotating}
\usepackage{longtable}
\usepackage{chngcntr}
\usepackage{changes}
\usepackage{pdflscape}
\usepackage{booktabs}
\usepackage{dcolumn}
\usepackage{gensymb}

\bibpunct{(}{)}{;}{a}{}{,} % to follow the A&A style

\makeatletter
\def\hlinewd#1{%
\noalign{\ifnum0=`}\fi\hrule \@height #1 %
\futurelet\reserved@a\@xhline}
\makeatother

\newfont{\gwpfont}{cmssq8 scaled 1000}

\begin{document}
   \title{Comparison of hydrostatic and dynamical masses of distant X-ray luminous galaxy clusters
   \thanks{Based on observations from the Very Large Telescope at Paranal, Chile}}

   \author{
          G. Fo\"ex\inst{1}
          \and
          H. B\"ohringer\inst{1}
          \and
          G. Chon\inst{1}
          }
   \institute{
          Max Planck Institute for Extraterrestrial Physics, Giessenbachstrasse, 85748 Garching, Germany
             }

   \date{Received ; accepted }

  \abstract  
  {A crucial ingredient to use galaxy clusters as cosmological probe, or to study the physics governing their formation and evolution, is a robust determination of their mass. Applying various estimators on well-defined cluster samples is a mandatory step to characterise their respective systematics.} 
  {Our main goal is to compare the results of three dynamical mass estimators to the X-ray hydrostatic values. Here we focus on massive galaxy clusters at intermediate redshifts $z\sim0.3$.}
  {We estimated dynamical masses with the virial theorem, the Jeans equation, and the caustic method using wide-field VIMOS spectroscopy; the hydrostatic masses were obtained previously from XMM-Newton observations. We investigated the role of colour selection and the impact of substructures on the dynamical estimators.}
  {The Jeans and caustic methods give consistent results, whereas the virial theorem leads to masses $\sim15\%$ larger. The Jeans, caustic, and virial masses are respectively $\sim20\%$, $\sim30\%$, and $\sim50\%$ larger than the hydrostatic values. Large scatters of $\gtrsim50\%$ are mainly due to the two outliers RXCJ0014 and RXCJ1347; excluding the latter increases the mass ratios by $\sim10\%$, giving a fractional mass bias significant at $\gtrsim2\sigma$. We found a correlation between the dynamical-to-hydrostatic mass ratio and two substructure indicators, suggesting a bias in the dynamical measurements. The velocity dispersions of blue galaxies are $\sim15\%$ ($\sim25\%$ after removing the substructures) larger than that of the red-sequence galaxies; using the latter leads to dynamical masses $\sim10\%-15\%$ smaller. Discarding the galaxies part of substructures reduces the masses by $\sim15\%$; the effect is larger for the more massive clusters, owing to a higher level of substructures. After the substructure analysis, the dynamical masses are in perfect agreement with the hydrostatic values and the scatters around the mean ratios are divided by $\sim2$. The mass bias is no longer significant, even after excluding RXCJ1347.}
  {}


   \maketitle

%
%________________________________________________________________

\section{Introduction}

In the era of precision cosmology, the cluster mass function occupies a central role owing to its dependence on the geometrical properties of the universe and the growth of structures (e.g. \citealt{vikhlinin09,mantz10a,allen11,bohringer14,planck24}). The main ingredient to derive cosmological constraints from galaxy clusters is a robust determination of their mass, which can only be achieved individually on small samples. The practical analysis of the cluster mass function thus relies on indirect estimates via mass-observable scaling relations predicted by the hierarchical model of structure formation \citep{kaiser86}. Considerable efforts have been made to calibrate such relations (e.g. \citealt{bohringer12} or \citealt{giodini13} for a review), since departures from the theoretical predictions provide insight on the non-gravitational physical processes at play in galaxy clusters (e.g. \citealt{voit05} for a review).

Estimating the mass of a cluster is a complicated task since its main component, the dark matter halo, is not directly observable. Instead, one can infer its mass via its gravitational lensing distortion on the shape of distant background galaxies, or via its effects on the other components of a cluster, such as the dynamics of the cluster galaxies or the X-ray emission of the intra cluster medium. Each of these approaches requires various hypothesis and suffers from different systematics. For instance, non-thermal pressure support, e.g. due to turbulence, cosmic rays, or feedback from active galactic nuclei, can lead to a substantial negative bias of $10\%-30\%$ in the hydrostatic estimates (e.g. \citealt{lau09,rasia12,nelson14}). Weak gravitational lensing does not assume dynamical equilibrium but is subject to projection effects, in particular from the triaxial shape of the dark matter halo, which can lead to overestimated masses by up to $\sim40\%$ (e.g \citealt{corless09,becker11,feroz12}).

The comparison of different mass estimators has been the subject of many studies, in particular between X-ray hydrostatic and lensing measurements. The topic regained interest with the tension found between the cosmological constraints derived by the Planck Collaboration from cluster counts \citep{bohringer14,planck24} and the cosmic microwave background \citep{planck13}. This discrepancy could possibly be resolved by a $\sim40\%$ hydrostatic mass bias (the Planck masses were derived from a Sunyaev-Zeldovich mass proxy calibrated with hydrostatic masses). Despite a significant amount of work focussing on this hydrostatic bias, no agreement has yet been reached regarding its level. Among the most recent studies, we can mention \cite{donahue14}, \cite{vdlinden14}, \cite{sereno15}, \cite{hoekstra15}, \cite{penna-lima16}, and \cite{sereno17} who found weak-lensing masses larger than the hydrostatic estimates by $\sim20\%-30\%$. On the other hand, \cite{gruen14}, \cite{israel14}, \cite{applegate16}, or \cite{smith16} obtained results consistent with a vanishing bias. Detailed comparisons between lensing and X-ray masses have also highlighted a radial (e.g. \citealt{mahdavi08,donahue14}), mass (e.g. \citealt{hoekstra15}), and redshift dependence (e.g. \citealt{sereno15,smith16}) of the hydrostatic bias. 

This variety of results suggests that, in addition to likely selection effects between the different samples (e.g. \citealt{sereno15}), the systematics of the X-ray and lensing techniques are not yet fully controlled. This seems to be particularly true for lensing estimates, as shown with the discrepant results found by different teams for common cluster subsamples (e.g. \citealt{okabe16}). Therefore, it is necessary to employ other estimators to further constrain a possible hydrostatic or lensing mass bias. An interesting alternative is to use the dynamics of cluster galaxies. For instance, \cite{rines15} compared the scaling relation between Planck masses and galaxy velocity dispersions, ruling out the large bias of $\sim40\%$ required by the Planck cosmological results. \cite{maughan16} applied the caustic method of \cite{diaferio97} to infer the mass of 16 massive clusters, finding also no evidence for an hydrostatic bias since their X-ray masses are actually $\sim20\%$ larger than the caustic estimates.

Our goal is to explore this route, alternative to the widely used lensing technique, to constrain the hydrostatic mass bias. To do so, we present here the dynamical analysis of ten galaxy clusters among the most luminous in X-rays at intermediate redshifts. To assess possible systematics in the dynamical mass estimates, we used three approaches: the virial theorem, the caustic method, and the Jeans equation. The latter is used here for the first time to derive statistical constraints on the hydrostatic bias; our results based on the caustic masses are also among the very first of their kind. Thanks to a detailed analysis of each cluster, made possible by the combination of wide field VIMOS spectroscopy and WFI photometry, we put a particular attention to the treatment of substructures. 

The paper is organised as follows. In Section 2, we introduce the cluster sample and describe the data sets used for this study. We outline the three dynamical mass estimators in Section 3 and we describe the substructure analysis in Section 4. The comparison of the mass estimators is presented in Section 5, before concluding in Section 6. We briefly discuss the clusters individually in the Appendix, which also contains some intermediate results of the dynamical and substructure analyses. Our results are scaled to a flat, $\Lambda$CDM cosmology with $\Omega_{m}=0.3$, $\Omega_{\Lambda}=0.7$ and a Hubble constant $H_{0}=70\,\mathrm{km\,s^{-1}\,Mpc^{-1}}$.

\section{Data: description and reduction}

\subsection{The DXL sample}

Thirteen medium distant X-ray galaxy clusters with luminosities $L_X^{\mathrm{bol}}=0.5-4\times10^{45}\mathrm{erg\,s^{-1}}$ were selected from the ROSAT-ESO Flux Limited X-ray survey (REFLEX, \citealt{bohringer01,bohringer04}) to form a statistically complete sample (DXL; see e.g. \citealt{zhang04b} for more details). It contains the clusters that are the most X-ray luminous in the redshift interval $z=0.26-0.31$ and it covers a mass range $M_{500}=0.48-1.1\times10^{15}\mathrm{M_{\odot}}$ \citep{zhang06}. Its volume completeness can be estimated with the well known selection function of the REFLEX survey \citep{bohringer04}. The most luminous clusters of the original DXL sample (ROSAT luminosities $L_X>10^{45}\mathrm{erg\,s^{-1}}$ in the (0.1-2.4 keV) band) were selected for a wide-field photometric and spectroscopic follow-ups, allowing for a comprehensive analysis of their properties; the DXL cluster RXCJ2011.3-5727 was also included despite its luminosity of $L_X=6.7\times10^{44}\mathrm{erg\,s^{-1}}$. Three clusters were added to the spectro-photometric campaign to cover a larger redshift range: RXCJ1206.2-0848 and RXCJ1347.5-114 at redshifts $z\sim0.45$, and RXCJ0225.9-4154 at redshift $z\sim0.22$. They were also selected according to their high X-ray luminosities, the former two being the most luminous clusters at high redshifts and the latter among the four most luminous in the redshift range $z\sim0.2-0.22$.

The X-ray properties of the DXL clusters are presented by \cite{zhang04b,zhang05,zhang06} and \cite{finoguenov05}, along with results on the calibration of scaling relations. \cite{braglia07,braglia09} and \cite{pierini08} studied the galaxy content of RXCJ0014.3-3022, RXCJ0232.2-4420, and RXCJ2308.3-0211, focusing on the star formation activity as function of environment and the properties of the diffuse stellar emission around the brightest cluster galaxies. \cite{ziparo12} and \cite{foex17} conducted a detailed analysis of the structure and dynamical state of RXCJ0225.9-4154, RXCJ0528.9-3927, RXCJ1131.9-1955, and RXCJ2308.3-0211.

\begin{table*}
\centering 
\begin{threeparttable}
\caption{Presentation of the sample and data sets.}
\label{table:sample}
\begin{tabular}{l c c c c c c c c c c}
\hline\hline\noalign{\smallskip}
Cluster & RA & Dec & z & WFI bands & $N_z$ &  Alternative name\\
   & (J2000) & (J2000) & & & &\\
\noalign{\smallskip}\hline\noalign{\smallskip}
RXCJ0014.3-3023 & 00:14:21.1 & -30:23:51.7 & $0.3058\pm0.0003$ & BVR & 430 & Abell 2744\\
RXCJ0225.9-4154 & 02:25:52.8 & -41:54:53.3 & $0.2189\pm0.0003$ & BVRI & 228 & Abell 3016\\
RXCJ0516.6-5430 & 05:16:37.3 & -54:31:30.4 & $0.2947\pm0.0005$ & BVRI & 108 & Abell S0520\\
RXCJ0528.9-3927 & 05:28:52.8 & -39:27:52.2 & $0.2837\pm0.0003$ & BVR & 219 & \\
RXCJ0658.5-5556 & 06:58:37.5 & -55:57:25.2 & $0.2965\pm0.0004$ & BVI & 251 & Bullet cluster\\
RXCJ1131.9-1955 & 11:31:55.3 & -19:53:33.4 & $0.3050\pm0.0003$ & BVR & 232 & Abell 1300\\
RXCJ1206.2-0848 & 12:06:13.4 & -08:47:58.9 & $0.4398\pm0.0002$ & BVRI & 562 & MACS J1206\\
RXCJ1347.5-1144 & 13:47:30.5 & -11:45:07.2 & $0.4503\pm0.0003$ & BVI & 212 & MACS J1347\\
RXCJ2011.3-5725 & 20:11:27.8 & -57:25:17.8 & $0.2780\pm0.0004$ & BV & 71 & \\
RXCJ2308.3-0211 & 23:08:21.8 & -02:11:19.7 & $0.2969\pm0.0003$ & BRV & 307 & Abell 2537\\
\noalign{\smallskip}\hline
\end{tabular}
    \begin{tablenotes}
      \small
      \item Columns: (1) Cluster name. (2,3) Equatorial coordinates of the surface density peak of spectroscopic members. (4) Spectroscopic redshift estimated in this work. (5) Available WFI pass bands. (6) Number of cluster members with a spectroscopic redshift. (7) Alternative name.
    \end{tablenotes}
  \end{threeparttable}
\end{table*}

\subsection{Optical spectroscopy}

Multi-object spectroscopy observations were carried out with the VIMOS instrument mounted at the Nasmyth focus B of VLT-UT3 {\it Melipal} at Paranal Observatory (ESO), Chile. The VIMOS provides an array of four identical CCDs separated by a $2'$ gap, each with a field of view (FOV) of $7\,\times\,8\,\mathrm{arcmin^{2}}$ and a $0.205''$ pixel resolution. Each cluster was observed at three different pointings, covering an extended region along the major axis of the cluster shape (as observed in X-rays) and overlapping over its core. Given the size of VIMOS' total FOV, the observations span a roughly rectangular area of $9\times5$ Mpc$^2$ at $z=0.3$ (see e.g. Fig. 1 in \citealt{braglia09}). The selection of targets was done only on the basis of their $I$-band luminosity, in order to avoid any colour bias for the comparative analysis of passive and star-forming galaxies.

The spectra were obtained with the low resolution LR-Blue grism, which provides a spectral coverage from 3700 to 6700 $\AA$, has a spectral resolution of about 200 for $1''$ width slits, and does not suffer from fringing. Moreover, it allows up to four slits in the direction of dispersion, thus significantly increasing the number of targets per mask. For a galaxy at $z\sim0.3$, this grism covers important spectral features such as the $\mathrm{[OII]},\,\mathrm{[OIII]},\,\mathrm{H}_{\beta},\,\mathrm{H}_{\delta}$ emission lines, the $\mathrm{CaII_{H+K}}$ absorption lines and the $4000\,\AA$ break, providing reliable redshift estimates up to $z\sim0.8$. The reduction of VIMOS spectra was performed in a standard way with the {\sc VIPGI} software \citep{scodeggio05}.

To estimate spectroscopic redshifts (hereafter $z_{\mathrm{spec}}$), we first ran the EZ tool \citep{garilli10} in blind mode, restricted to $z\in[0-2]$. In the second step, we reviewed the spectra by eye, and used VIPGI for the manual detection and fit of spectral features in case of probable misidentification. Since EZ does not provide redshift errors, we relied on repeated observations of the same object to estimate a typical uncertainty. We found an average value $\delta_{cz}\sim300\,\mathrm{km\,s^{-1}}$ with variations of $\sim50\,\mathrm{km\,s^{-1}}$ from cluster to cluster; velocity dispersions were corrected accordingly, following the prescription of \cite{danese80}.

Some of the spectroscopic data have already been used in previous works, e.g. for RXCJ0014 \citep{braglia09}, RXCJ1131 \citep{ziparo12}, or RXCJ1206 \citep{biviano13}. However, we re-reduced them for consistency with the other clusters. We completed our data set with redshifts available in the literature. For RXCJ1347, we also reduced additional VIMOS masks found in the ESO archive, from the programs 186.A-0798 (PI: P. Rosati) and 090.A-0958 (PI: A. Von der Linden). 

\subsection{Optical imaging}

The photometric follow-up of the clusters was conducted with the Wide Field Imager (WFI, \citealt{baade99}) mounted on the Cassegrin focus of the ESO/MPG 2.2-m telescope at La Silla, Chile. The WFI is a mosaic camera composed of $4\times2$ CCD chips, each made of $2048\times4096$ pixels with an angular resolution of $0.238"$/pixel. The total FOV is $34'\times33'$, which fully encompasses the region observed with VIMOS. The data reduction was performed with the THELI pipeline \citep{schirmer13}, which performs the basic pre-processing steps (bias subtraction, flat-fielding, background modelling and sky subtraction), and uses third party softwares for the astrometry ({\sc Scamp}, \citealt{bertin06}) and the co-addition of mosaic observations ({\sc SWarp}, \citealt{bertin10}). The photometry was made with {\sc SExtractor}. Stars, galaxies, and false detections were sorted according to their position in the magnitude-central flux diagram, their size with respect to that of the PSF, and their stellarity index (CLASS\_STAR parameter). Luminosities were estimated from the MAG\_BEST parameter, whereas colours were computed with MAG\_APER, measured in a fixed aperture of 3''.

The clusters were observed in at least three of the four (B,V,R,I) pass bands (only two for RXCJ2011.3-5725), allowing us to compute photometric redshifts (hereafter $z_{\mathrm{phot}}$). Our approach, which is described in details in \cite{foex17}, relies on the simple technique of the "k-nearest neighbour" fitting \citep{altman92}, performed in colour space. For each cluster, we divided the sample of $z_{\mathrm{spec}}$ into training and testing sets, and we assessed the accuracy of $z_{\mathrm{phot}}$ with the fraction of catastrophic errors $\eta=|z_{\mathrm{phot}}-z_{\mathrm{spec}}|/(1+z_{\mathrm{spec}})>0.15$, and the redshift accuracy $\sigma_{z}=1.48\times \mathrm{med}[|z_{\mathrm{phot}}-z_{\mathrm{spec}}|/(1+z_{\mathrm{spec}})]$ \citep{ilbert06}. For most clusters, we obtained $\eta$ and $\sigma_{z}$ within 0.05-0.1 (see top panel of Fig. \ref{fig:zphot}). The variability of these values between clusters is mostly due to the number of available $z_{\mathrm{spec}}$, i.e. a larger number implies a better sampling of the colour space, hence a better accuracy.

\subsection{Selection of cluster members}
\label{sec:cat}
\begin{figure}
\center
\includegraphics[width=6.5cm, angle=-90]{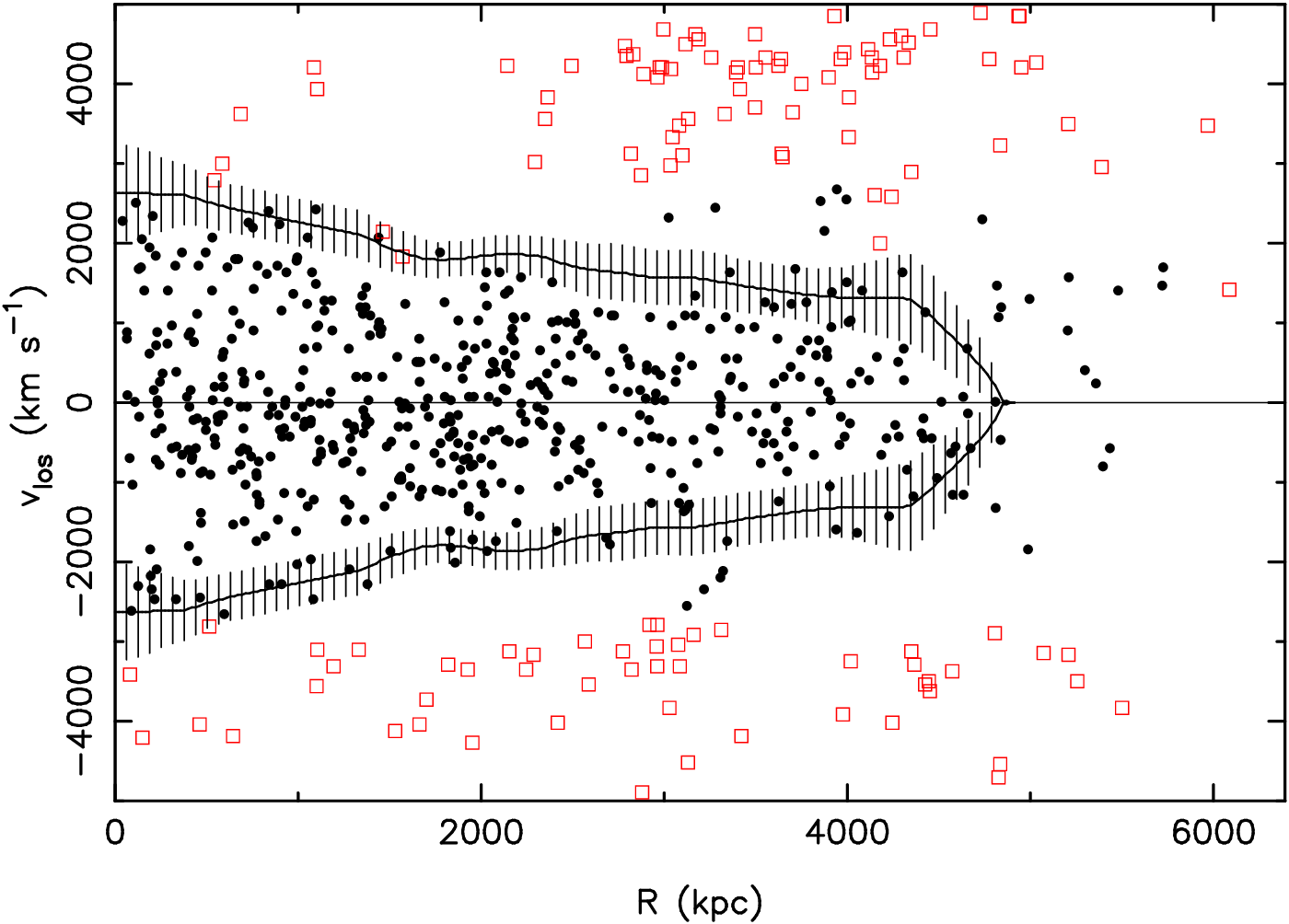}
\caption{Projected-phase space diagram of RXCJ1206. Black points are galaxies identified as cluster members with our method and red squares are field galaxies. The two curves show the caustic amplitude and its $1\sigma$ uncertainty (see Sect. \ref{sec:caustics}).}
\label{fig:PPS1206} 
\end{figure}

To select the cluster members from the spectroscopic data, we opted for an iterative $3\sigma$ clipping scheme, combined with an iterative radial binning in the projected-phase space (hereafter PPS). This method extends the approach originally proposed by \cite{yahil77}, by accounting for radial variations in velocity dispersion; more details of our implementation are given in \cite{foex17}. We used the spectroscopic members to compute the clusters' redshift with the robust biweight estimator of \cite{beers90}. The number of cluster members and the clusters' redshifts, along with their $1\sigma$ confidence interval obtained from bootstrapping, are given in Table \ref{table:sample}. We present in Figure \ref{fig:PPS1206} the location of galaxies in PPS for the cluster RXCJ1206 (see Fig. \ref{fig:PPS_all} for the other clusters).

\begin{figure}
\center
\includegraphics[width=6.5cm, angle=-90]{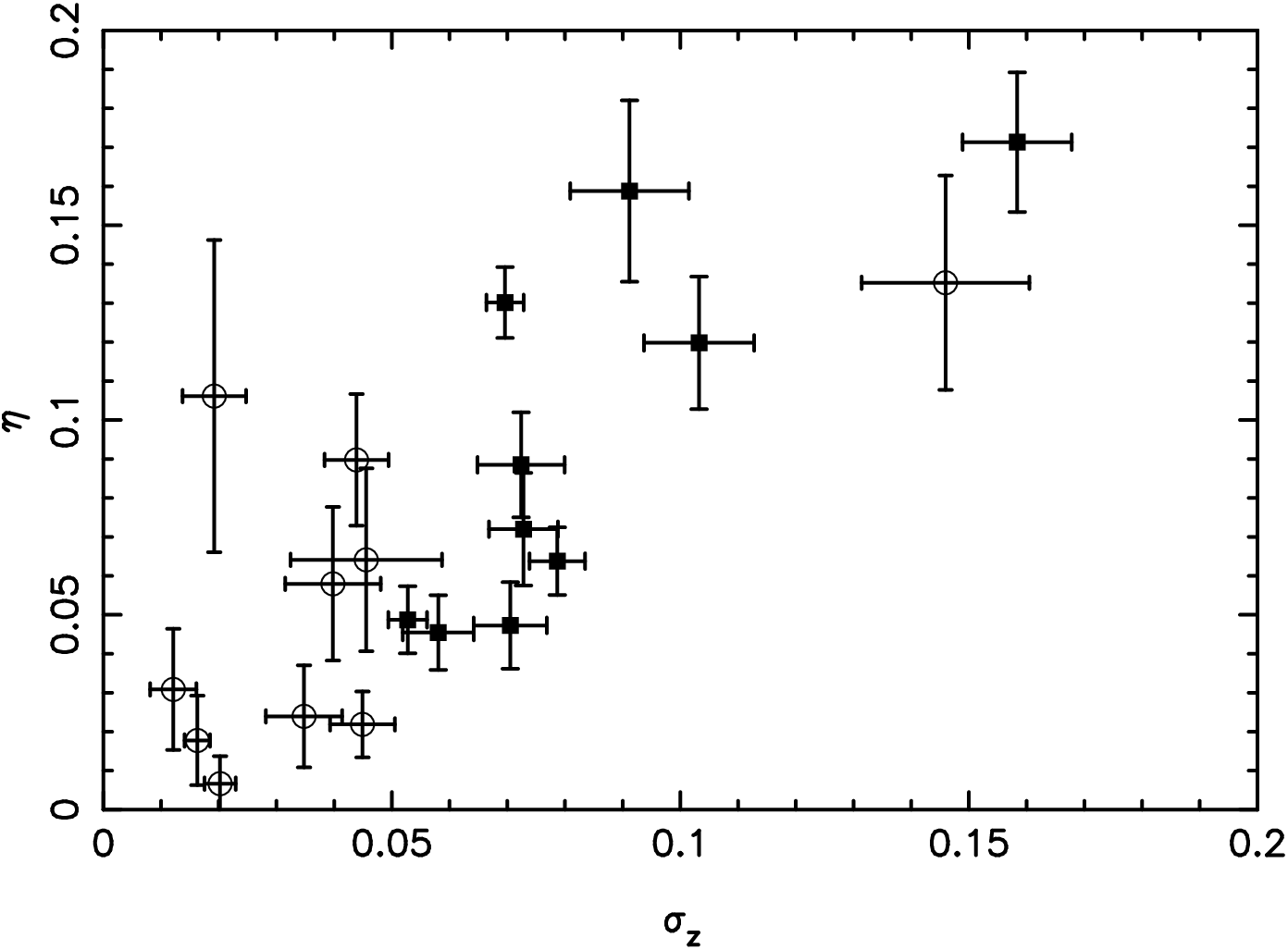}\\
\includegraphics[width=6.5cm, angle=-90]{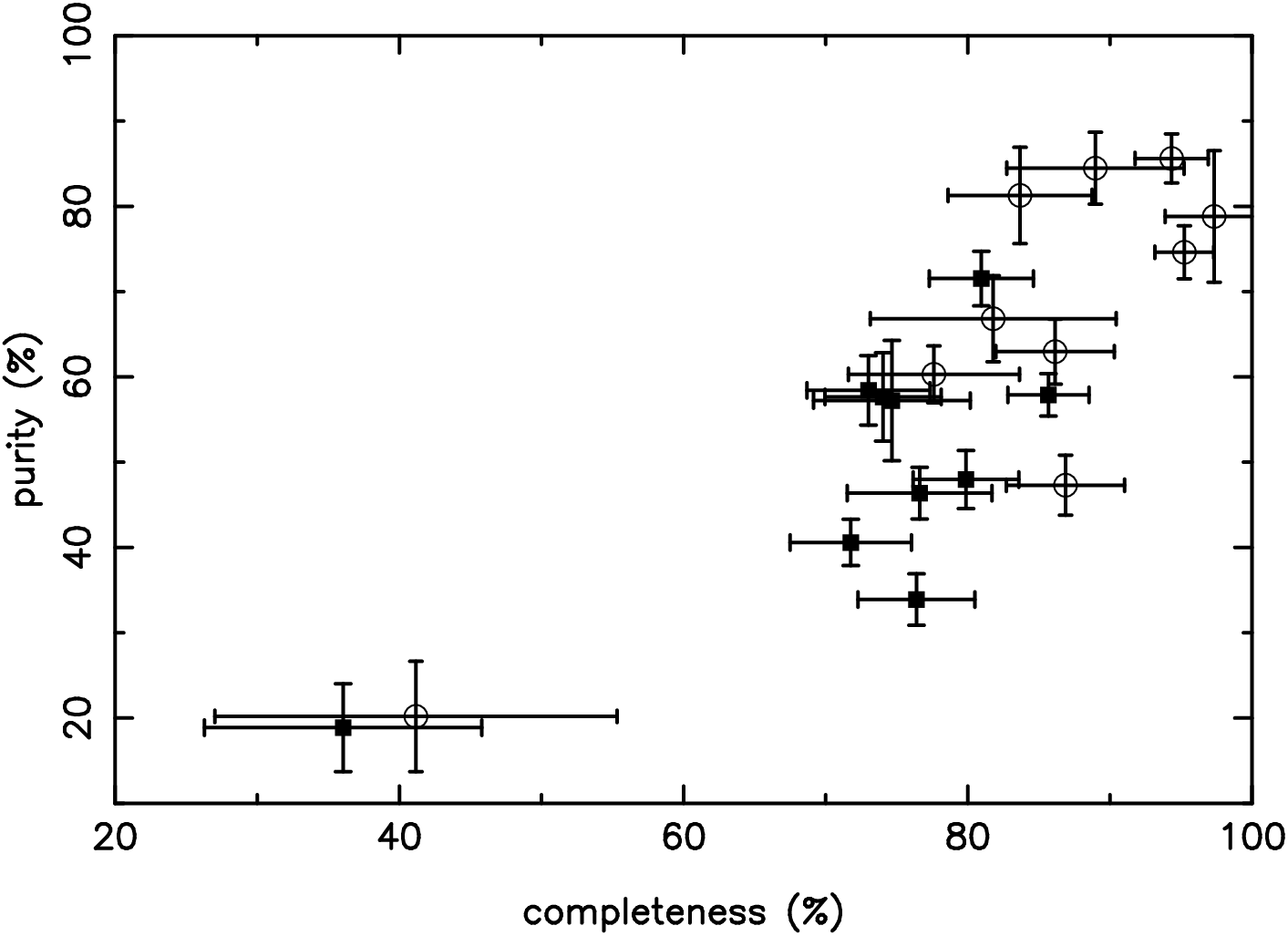}
\caption{Top panel: photometric redshift accuracy $\sigma_{z}$ versus fraction of catastrophic errors $\eta$. Bottom panel: completeness versus purity of the $z_{\mathrm{phot}}$ catalogues. In both panels, filled squares are for the whole population, whereas empty circles are for the red-sequence galaxies only. Error bars were estimated by repeating the measurements on 100 randomly-picked training/testing samples. Notice that the worst values are for RXCJ2011, which was observed only in two filters.}
\label{fig:zphot} 
\end{figure}

For the galaxies with only a $z_{\mathrm{phot}}$ estimate, we proceeded as follows. First, we discarded the galaxies with $|z_{\mathrm{phot}}-\mu_c|>\mathrm{max}[0.1,3\sigma_{\mathrm{c}}]$, where $\mu_c$ represents the mean photometric redshift of the spectroscopically confirmed members and $\sigma_{\mathrm{c}}$ its scatter. To improve the selection, we then removed galaxies having $\sigma_{\mathrm{z,spec}}>0.1$, where $\sigma_{\mathrm{z,spec}}$ is the dispersion in $z_{\mathrm{spec}}$ of the kNN ten nearest neighbours. These selection criteria lead to a typical completeness of $\sim70\%-80\%$ and purity of $\sim40\%-60\%$ (see bottom panel of Fig. \ref{fig:zphot}), values that were estimated by dividing the sample of $z_{\mathrm{spec}}$ into training and testing sets. For each testing set, the number of spectroscopic members surviving the $z_{\mathrm{phot}}$ selection criteria was used to defined the purity, as the proportion with respect to the total number of galaxies in the final sample, and the completeness, as the proportion with respect to the number of spectroscopic members present in the initial sample.

Finally, we merged for each cluster the spectroscopic and photometric catalogues, giving priority to the spectroscopic classification when possible. From these combined catalogues, we fitted the clusters' red sequence with a $2\sigma$ clipping method in the magnitude-colour diagram. The locus and scatter, $\sigma_{\mathrm{RS}}$, of the red sequence were used to divide the catalogues into two broad populations: the red-sequence galaxies, i.e. those having a colour residual within $3\sigma_{\mathrm{RS}}$, and the blue members. Additionally, the combined catalogues were cut to a limiting magnitude $m\le m^{*}+3$ in order to reduce the contamination by faint background galaxies. 

Assuming that the spectroscopic samples do not contain interlopers, the purity of the combined catalogues are typically increased by $\sim10\%$ as compared to the photometric catalogues. We can also note in Figure \ref{fig:zphot} that the completeness and purity of the red-sequence galaxies are $\sim10\%$ better than the values obtained for the full population. If we assume that the contamination by field galaxies is constant over the field of view, we can use the fraction of spectroscopic members in the combined catalogues to estimate the completeness of the spectroscopic survey. It decreases smoothly from $\sim0.5\pm0.1$ at small radii to $\sim0.3\pm0.1$ ($\sim0.1\pm0.1$) at $R_{200}$ ($1.5R_{200}$) for galaxies with magnitudes $m_{BCG}<m<m*+3$. Finally, we can mention the fraction of red-sequence galaxies within $R_{200}$. With average values of $\sim0.4\pm0.1$ for the photometric catalogues, $0.5\pm0.1$ for the combined catalogues, and $\sim0.6\pm0.1$ for the spectroscopic members, we can see that the spectroscopic data are only slightly biased towards the population of early-type galaxies.

\section{Dynamical masses}

\subsection{Jeans equation}
\label{sec:jeans}

Assuming stationarity, the absence of streaming motions, and spherical symmetry, the first velocity moment of the collisionless Boltzmann equation gives the Jeans equation: 
\begin{equation}
\frac{\mathrm{d}p(r)}{\mathrm{d}r}+2\frac{\beta(r)}{r}p(r)=-\nu(r)\frac{GM(r)}{r^2},
\end{equation}
where $p(r)=\nu(r)\sigma_r^2(r)$ is the radial dynamical pressure, $\nu(r)$ the space density of the tracer used to observe the system (i.e. galaxies), $\sigma_r(r)$ the radial velocity dispersion of the tracer, and $M(r)$ the total mass profile. The tracer's velocity anisotropy profile, $\beta(r)$, reads
\begin{equation}
\beta(r)=1-\frac{\sigma_\theta^2(r)}{\sigma_r^2(r)}.
\end{equation}
It varies from $\beta\rightarrow-\infty$ for circular orbits to $\beta=1$ for radial orbits; $\beta=0$ for an isotropic velocity field.

The Jeans equation is a linear first-order differential equation, which under the boundary condition $p(r)\rightarrow0$ when $r\rightarrow\infty$, has the general solution \citep{vandermarel94}:
\begin{equation}
p(r)=\int_r^\infty\exp\left[2\int_r^s\beta(t)\frac{\mathrm{d}t}{t}\right]\nu(s)\frac{GM(s)}{s^2}\mathrm{d}s.
\end{equation}
The tracer's density profile, $\nu(r)$, can be inferred from the observed surface density $\Sigma(R)$ via Abel de-projection, and a third equation links the projected dynamical pressure $P(R)=\Sigma(R)\sigma_P^2(R)$ (where $\sigma_P(R)$ is the observed line-of-sight velocity dispersion at projected radius $R$) to the dynamical pressure \citep{binney82}. However, since the number of unknowns exceeds the number of equations, one needs to make further assumptions to break the well-known degeneracy between the mass and anisotropy profiles (e.g. \citealt{merritt87,merrifield90,vandermarel00}).

\begin{figure}
\center
\includegraphics[width=6.5cm, angle=-90]{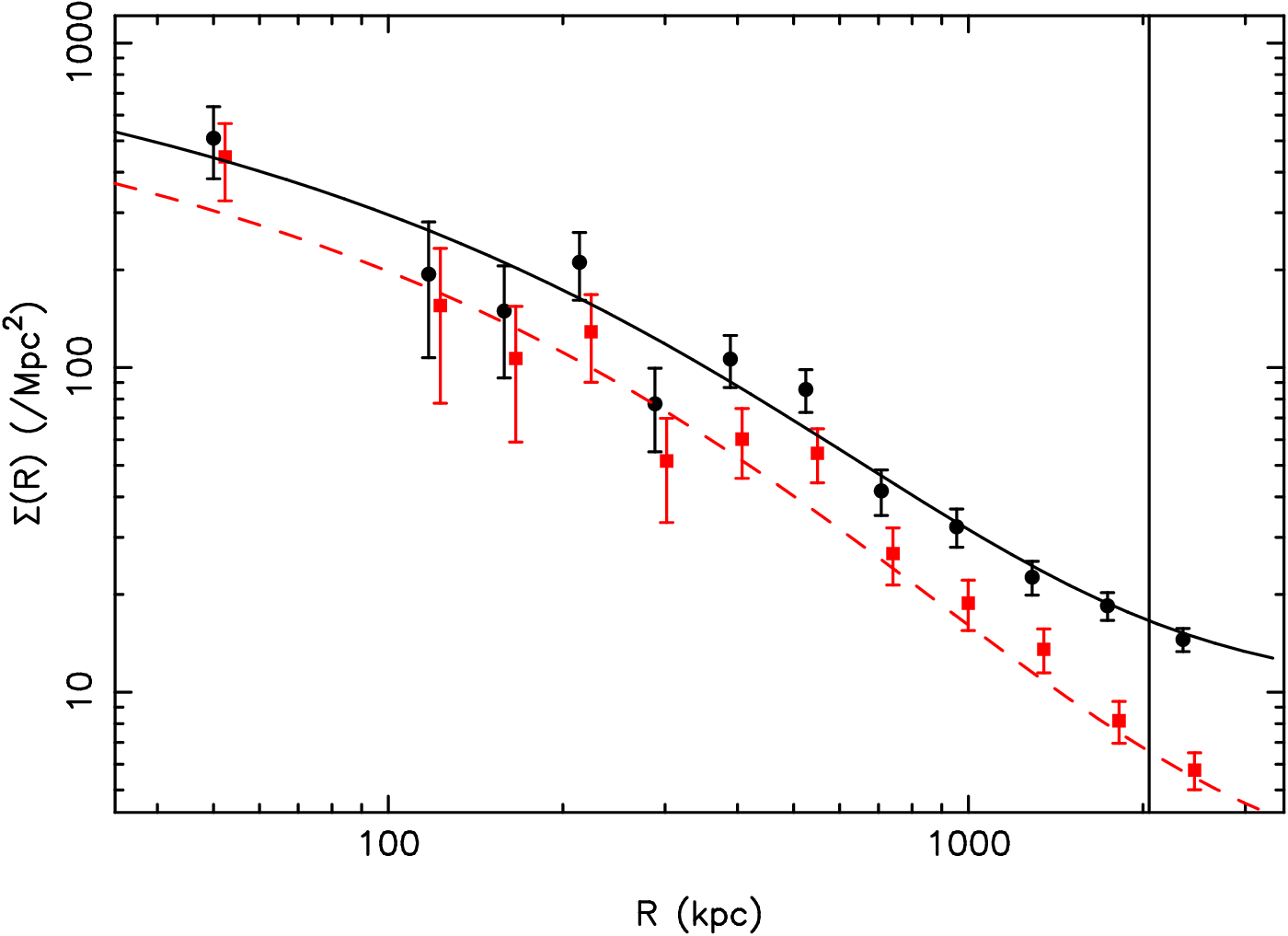}
\caption{Galaxy surface density profile of RXCJ1206 (black points). The black solid curve shows the best-fit NFW model plus a constant background. The red-dashed curve and red points are for the red-sequence galaxies. The vertical line shows $R_{200}$.}
\label{fig:surf_dens} 
\end{figure}

Here we have chosen to use a forward fitting approach, following closely the prescription given by \cite{mamposst}. The method consists in using a maximum likelihood estimator to find the best-fit parameters of analytical $M(r)$, $\nu(r)$, and $\beta(r)$ profiles. In our case, however, we chose to derive the density profile $\nu(r)$ prior to the dynamical analysis since our spectroscopic catalogues suffer from a radial-dependent completeness. We used the combined catalogues defined in \ref{sec:cat} to fit the observed surface density profiles $\Sigma(R)$ by the sum of a background contribution plus the projection of a spherical NFW \citep{navarro97} or King distribution; the model giving the smallest $\chi^2$ was chosen (an example is given in Fig. \ref{fig:surf_dens}; see Fig. \ref{fig:sigma_prof_all} for the full sample). With the characteristic radius $r_c$ of the tracer's density in hand, the best-fit parameters for $M(r)$ and $\beta(r)$ are obtained by minimising
\begin{equation}
-\ln\mathcal{L}=-\sum\left[\ln g(R_i,\upsilon_{z,i}|\bm{\theta})-\ln\Sigma(R_i|r_c)\right],
\end{equation}
where $\bm{\theta}$ is the vector containing the free parameters of the mass and anisotropy models.
Assuming a gaussian 3D velocity distribution, the galaxy density in PPS is given by \citep{mamposst}: 
\begin{equation}
g(R,\upsilon_z)=\sqrt{\frac{2}{\pi}}\int_R^\infty\frac{\nu(r)}{\sigma_z(R,r)}\exp\left[-\frac{\upsilon_z^2}{2\sigma_z^2(R,r)}\right]\frac{r\mathrm{d}r}{\sqrt{r^2-R^2}},
\end{equation}
with
\begin{equation}
\sigma_z^2(r,R)=\left[1-\beta(r)\left(\frac{R}{r}\right)^2\right]\sigma_r^2(r).
\end{equation}
The mass profile enters the above equation via the general solution of the dynamical pressure, $\sigma_r^2(r)=p(r)/\nu(r)$.

For simplicity, we restricted our analysis to the combination of a NFW mass profile with three different anisotropy models: a constant anisotropy (hereafter "C" model), the \cite{mamon05} model (hereafter "ML"),
\begin{equation}
\beta_{\mathrm{ML}}(r)=\frac{1}{2}\frac{r}{r+r_\beta},
\end{equation}
and the simplified \cite{tiret07} model (hereafter "T"),
\begin{equation}
\beta_{\mathrm{T}}(r)=\beta_\infty\frac{r}{r+r_s},
\end{equation}
where $r_s$ is the scale radius of the NFW mass profile. Since we do not assume that "light traces mass", the NFW scale radius can be different than the characteristic radius $r_c$ of the galaxy spatial distribution.

\begin{figure}
\center
\includegraphics[width=6.5cm, angle=-90]{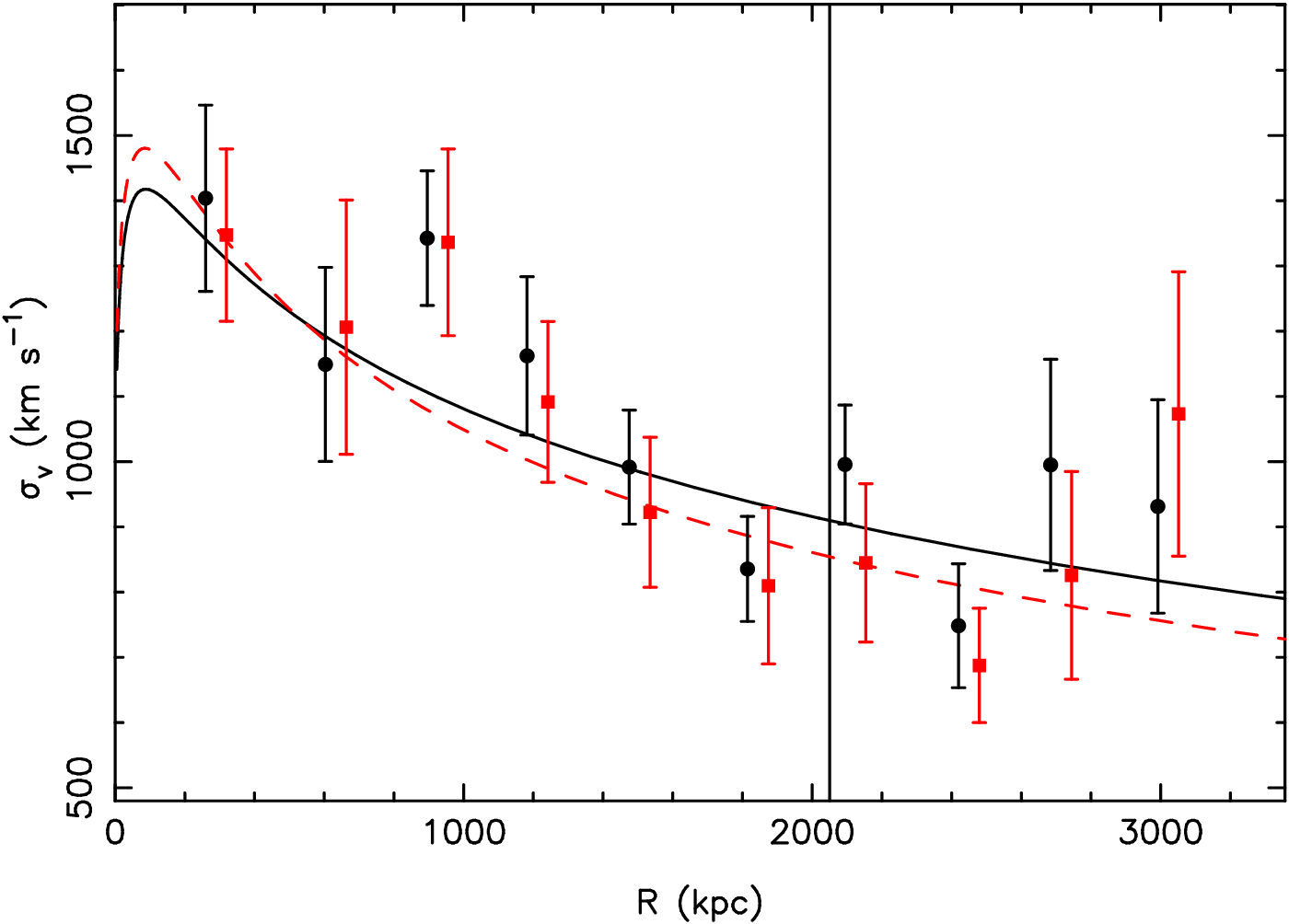}
\caption{Velocity dispersion profile of RXCJ1206. The black solid curve shows the Jeans prediction given the best-fit parameters for the whole population (black points). The red dashed curve is the solution obtained using the red-sequence galaxies only (red points). We note here that we did not fit these profiles. The vertical line marks $R_{200}$.}
\label{fig:dv_prof} 
\end{figure}

We performed the minimisation of $-\ln\mathcal{L}$ with the Powell algorithm (e.g. \citealt{NRbook}), and kept the combination $(M_{200},c_{200},\beta)$ that gives the smallest value. The parameter space was restricted to $c_{200}\in[2,20]$, $\beta_C\in[0,1]$ (C model), $r_\beta\in[0,5]$ Mpc (ML model), and $\beta_\infty\in[0,1]$ (T model). Only galaxies within the radius $R_{200}$ estimated from the virial theorem (see below) were used to compute the likelihood. The $1\sigma$ confidence intervals on the best-fit parameters were estimated by bootstrapping the catalogue of galaxies. We accounted for the uncertainties in $r_c$ by sampling its distribution, assumed to be gaussian, for each bootstrap realisation. In Figure \ref{fig:dv_prof}, we compare the projected velocity dispersion profile derived from the Jeans best-fit model to the observed profile for RXCJ1206 (see Fig. \ref{fig:vdisp_all} for the other clusters).

\subsection{Virial theorem}

Our second dynamical mass estimator is based on the scalar virial theorem (e.g. \citealt{limber60,heisler85,merritt88}):
\begin{equation}
\label{eq:M_V}
M_V=\frac{3\pi}{G}\sigma_{P}^{2}R_{PH}.
\end{equation}
The projected (line-of-sight) velocity dispersion, $\sigma_P$, and the projected harmonic mean radius, $R_{PH}$, are given by:
\begin{subequations}
\begin{align}
\sigma_P^2 &=\frac{\sum_{i}{\upsilon_i^2}}{N-1},\\
R_{PH} &=\frac{N(N-1)}{2\sum_{i>j}{R_{ij}^{-1}}},
\end{align}
\end{subequations}
with $N$ the number of galaxies, $\upsilon_i$ the rest-frame line-of-sight velocity, and $R_{ij}$ the angular-diameter distance between galaxy pairs. In practice, we used the robust biweight estimator of \cite{beers90} to evaluate $\sigma_P$. We used bootstrap realisations of the galaxy catalogue to estimate $1\sigma$ confidence intervals on the mass and velocity dispersion.

Similarly to the Jeans analysis, the virial theorem is based on the hypothesis of dynamical equilibrium and sphericity. Interestingly, its application does not require knowledge of the velocity field anisotropy since one always has $3\sigma_P^2=\sigma_r^2+\sigma_\phi^2+\sigma_\theta^2$ for a spherical system. However, the virial theorem relies on the additional assumptions that the galaxies have the same spatial and velocity distribution as dark matter particles, and that all galaxies have the same mass. The latter approximation can be justified by the lack of observational evidence of a strong luminosity/mass segregation in the galaxy population (e.g. \citealt{adami98b,biviano02}). However, dark matter haloes are well represented by a cuspy NFW profile, whereas cluster members have typically a cored King-like spatial distribution. Numerical simulations have also shown that a velocity bias exists between galaxies and dark matter particles (e.g. \citealt{berlind03,biviano06,munari13}). Despite these various approximations, the virial theorem has proven to be a good estimator (e.g. \citealt{biviano06}), thus it has been widely used owing to its simple and direct implementation.

The theorem, as expressed above, makes the implicit assumption that the system is isolated and observed in its entirety. However, galaxy clusters are embedded in dense environments, continuously accreting matter from their surroundings. Therefore, the virialised region of a cluster cannot be considered as an isolated system. Consequently, an additional correction must be applied to the virial theorem, the so-called surface pressure term (hereafter SPT; see e.g. \citealt{binney87,carlberg96}). Neglecting the dynamical pressure from the radial infall of matter leads to virial masses typically overestimated by $\sim15\%-20\%$ \citep{carlberg97,girardi98,biviano06}. When the spectroscopic survey does not fully cover the virialised region, a larger correction should be used, since the velocity dispersion is in general decreasing with radius. Under the hypothesis that the galaxies and dark matter share the same density profile, $\rho(r)$, the corrected virial mass can be expressed as \citep{girardi98}:
\begin{equation}
M_{CV}=M_V\left\{1-4\pi b^3\frac{\rho(b)}{\int_0^b4\pi r^2\rho(r)\mathrm{d}r}\left[\frac{\sigma_r(b)}{\sigma(<b)}\right]^2\right\},
\end{equation}
where the projected radius $b$ is the maximal extent of the spectroscopic observation (or the aperture radius within which one wishes to estimate the mass), and $\sigma(<b)$ is the integrated velocity dispersion within $b$.

For an isotropic velocity field of decreasing dispersion with radius, one has $\sigma^2(<r)=3\sigma_r^2(<r)\ge3\sigma_r^2(r)$. In that case, the term involving the velocity dispersion is at most equal to $1/3$ ($1$ for radial orbits, $0$ for circular ones), which is the value that we used to estimate $M_{CV}$. For the term involving the density $\rho(r)$, we used the characteristic radius $r_c$ obtained previously from the fit of the galaxy surface density profile by either a NFW or King model. For the clusters analysed in this work, we found SPT corrections in the range $20\%-30\%$ at $R_{200}$.

In \cite{foex17} we outlined a procedure to estimate the virial mass (or $M_{200}$) without prior knowledge of the virial radius (or $R_{200}$). We followed this approach to derive the radii $R_{200}$ used to select the galaxies from which the likelihood of the Jeans analysis was computed. Otherwise, we ran the virial theorem within a given aperture, e.g. $R_{200}$ from the Jeans analysis or the X-ray $R_{500}$, in order to make meaningful comparisons between the different mass estimators. 

\subsection{Caustics}
\label{sec:caustics}

In contrast with the two previous mass estimators, the caustic approach \citep{diaferio97,diaferio99} does not rely on the hypothesis of dynamical equilibrium. It is based on the simple consideration that particles gravitationally bound to a system cannot have a velocity larger than the escape velocity $\upsilon_{esc}^2(r)=-2\phi(r)$. For a realistic mass distribution, the gravitational potential $\phi(r)$ is an increasing function of radius. Therefore, the maximum allowed velocity decreases with radius, producing a trumpet-like pattern in PPS that is delimited by the upper and lower caustics. Accounting for projection effects, that is taking the average line-of-sight component of the escape velocity, one can obtain a simple relation that links the caustic amplitude, $\mathcal{A}(r)$, to the mass profile \citep{diaferio97}:
\begin{equation}
GM(r)=\int_0^r\mathcal{A}^2(s)\mathcal{F}_\beta(s)\mathrm{d}s.
\end{equation}
The projection along the line of sight is captured within the filling factor $\mathcal{F}_\beta(r)$:
\begin{equation}
\mathcal{F}_\beta(r)=-2\pi G\frac{\rho(r)r^2}{\phi(r)}\left(\frac{3-2\beta(r)}{1-\beta(r)}\right),
\end{equation}
where $\rho(r)$ and $\beta(r)$ are the mass density and anisotropy profiles, respectively.

\begin{figure}
\includegraphics[width=6.5cm, angle=-90]{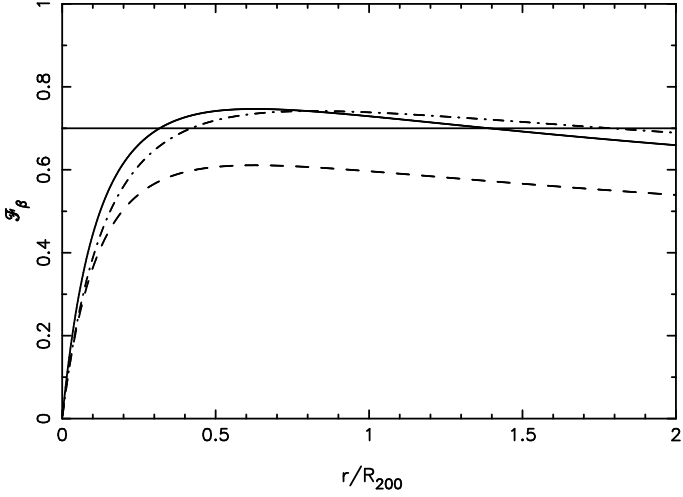}
\caption{Radial variation of the filling factor $\mathcal{F}_\beta(r)$ for a NFW mass profile with $M_{200}=10^{15}\mathrm{M_\odot}$ and $c_{200}=4$. The horizontal line shows the approximation used in this work, $\mathcal{F}_\beta(r)=0.7$. The dashed curve corresponds to a constant anisotropy $\beta=0$, the solid curve is for $\beta=0.5$, while the dot-dashed curve is obtained with a ML profile characterised by $r_\beta=0.2R_{200}$.}
\label{fig:Fbeta}
\end{figure}

Similarly to the Jeans analysis, a degeneracy between the mass and anisotropy profiles prevents estimating a cluster mass from spectroscopic observations alone. Therefore, it is customary to assume a constant value for the filling factor. \cite{serra11} showed that setting ${F}_\beta(r)=0.7$ provides a good approximation, allowing for the recovery of true masses within $10\%$ at radii larger than $\sim0.6R_{200}$. As shown in Figure \ref{fig:Fbeta} for a NFW mass profile with $M_{200}=10^{15}\mathrm{M_\odot}$ and $c_{200}=4$, this approximation is roughly equivalent to assuming a constant anisotropy $\beta\approx0.5$, or a $\beta_{\mathrm{ML}}(r)$ profile with $r_\beta\approx0.2R_{200}$. At small radii, the filling factor takes smaller values, hence caustic masses obtained with this method are typically overestimated. In principle one could use the $\beta(r)$ profile derived from the Jeans analysis to obtain a more robust mass profile (e.g. \citealt{biviano13}). However, we chose to set a constant $\mathcal{F}_\beta(r)=0.7$, to avoid introducing too much correlation between the two mass estimators, and to facilitate direct comparisons with previous works.   

Our implementation of the caustic method follows closely the prescription given by \cite{diaferio99}. The first step consists in estimating the galaxy density in PPS, after an appropriate rescaling of the coordinates (we chose a scaling factor $q=\sigma_\upsilon/\sigma_r=25$). The smooth density map, $f_q(r,\upsilon)$, was obtained with the DEDICA algorithm \citep{pisani96}, which uses an iterative and adaptive kernel estimator. In the second step, we determined the upper and lower caustics as the {\it loci} of pairs $(r,\upsilon)$ where $f_q(r,\upsilon)=\kappa$. To determine the adequate iso-density level $\kappa$ that sets the caustic amplitude, we can assume that the virial condition $\langle\upsilon_{\mathrm{esc}}^2\rangle_{R}=4\langle\upsilon^2\rangle_R$ holds within the central region, defined here by $R=R_{200}$ from the virial theorem. Since this condition remains valid with line-of-sight velocities under the hypothesis of isotropic orbits, the density level $\kappa$ is obtained as the root of the equation $\langle\upsilon_{\mathrm{esc}}^2\rangle_{\kappa,R}-4\sigma_P^2(<R)$=0, with $\langle\upsilon_{\mathrm{esc}}^2\rangle_{\kappa,R}=\int_0^R\mathcal{A}_\kappa^2(r)\varphi(r)\mathrm{d}r/\int_0^R\varphi(r)\mathrm{d}r$ and $\varphi(r)=\int f_q(r,\upsilon)\mathrm{d}\upsilon$.

\begin{figure}
\center
\includegraphics[width=6.5cm, angle=-90]{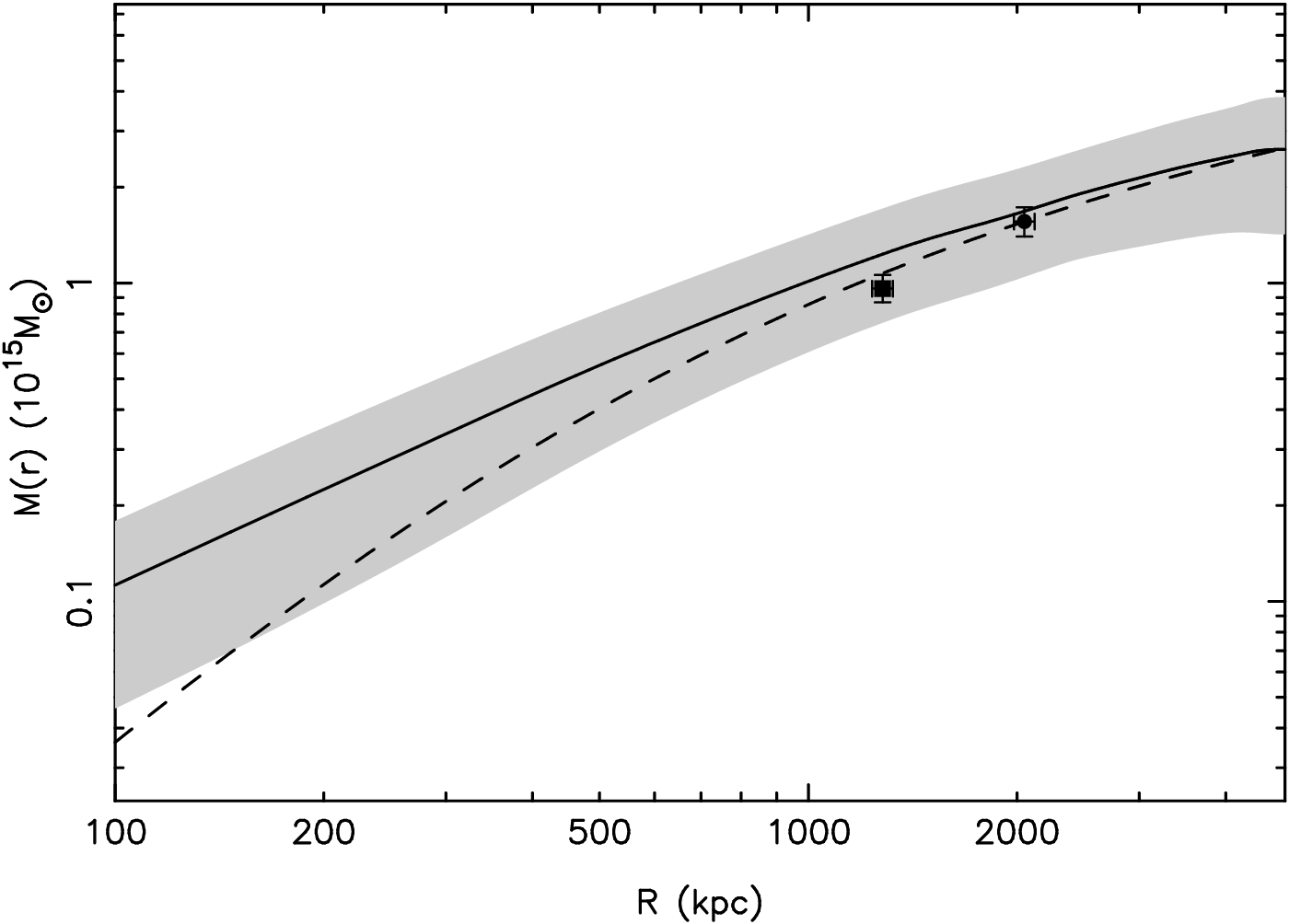}
\caption{Caustic mass profile of RXCJ1206 (the shaded area delimits the $1\sigma$ error). The dashed curve corresponds to the NFW mass profile derived from the Jeans analysis; it passes through the point $(R_{200},M_{200})$. The second point, at a smaller radius, shows the couple $(R_{500},M_{500})$ derived from the X-ray analysis.}
\label{fig:Mc} 
\end{figure}

At each radius, the caustic amplitude was set to $\mathcal{A}(r)=\mathrm{min}[|\mathcal{V}^-|,\mathcal{V}^+]$, where $\mathcal{V}^{\pm}$ are the lower and upper caustics. This approach, as compared to taking their average value, reduces the contamination by interlopers, and limits the impact of massive high-velocity substructures. The last step of the algorithm consists in computing the logarithmic derivative $\mathrm{d}\ln{\mathcal{A}}/\mathrm{d}\ln{r}$, which should be $\lesssim1/4$ for a typical cluster. Following the prescription by \cite{serra11}, we replaced $\mathcal{A}(r)$ with a new value yielding $\mathrm{d}\ln{\mathcal{A}}/\mathrm{d}\ln{r}=1/4$ whenever the original derivative was superior than two (e.g. due to interlopers or substructures). Finally, the error on the caustic amplitude was estimated as $\delta\mathcal{A}(r)/\mathcal{A}(r)=\kappa/\mathrm{max}[f_q(r,\upsilon)]$, where $\mathrm{max}[f_q(r,\upsilon)]$ is the maximum density along the $\upsilon$-axis at fixed $r$ \citep{diaferio99}. \cite{serra11} found that this recipe gives a $50\%$ confidence level on the true amplitude (see their Fig. 16). Therefore, we multiplied it by 1.4 to get a $1\sigma$ error before propagating it to the mass profile. The Figure \ref{fig:Mc} presents the caustic mass profile of RXCJ1206 (see Fig. \ref{fig:Mr_all} for the other clusters).

%%%%%%%%%%%%%%%%%%%%%%%%%%%%%%%%
%                          --------- mass table -----------
%%%%%%%%%%%%%%%%%%%%%%%%%%%%%%%%

\begin{table*}
\centering 
\begin{threeparttable}
\caption{Dynamical mass estimates.}
\label{table:mass}
\begin{tabular}{l c c c c c c c c}
\hline\hline\noalign{\smallskip}
Cluster & $M_{200}$ & $c_{200}$ & $R_{200}$ & $A_\beta$ & $r_c$ & $M_V(R_{200})$ & $\sigma_P(R_{200})$ & $M_c(R_{200})$\\
   & $(10^{15}\mathrm{M_\odot})$ & & (Mpc) & & (Mpc) & $(10^{15}\mathrm{M_\odot})$ & $(\mathrm{km\,s^{-1}})$ & $(10^{15}\mathrm{M_\odot})$\\
\noalign{\smallskip}\hline\noalign{\smallskip}
RXCJ0014 & $2.99_{-0.30}^{+0.32}$ & $9.0_{-1.2}^{+1.9}$ & $2.68_{-0.09}^{+0.09}$ & $0.8_{-0.1}^{+0.1}$ (T) & $0.41\pm0.06$ (K) & $3.35_{-0.23}^{+0.25}$ & $1472_{-52}^{+52}$ & $2.46\pm1.17$ \\[3pt]
RXCJ0225 & $0.99_{-0.21}^{+0.20}$ & $<3.8$ & $1.91_{-0.14}^{+0.12}$ & $>0.73$ (ML) & $0.69\pm0.34$ (K) & $1.38_{-0.14}^{+0.14}$ & $928_{-48}^{+49}$ & $1.02\pm0.45$ \\[3pt]
RXCJ0516 & $1.57_{-0.28}^{+0.31}$ & $<2.7$ & $2.17_{-0.14}^{+0.13}$ & $>0.56$ (ML) & $1.25\pm0.33$ (N) & $2.02_{-0.31}^{+0.33}$ & $1223_{-82}^{+80}$ & $1.86\pm0.94$ \\[3pt]
RXCJ0528 & $0.90_{-0.13}^{+0.14}$ & $<2.0$ & $1.81_{-0.09}^{+0.09}$ & $<0.05$ (C) & $0.75\pm0.15$ (K) & $1.00_{-0.11}^{+0.13}$ & $899_{-53}^{+50}$ & $0.83\pm0.32$ \\[3pt]
RXCJ0658 & $1.98_{-0.28}^{+0.30}$ & $<4.3$ & $2.34_{-0.11}^{+0.11}$ & $0.3_{-0.1}^{+0.1}$ (C)& $0.44\pm0.16$ (N) & $1.99_{-0.20}^{+0.20}$ & $1154_{-55}^{+56}$ & $1.47\pm0.31$ \\[3pt]
RXCJ1131 & $1.51_{-0.19}^{+0.22}$ & $<2.0$ & $2.13_{-0.09}^{+0.10}$ & $<0.05$ (T) & $0.84\pm0.25$ (N) & $1.47_{-0.16}^{+0.17}$ & $1079_{-53}^{+50}$ & $1.41\pm0.62$ \\[3pt]
RXCJ1206 & $1.56_{-0.16}^{+0.17}$ & $5.0_{-1.5}^{+2.1}$ & $2.05_{-0.07}^{+0.07}$ & $0.11_{-0.08}^{+0.28}$ (ML) & $0.39\pm0.11$ (N) & $2.07_{-0.18}^{+0.18}$ & $1150_{-50}^{+49}$ & $1.67\pm0.63$ \\[3pt]
RXCJ1347 & $0.61_{-0.11}^{+0.15}$ & $<2.1$ & $1.49_{-0.09}^{+0.11}$ & $<0.1$ (T) & $1.28\pm0.46$ (N) & $0.57_{-0.09}^{+0.09}$ & $810_{-61}^{+62}$ & $0.76\pm0.26$ \\[3pt]
RXCJ2011 & $0.33_{-0.08}^{+0.11}$ & $<7.6$ & $1.29_{-0.12}^{+0.13}$ & $<0.22$ (ML) & $1.00\pm0.46$ (N) & $0.33_{-0.07}^{+0.08}$ & $756_{-76}^{+71}$ & $0.48\pm0.13$  \\[3pt]
RXCJ2308 & $1.84_{-0.19}^{+0.22}$ & $6.8_{-2.1}^{+3.1}$ & $2.29_{-0.08}^{+0.09}$ & $0.05_{-0.03}^{+0.13}$ (ML) & $0.39\pm0.08$ (N) & $2.42_{-0.21}^{+0.22}$ & $1161_{-51}^{+50}$ & $1.90\pm0.44$\\
\noalign{\smallskip}\hline
\end{tabular}
    \begin{tablenotes}
      \small
      \item Columns: (1) Cluster name. (2,3) NFW best-fit parameters derived from the Jeans analysis. (4) Corresponding NFW radius $R_{200}$. (5) Best-fit parameter of the $\beta$ model, i.e. $\beta_C$ for the C model, $\beta_\infty$ for the T model, and $r_\beta$ for the ML model (in units of Mpc for the latter). (6) Characteristic radius $r_c$ of the galaxy density profile, obtained from the fit of the surface density profile by either a King (K) or NFW (N) profile. (7) Dynamical mass derived with the virial theorem, applied within the Jeans $R_{200}$. (8) Projected velocity dispersion within the Jeans $R_{200}$. (9) Caustic mass within the Jeans $R_{200}$. Note: upper (lower) limits are given when the $1\sigma$ confidence interval of a parameter reaches the lower (upper) range of its allowed values.
    \end{tablenotes}
  \end{threeparttable}
\end{table*}

%%%%%%%%%%%%%%%%%%%%%%%%%%%%%%%%

\section{Substructures}

The strongest prerequisite for obtaining unbiased masses with the Jeans analysis and the virial theorem is to have a tracer that reached dynamical equilibrium. High-velocity substructures associated with mergers disturb the cluster's velocity field and increase its overall velocity dispersion, thus leading to overestimated masses. The caustic amplitude in PPS depends also, though indirectly, on the galaxy velocity dispersion $\sigma_P^2(<R)$. Thus we can expect the caustic estimator to fail in some cases, for instance when a major merger is taking place \citep{diaferio99}. Therefore, we need to quantify the degree of relaxation of the clusters, in order to assess the robustness of their mass estimates. To do so, we introduce in the following two substructure indicators: a photometric value that is based on the galaxy surface density excess with respect to its best-fit model, and a dynamical indicator that quantifies local deviations of the velocity distribution with respect to the overall dynamics. We also outline our procedure for the identification and removal of individual substructures.

\subsection{Galaxy surface density}
\label{sec:delta}

We constructed galaxy surface density maps, $\Sigma(x,y)$, from the photometric plus spectroscopic catalogues of cluster members. After smoothing with a gaussian kernel of width 100 kpc, the maps were fitted with a two-dimensional elliptical King profile:
\begin{subequations}
\begin{align}
& \Sigma_\mathrm{K}(x,y)=\frac{\Sigma_0}{1+(r/r_c)^2}+b,\\
& r^2=(x\cos\phi+y\sin\phi)^2+(y\cos\phi-x\sin\phi)^2/(1-e)^2.
\end{align}
\end{subequations}
The ellipticity and position angle $(e,\phi)$ describe the shape of the cluster. The centre was fixed at the highest density peak, corresponding to the origin of the coordinates. $r_c$ is the core radius, $\Sigma_0$ the central surface density, and $b\ge0$ a constant background contribution. Defining the residual $\delta_{i,j}=\Sigma(x_i,y_j)-\Sigma_\mathrm{K}(x_i,y_j)$ between the measured surface density and the best-fit model, we computed the photometric substructure indicator
\begin{equation}
\Delta=\frac{\sum_{i,j}\mathrm{max}[0,\delta_{i,j}]}{\sum_{i,j}\Sigma(x_i,y_j)}.
\end{equation}
The sum was limited to pixels within $R_{200}$ and we only considered positive excess in order to focus on deviations produced by the largest substructures. Uncertainties on $\Delta$ were obtained by propagating the Poisson noise of the surface density maps. Assuming that the number of galaxies in a substructure scales linearly with its mass, as does the richness of a cluster with its total mass, then $\Delta$ provides a crude estimate of the mass fraction contained in substructures.

\cite{weissmann13} analysed the X-ray morphology of 80 galaxy clusters with the classical power ratio $P3/P0_{max}$ and centre shifts $\omega_c$. They also introduced the maximum power ratio $P3/P0_{max}$, corresponding to the peak of the ratio profile within $0.3-1\,R_{500}$. Seven of our clusters are in their sample, therefore we can compare our photometric substructure indicator with their X-ray morphology estimators. The results are presented in Figure \ref{fig:Xsub}. The Pearson coefficients $\rho(\Delta,P3/P0_{max})=0.89$ and $\rho(\Delta,\omega_c)=0.83$ indicate a good correlation between the X-ray and optical morphologies.

\subsection{Dressler-Shectman test}
\label{sec:DS}

Our second indicator is based on the Dressler \& Shectman test (hereafter DS test, \citealt{dressler88}), which combines spatial and line-of-sight velocity information. For each of the $N$ galaxies in the spectroscopic sample, the local velocity, $\langle\upsilon\rangle_{\mathrm{loc}}$, and projected velocity dispersion, $\sigma_{\mathrm{loc}}$, are estimated with the $n_{NN}$ nearest neighbours (in projected distance). We used $n_{NN}=\sqrt N$, since it provides a better sensitivity to significant substructures while being less affected by Poisson noise (e.g. \citealt{silverman86}). The dynamical deviation of the $i-$th galaxy is quantified as:
\begin{equation}
\delta_i^2=\frac{n_{NN}+1}{\sigma_P^2}\left[(\langle\upsilon\rangle_{\mathrm{loc},i}-\langle\upsilon\rangle)^2+(\sigma_{\mathrm{loc},i}-\sigma_P)^2\right],
\end{equation}
where $\langle\upsilon\rangle$ and $\sigma_P$ are the global values, obtained with the $N$ galaxies. As pointed out by \cite{pinkney96}, gradients in the velocity dispersion profile can produce false positive detections of substructures. Therefore, we used a radial-dependent $\sigma_P$ (best-fit power law of the observed profile) as the 'global' value against which $\sigma_{\mathrm{loc}}$ is compared.

For each cluster, we ran the DS test on $10^4$ random realisations of the galaxy distribution, where positions were kept fixed but velocities shuffled, in order to erase any correlation between them. We used the resulting probability distribution of the deviation $\delta$ to specify the threshold $\delta_{min}$ such as $P(\delta>\delta_{min})=0.05$. Above this value, the local dynamics of a galaxy differs significantly from that of the cluster, thus it is likely to be part of a substructure. Finally, we defined our dynamical substructure indicator, $f_{\mathrm{DS}}$, to be the fraction of cluster members satisfying this selection criterion, i.e. $f_{\mathrm{DS}}=N(\delta_i>\delta_{min})/N$. As for the photometric indicator, we limited the analysis within $R_{200}$; uncertainties on $f_{\mathrm{DS}}$ were obtained assuming Poisson statistics. 

The values of the two substructure indicators are given in Table \ref{table:sub} (see also Fig. \ref{fig:PDsub}). A Pearson coefficient of $\rho(\Delta,f_{\mathrm{DS}})=0.69$ indicates a moderate correlation. Since both indicators are based on relative galaxy counts, we used their average value as a third indicator, hereafter $\Delta f$. All the clusters show a rather high degree of substructure with $\Delta>0.15$. RXCJ0014 presents the largest value for both indicators, while RXCJ1347 appears to be the most regular cluster in the sample.

\begin{table}[t]
\centering 
\begin{threeparttable}
\caption{Substructure indicators.}
\label{table:sub}
\begin{tabular}{l c c c}
\hline\hline\noalign{\smallskip}
Cluster & $\Delta$ & $f_{\mathrm{DS}}$ & $\Delta f$\\
\noalign{\smallskip}\hline\noalign{\smallskip}
RXCJ0014 & $0.33\pm0.01$ & $0.33\pm0.04$ & $0.33\pm0.02$\\
RXCJ0225 & $0.23\pm0.01$ & $0.14\pm0.02$ & $0.19\pm0.01$\\
RXCJ0516 & $0.28\pm0.01$ & $0.08\pm0.02$ & $0.18\pm0.01$\\
RXCJ0528 & $0.22\pm0.01$ & $0.03\pm0.01$ & $0.12\pm0.01$\\
RXCJ0658 & $0.23\pm0.01$ & $0.07\pm0.02$ & $0.15\pm0.01$\\
RXCJ1131 & $0.26\pm0.01$ & $0.17\pm0.03$ & $0.21\pm0.01$\\
RXCJ1206 & $0.21\pm0.01$ & $0.12\pm0.02$ & $0.16\pm0.01$\\
RXCJ1347 & $0.16\pm0.01$ & $0.03\pm0.01$ & $0.10\pm0.01$\\
RXCJ2011 & $0.19\pm0.03$ & $0.12\pm0.03$ & $0.15\pm0.02$\\
RXCJ2308 & $0.20\pm0.01$ & $0.16\pm0.02$ & $0.18\pm0.01$\\
\noalign{\smallskip}\hline
\end{tabular}
    \begin{tablenotes}
      \small
      \item Columns: (1) Cluster name. (2) Photometric substructure indicator. (3) Dynamical substructure indicator. (4) Dynamical substructure indicator. Their values were estimated within $R_{200}$.
    \end{tablenotes}
  \end{threeparttable}
\end{table}

\subsection{Identification and removal of substructures}
\label{sec:sub}

To identify substructures, we first made use of the DS test results to locate groups of galaxies having $\delta>\delta_{min}$. Following \cite{foex17}, we ran the DS test using separately the deviations in the local mean velocity and those in the local velocity dispersion. For each group, we computed its rest-frame velocity, velocity dispersion, average position, and projected shape (using the moment approach, e.g. \citealt{carter80}). These values were used as initial guess for the 3D version of the Kaye's mixture model algorithm (KMM, \citealt{ashman94}), which is a typical iterative expectation-maximisation algorithm for the modelling of a mixture of Gaussian distributions. The KMM algorithm was developed initially to separate different components in velocity space. However, it is straightforward to include spatial information by using multivariate distributions (more details are given in \citealt{foex17}). Starting from the initial guess describing the parameters of each component, the KMM algorithm estimates the probability that a given galaxy belongs to a given component. It then partitions the galaxies, and re-estimates the parameters of each component before the next iteration. Once the algorithm has converged, it provides the list of galaxies associated with the cluster main body and with each additional substructure. 

We also used the galaxy surface density maps to identify substructures not detected with the DS test. Here we focused on the spatial distribution of red-sequence galaxies, in particular the bright ones with $m<m^*+1$, since the $z_{phot}$ accuracy, completeness, and purity are higher for this galaxy population (see Fig. \ref{fig:zphot}). We located the largest over-density peaks, and defined the substructures as elliptical regions around these peaks. 

\begin{figure}
\includegraphics[width=8cm, angle=-90]{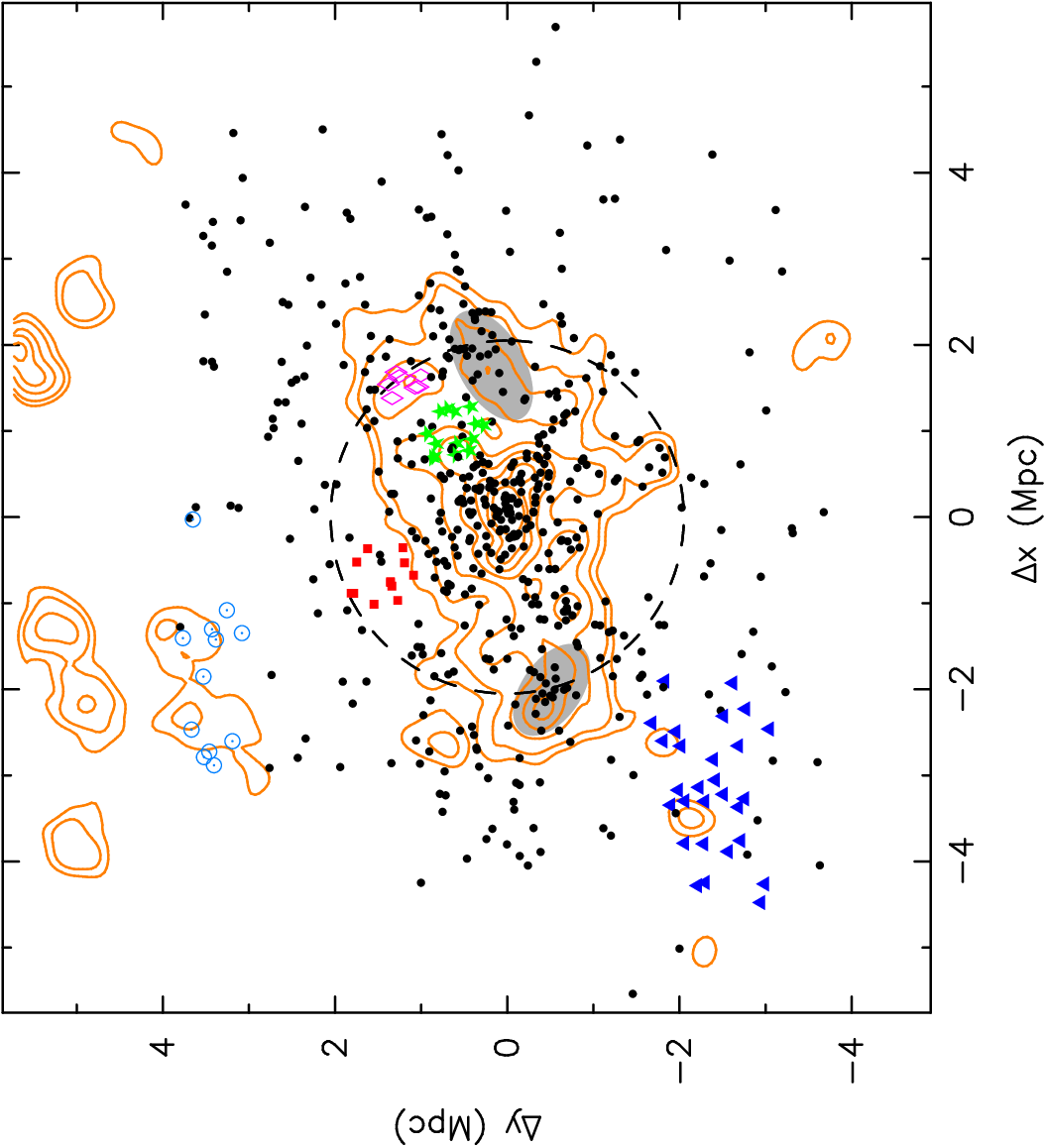}
\caption{Structure of RXCJ1206. The different symbols show the KMM partitions (the black dots are for the main body of the cluster). The orange contours trace the surface density of the red-sequence galaxies; they start at $5\sigma$ above the mean background density. The dashed circle has a radius $R_{200}$. The light-grey ellipses trace the additional substructures not detected by the DS test; the galaxies inside them were also excluded for the updated dynamical analysis (see Sect. \ref{sec:newmass}).}
\label{fig:KMM1206}
\end{figure}

Finally, we combined the two previous steps to select the galaxies that most likely trace the relaxed component of a cluster. We kept the main KMM partition and discarded the galaxies located within the additional substructures defined from the galaxy surface density maps. An example is provided in Figure \ref{fig:KMM1206} (see Figs. \ref{fig:KMM_1}-\ref{fig:KMM_3} for the other clusters). It shows the isopleths of the galaxy surface density (combined catalogues), the spatial distribution of the spectroscopically confirmed cluster members, highlighting those associated with a KMM partition, and the ellipses englobing the extra substructure candidates.

To summarise, we excluded the galaxies which are part of substructures identified due to their specific dynamics or as over-densities in the galaxy surface density. The remaining spectroscopic members were then used to estimate the dynamical mass of the relaxed component of the clusters (see Section 5.5).

\section{Comparison of the different mass estimators}

The three dynamical estimators make use of the same data set, however, they rely on different hypothesis and simplifications. Therefore, we start by looking for differences in their results. We constrain the scaling relation between mass and velocity dispersions, then we examine how the dynamical estimators compare with the X-ray hydrostatic masses. Finally, we investigate how the dynamical measurements are affected by using only the red-sequence galaxies, or by excluding galaxies located in substructures. The results are summarised in Table \ref{table:comp}.

\subsection{Dynamical masses}

As mentioned previously, we used the radius $R_{200}$ derived form the Jeans analysis as the aperture within which we applied the virial theorem, giving the mass $M_V(R_{200})$. We did not fit the caustic mass profile with an NFW model, or directly estimate a non-parametric $M_{200}$. Instead, we simply considered the caustic mass at this radius, labelled $M_c(R_{200})$ hereafter. While this approach introduces a correlation between the different estimators, it has the advantage of comparing masses within the same physical radius. For each pair of estimators, we estimated the arithmetic mean ratio, its standard deviation, and the geometric mean ratio, which is equivalent to the best-fit intercept of a linear regression of slope 1 in log-log space. The latter has been used for instance by \cite{smith16} to infer the hydrostatic mass bias with respect to lensing masses. We also estimated the logarithmic orthogonal scatter, $\sigma_\perp$, of the points around this best-fit linear regression. Errors on each parameter were obtained from bootstrap realisations of the cluster catalogue. Given the limited number of objects and the small mass range, we did not try to fit a slope for the regression in log-log space. The individual cluster masses are given in Table \ref{table:mass} and the relation between the different estimators are summarised in Table \ref{table:comp}.

\begin{figure}
\includegraphics[width=6.5cm, angle=-90]{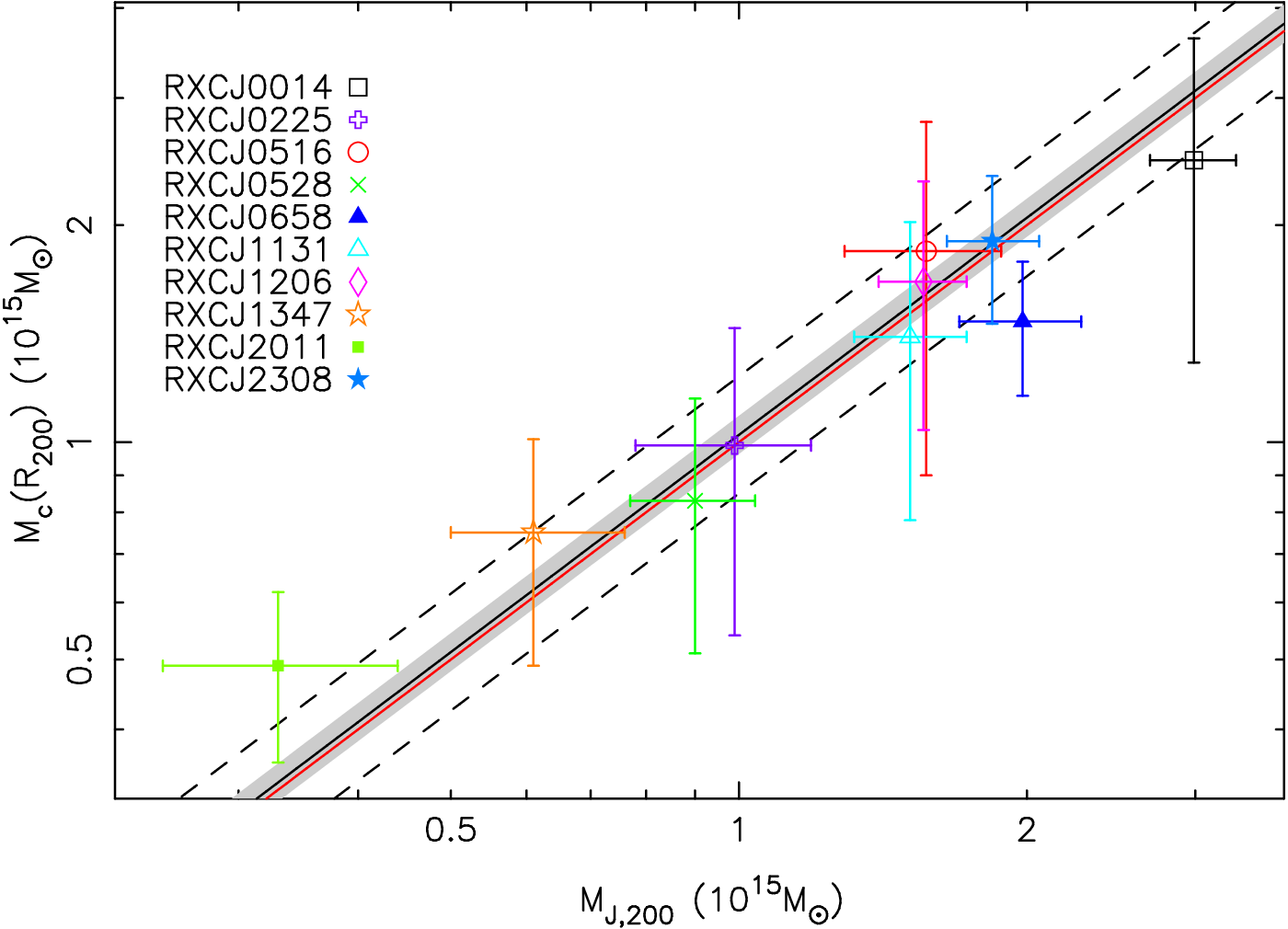}\\[4pt]
\includegraphics[width=6.5cm, angle=-90]{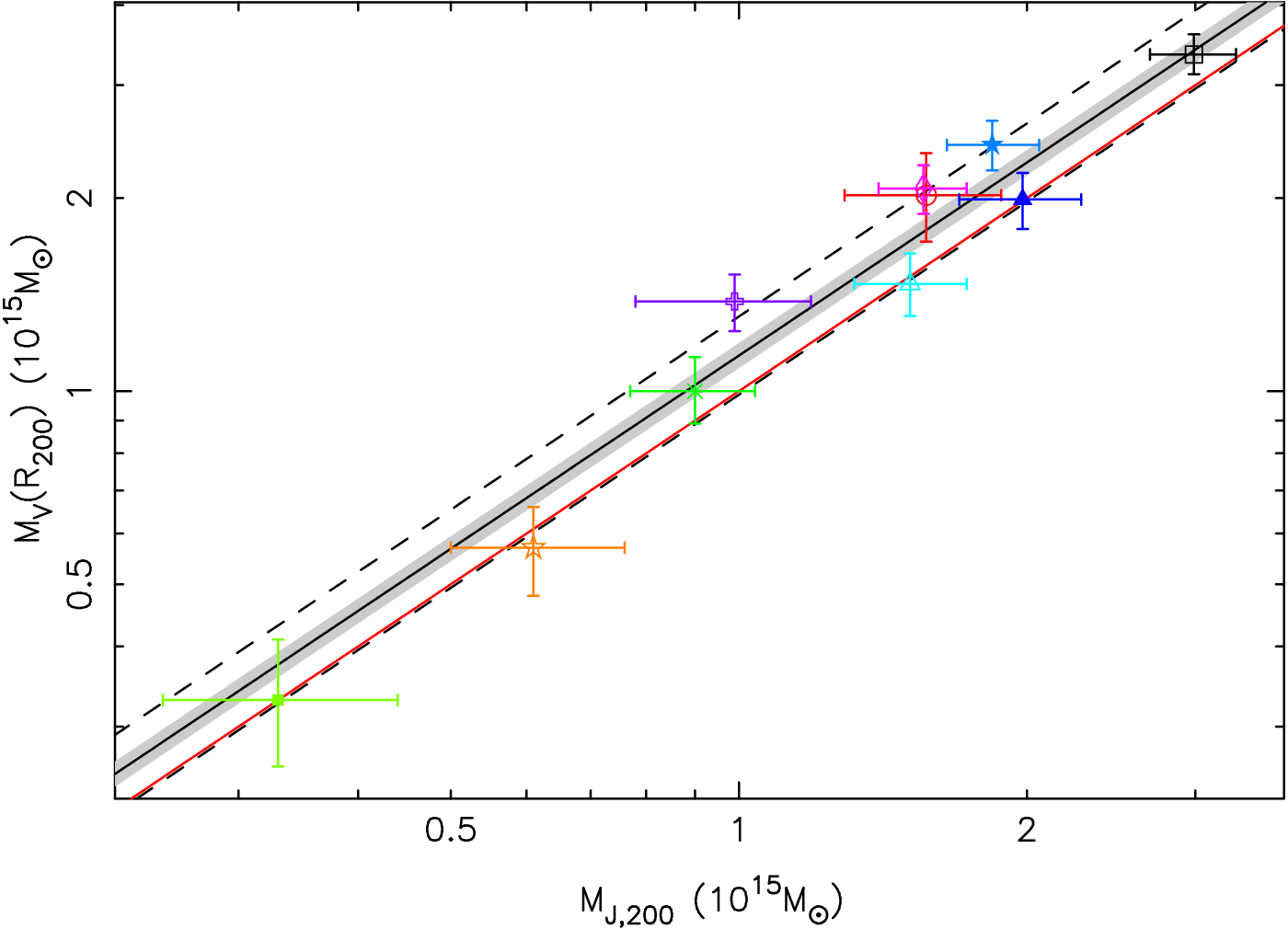}
\caption{Comparison of the Jeans, caustic (top panel), and virial (bottom panel) estimators. In both panels, the red line represents equality. The black line has an intercept corresponding to the geometrical mean of the mass ratio; the shaded area traces its $1\sigma$ uncertainty estimated from bootstrapping. The dashed lines indicate the orthogonal scatter $\sigma_\perp$ of the clusters around the mean ratio.}
\label{fig:JVc}
\end{figure}

The top-panel of Figure \ref{fig:JVc} presents the comparison between the Jeans and caustic estimators. The average mass ratio is compatible with 1, indicating the absence of systematics between the two estimators. This suggests that having used a constant $\mathcal{F}_\beta=0.7$ was, on average, a valid approximation. 

The Jeans and virial estimators (bottom panel of Fig. \ref{fig:JVc}) have a mass ratio larger than 1 at the $3\sigma$ level, the latter giving masses larger by $\sim15\%$. One possibility to explain this bias is the contamination by interlopers. Working with N-body numerical simulations, \cite{biviano06} have shown that the projected harmonic radius, $R_{PH}$, tends to be overestimated because of to this contamination. They found that the virial estimator gives masses that are typically overestimated by $\sim10\%$, which is close to the value found here. Another possibility could be an underestimation of the SPT. As mentioned previously, we made the hypothesis of isotropy to correct the $M_V$ masses. The results of the Jeans analysis suggest that the clusters' velocity field has some degree of anisotropy, so we can assume that the SPTs were indeed underestimated, leading to overestimated $M_V$ masses.

\subsection{Scaling relation $M-\sigma_P$}

\begin{figure}
\includegraphics[width=6.5cm, angle=-90]{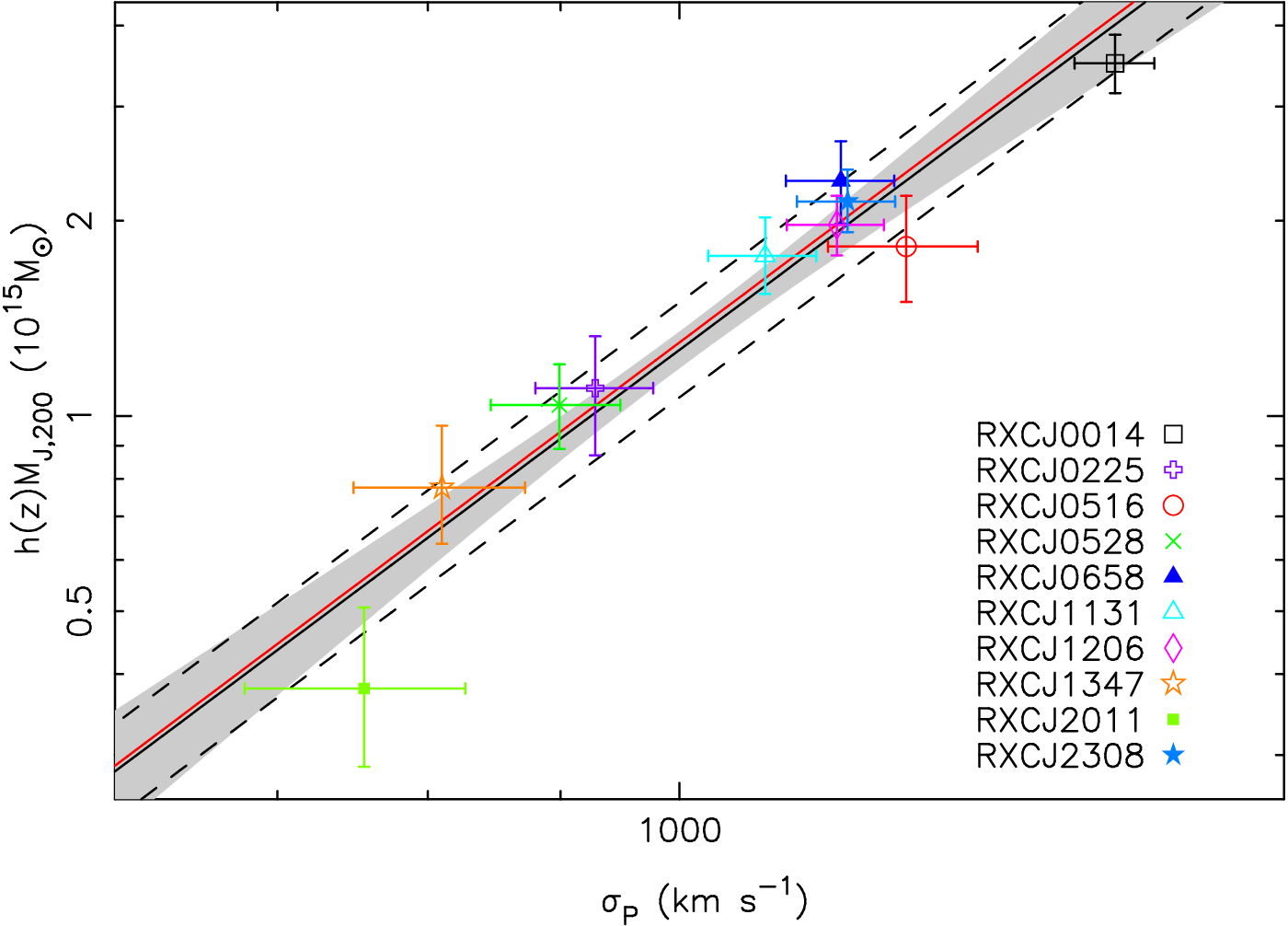}\\[4pt]
\includegraphics[width=6.5cm, angle=-90]{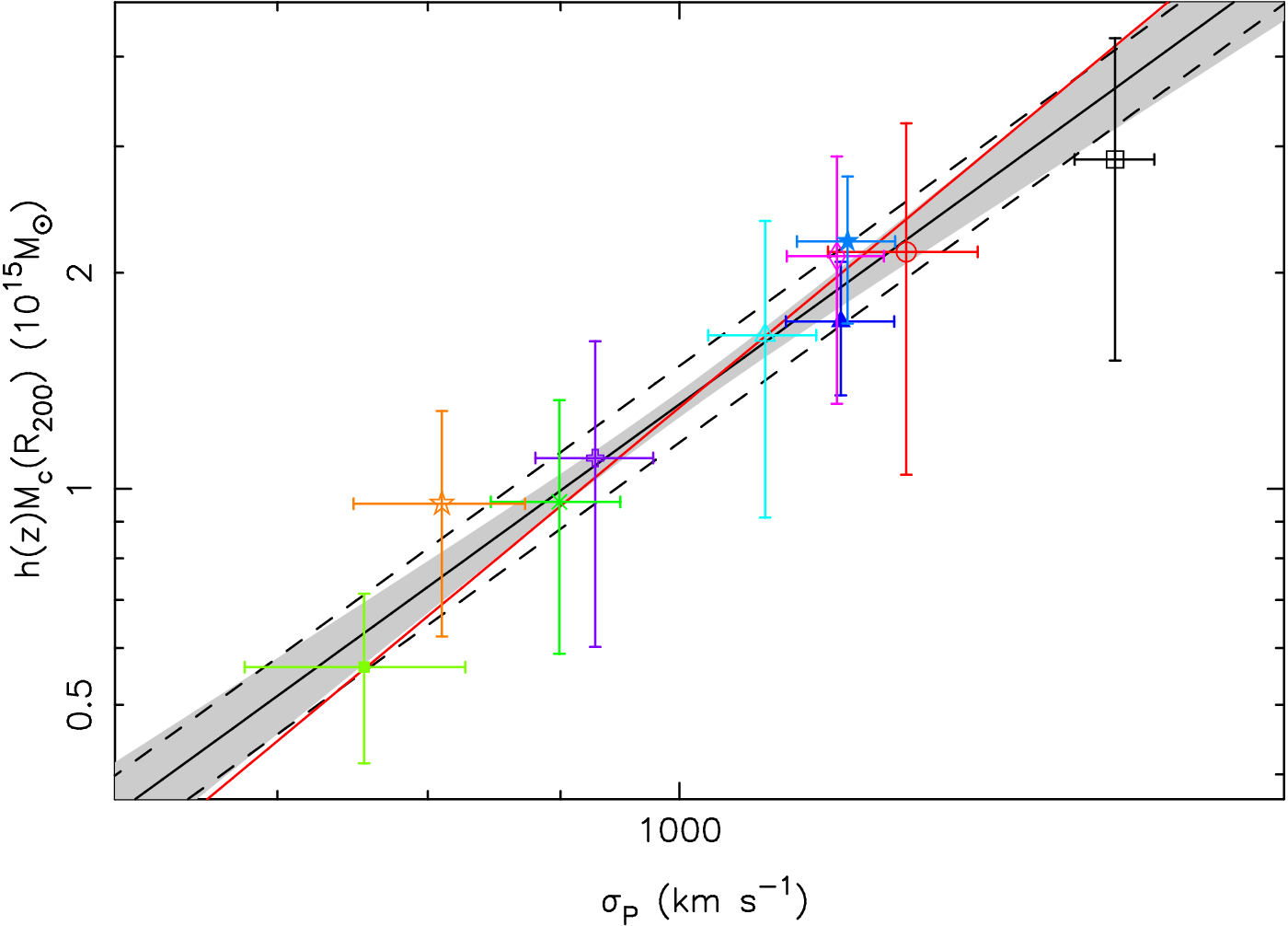}
\caption{Calibration of the mass-velocity dispersion scaling relation with the Jeans (top panel) and caustic (bottom panel) masses. The black line represents the best-fit regression line and the shaded area gives its statistical uncertainty. The dashed lines delimit the scatter in mass at fixed velocity dispersion. The best-fit relation obtained by \cite{biviano06} is indicated by the red line.}
\label{fig:Mvdisp}
\end{figure}

For a virialised cluster, the gravitational potential energy scales with its kinetic energy, thus one has $M/R\propto\sigma_\upsilon^2$, where $\sigma_\upsilon$ is the 3d velocity dispersion of the dark matter particles. Using the definition of a cluster mass, $M_\Delta=4/3\pi R_\Delta^3\Delta\rho_c$, one obtain the simple scaling relation $h(z)M_\Delta\propto\sigma_\upsilon^{3}$, where the reduced Hubble "constant" $h(z)=H(z)/H_0$ comes from the definition of the critical density $\rho_c(z)\propto h^2(z)$. Since this scaling law has been validated at high accuracy with numerical simulations (e.g. \citealt{evrard08}), it offers an interesting consistency check for our dynamical measurements. The Figure \ref{fig:Mvdisp} (top panel) presents the results obtained for the Jeans $M_{200}$ and the velocity dispersion estimated within the corresponding $R_{200}$. Here we used the BCES(X|Y) estimator of \cite{akritas96} to fit the regression $\log{[h(z)M_{200}]}=A\log\sigma_P+B$, treating the velocity dispersion as the response variable. With a slope $A=2.99\pm0.41$, the agreement with the theoretical prediction is excellent. It is also interesting to see that the best-fit normalisation matches very well that of \cite{biviano06} after a rescaling to a redshift $z=0.3$ and setting $\sigma_\upsilon=\sqrt{3}\sigma_P$ (their normalisation was obtained assuming a constant slope of three). The scatter in $\log{[h(z)M_{200}]}$ at fixed velocity dispersion is $\sigma_{\log{\mathrm{M}}}=0.07\pm0.02$. The calibration of the scaling with caustic masses gives similar results, see bottom panel of Fig. \ref{fig:Mvdisp}. Despite a shallower slope $A=2.63\pm0.35$, the best-fit regression agrees, within its errors, with the relation of \cite{biviano06} over the mass range probed by our sample. Here the scatter in mass is $\sigma_{\log{\mathrm{M}}}=0.06\pm0.01$.

We also calibrated the $\sigma_P(M_{200})$ scaling relation with the Jeans masses. We obtained a slope $A=0.303\pm0.041$ and a normalisation $10^B=1042\pm18\,\mathrm{km\,s^{-1}}$ for masses in units of $10^{15}\,h^{-1}\mathrm{M_\odot}$. While the former agrees well with the results of \cite{evrard08}, our normalisation is slightly smaller than their value $10^B=1082.9\pm4.0\,\mathrm{km\,s^{-1}}$. {Since the scaling relations of \cite{biviano06} or \cite{evrard08} were obtained with dark matter particles, our results tend to exclude a strong dynamical segregation between galaxies and dark matter.}

\subsection{X-rays vs. dynamics}
\label{sec:xray}

\begin{figure}
\includegraphics[width=6.5cm, angle=-90]{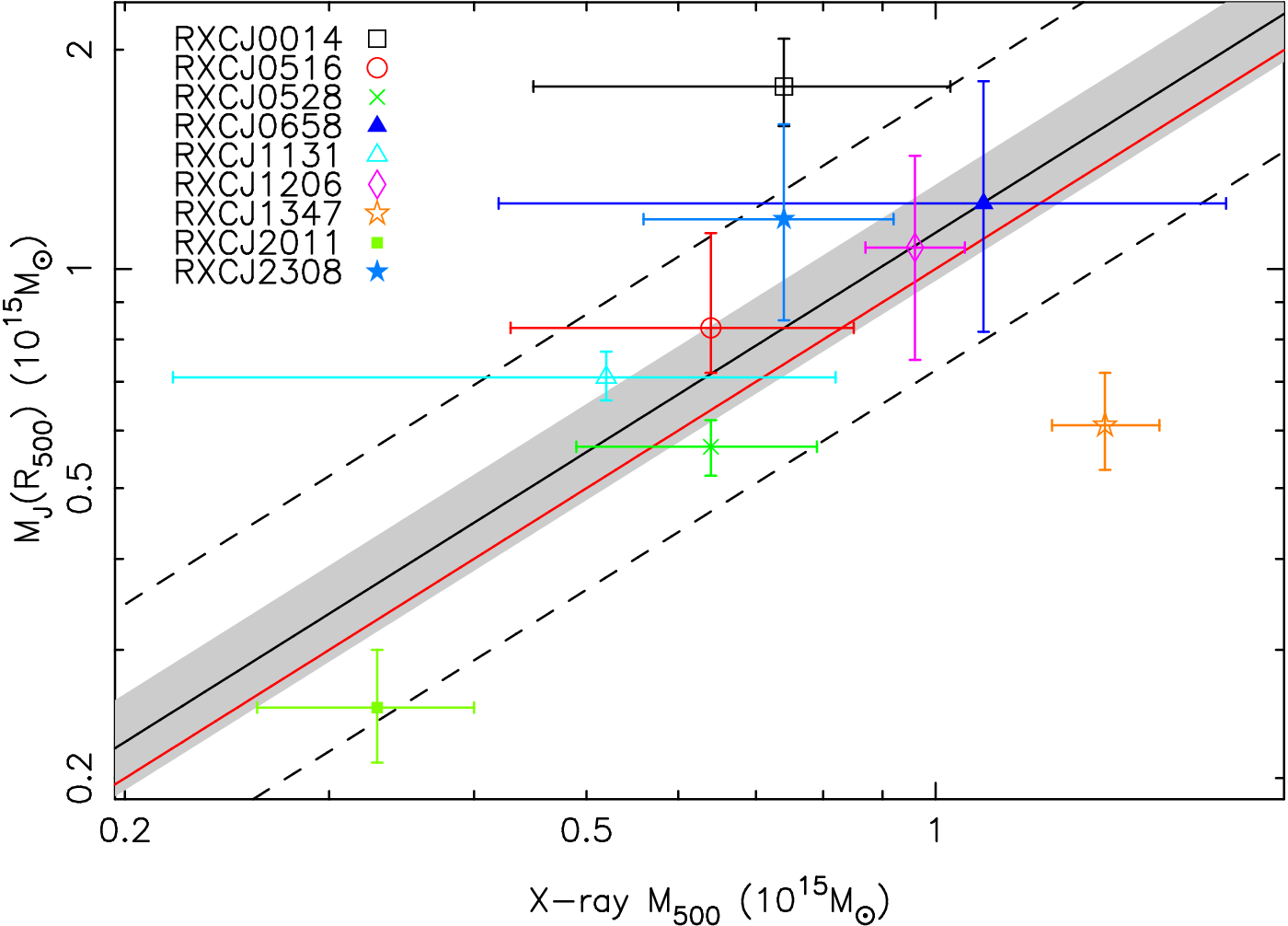}\\[4pt]
\includegraphics[width=6.5cm, angle=-90]{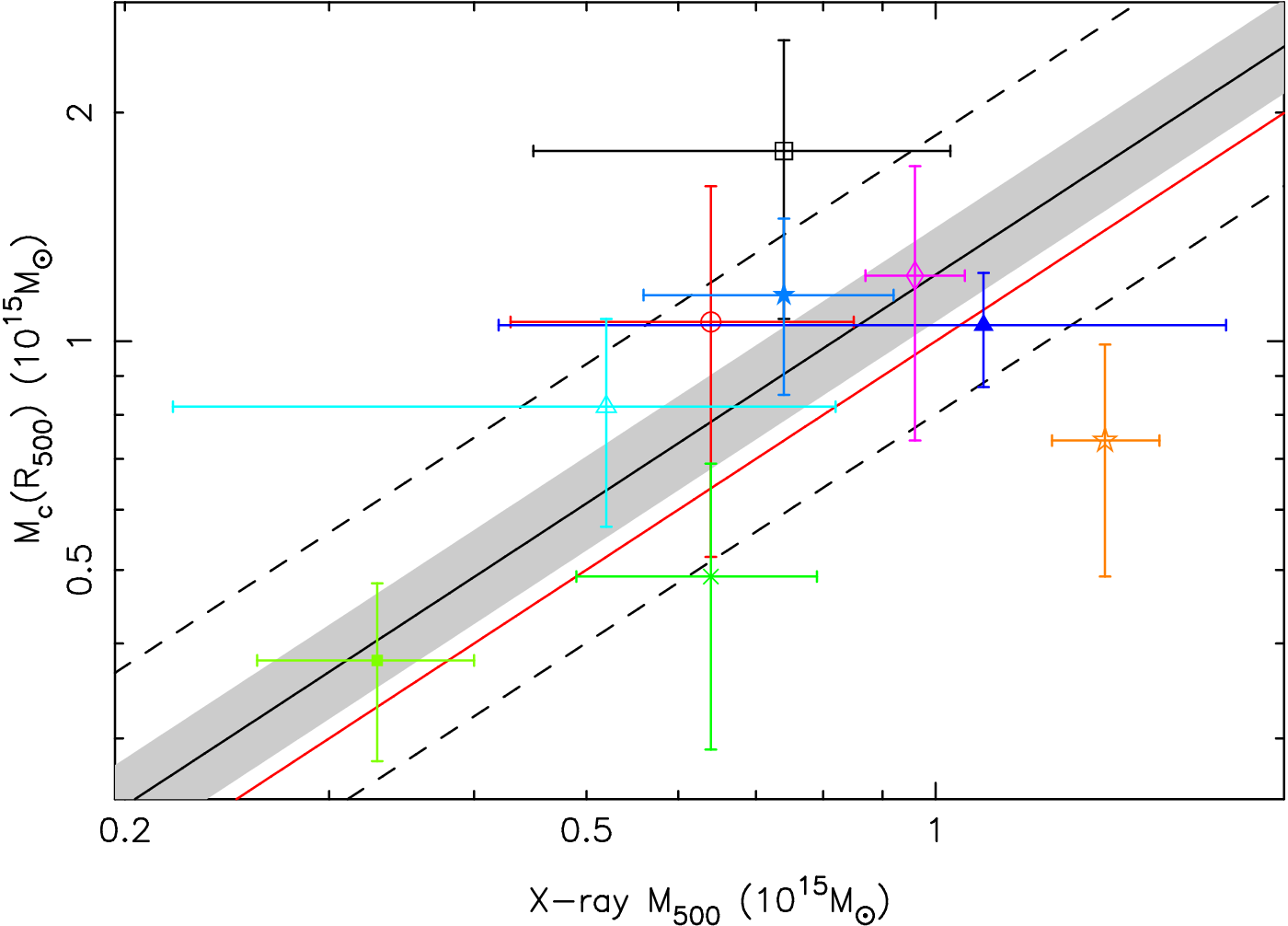}
\caption{Comparison of the hydrostatic, Jeans (top panel), and caustic (bottom panel) masses. See Fig. \ref{fig:JVc} for the legend.}
\label{fig:MJC_X}
\end{figure}

We now turn to the comparison between our dynamical masses and the hydrostatic estimates derived from XMM-Newton observations. The latter were taken from \cite{zhang06}, from \cite{foex12} for RXCJ1206 and RXCJ1347, and from \cite{mantz10} for RXCJ2011. The comparison is done within the same aperture, the X-ray NFW radius $R_{500}$. We interpolated the Jeans best-fit NFW model to $R_{500}$, while we ran the virial theorem using only the galaxies within this aperture (a new SPT was estimated accordingly); the caustic mass profile provides directly $M_c(R_{500})$. The results are presented in Figure \ref{fig:MJC_X} for the Jeans and caustic estimators.

For the Jeans, virial theorem, and caustic estimators, we obtained a mass ratio $\langle M_J/M_{HE}\rangle=1.22\pm0.18$, $\langle M_V/M_{HE}\rangle=1.51\pm0.26$, and $\langle M_c/M_{HE}\rangle=1.32\pm0.18$, respectively. The dynamical masses are on average larger than the hydrostatic values, however, the results do not provide a significant evidence for a bias between the hydrostatic and dynamical estimators ($<2\sigma$). The clusters present a rather large scatter around the mean ratio, $\sigma\sim50\%$ for the Jeans an caustic estimators, $\sigma\sim75\%$ for the virial theorem. We can also note that the geometric means of the mass ratios are slightly smaller than their arithmetic values, hence the intercepts of the regression lines in log-log space are closer to a vanishing fractional bias ($\lesssim1\sigma$).

\begin{figure}
\includegraphics[width=6.5cm, angle=-90]{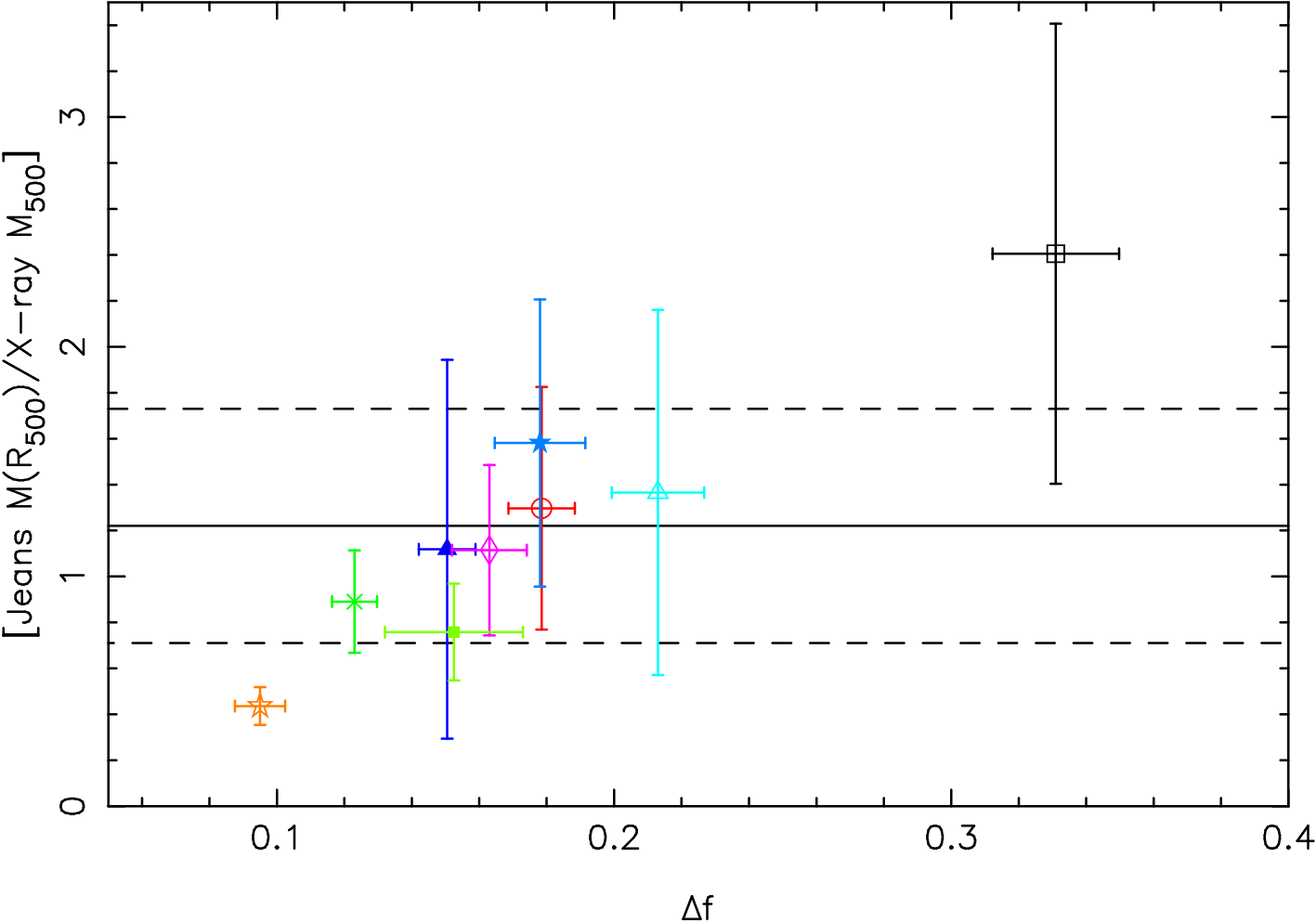}\\[4pt]
\includegraphics[width=6.5cm, angle=-90]{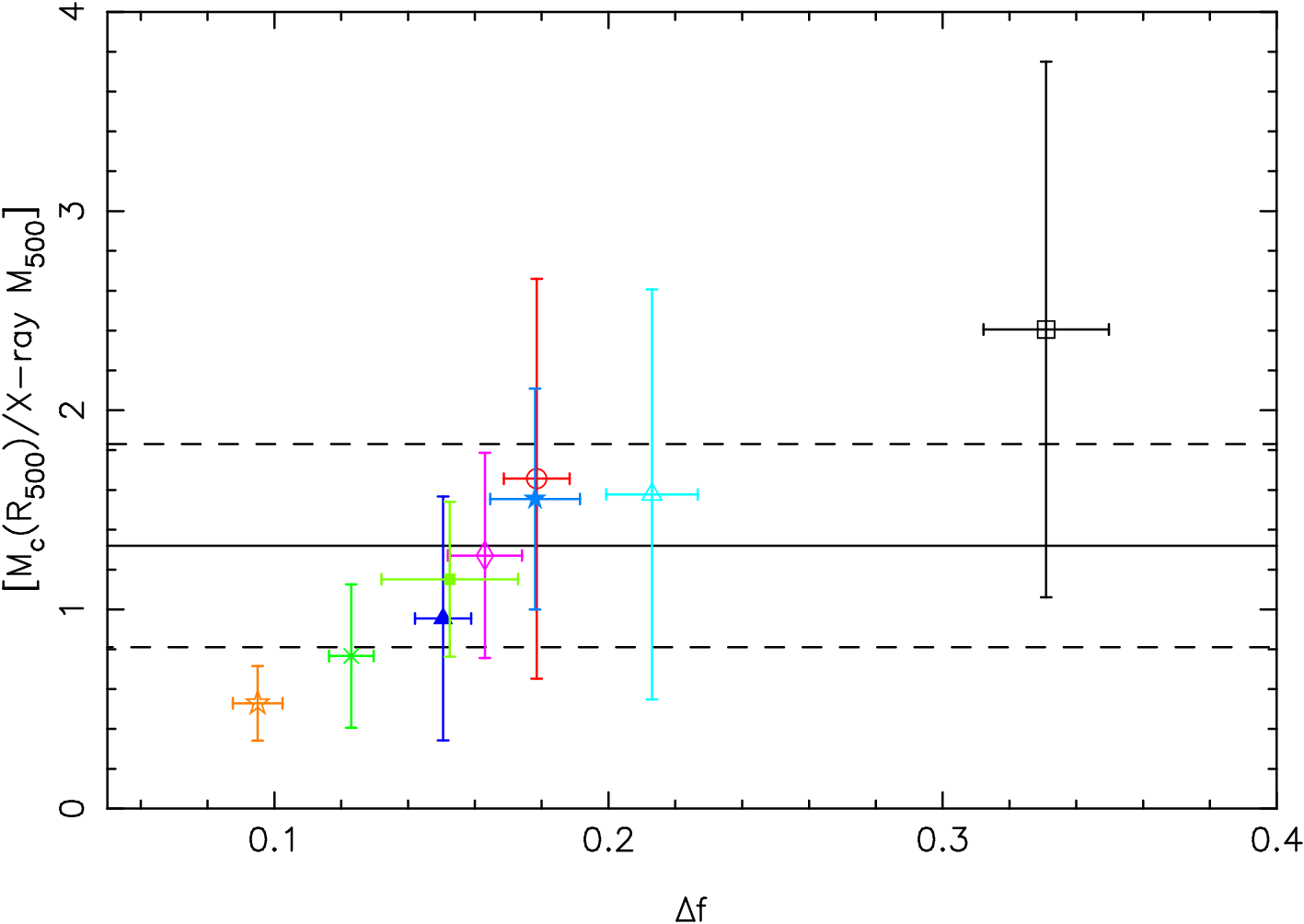}
\caption{Comparison between the Jeans-to-hydrostatic (top panel) and caustic-to-hydrostatic (bottom panel) mass ratio and the substructure indicator $\Delta f=(\Delta+f_{\mathrm{DS}})/2$. The horizontal lines show the average mass ratios and their standard deviation. In both cases, the Pearson correlation coefficient is $\rho=0.95$.}
\label{fig:ratio_sub}
\end{figure}

The standard deviation of the mass ratios and the residual scatters $\sigma_\perp$, are $\sim2-3$ times larger than the values obtained when comparing the dynamical estimators to each other. This is not surprising, since the latter involve similar assumptions and make use of the same data set, whereas the hydrostatic estimator is based on different physical mechanisms and hypothesis. In particular, one can expect the dynamical estimators to be more affected by substructures in the galaxy distribution. As a matter of fact, we do obtain strong correlations between the dynamical-to-hydrostatic mass ratio and the value of the substructure indicators (see Fig. \ref{fig:ratio_sub}). For the Jeans estimator, the Pearson coefficients are $\rho(M_J/M_{HE},\Delta)=0.84$ and $\rho(M_J/M_{HE},f_{\mathrm{DS}})=0.90$. Defining the average value $\Delta f=(f_{\mathrm{DS}}+\Delta)/2$, we obtain a correlation $\rho(M_J/M_{HE},\Delta f)=0.95$. Using the caustic masses (bottom panel of Fig. \ref{fig:ratio_sub}) or those from the virial theorem leads to very similar correlation coefficients (see Table \ref{table:corr}). These results suggest that the dynamical state of a cluster, as traced by its substructure content, is a potential source of bias for dynamical mass measurements.

Finally, we can note that the large scatter of the mass ratio is mainly driven by two outliers, RXCJ0014 and RXCJ1347, whose dynamical masses are respectively $\sim2$ times larger and smaller than their hydrostatic counterpart. While this is not a surprise for RXCJ0014 given its level of substructures, the mass ratio of RXCJ1347 is more intriguing, since its dynamical and morphological properties are rather simple (more details given in the Appendix). Therefore, we can suppose that the X-ray analysis is failing at estimating the correct mass of this particular cluster. If we exclude it from the sample, we end with average mass ratios larger by $\sim10\%$: $\langle M_J/M_{HE}\rangle=1.32\pm0.17$, $\langle M_c/M_{HE}\rangle=1.41\pm0.17$, and $\langle M_V/M_{HE}\rangle=1.65\pm0.25$; the rms standard deviations are reduced by $\sim10\%$. In that case, we obtain a fractional mass bias that is significant at the $\gtrsim2\sigma$ level (the bias has a similar significance when using the geometrical means).

\subsection{Red-sequence galaxies}
\label{sec:RS}

When making a mask for MOS observations, it is tempting to focus on red-sequence galaxies, since they provide a higher success rate for selecting cluster members. From the dynamical point of view, one can also argue that a high-velocity galaxy belonging to the red sequence has a smaller risk of being an interloper than a blue galaxy observed with the same rest-frame velocity. In other words, one can expect a smaller contamination by interlopers for this galaxy population, hence more accurate dynamical mass estimates (e.g. \citealt{biviano06,saro13}). It has also been argued that early-type galaxies are a better tracer of the relaxed component of a cluster, whereas the late-type population has not yet fully reached equilibrium because of galaxies on radial orbits falling into the cluster for the first time (e.g. \citealt{biviano92,colless96,adami98b,biviano04,barsanti16}; however see e.g. \citealt{rines05,rines13,girardi15} for opposite claims). On the other hand, massive elliptical galaxies are more likely to be affected by dynamical friction (e.g. \citealt{merritt85}). Since this mechanism is not taken into account during the Jeans analysis or with the virial theorem, it may be a possible source of bias when estimating masses with the red-sequence galaxies only.

\begin{figure}
\includegraphics[width=6.5cm, angle=-90]{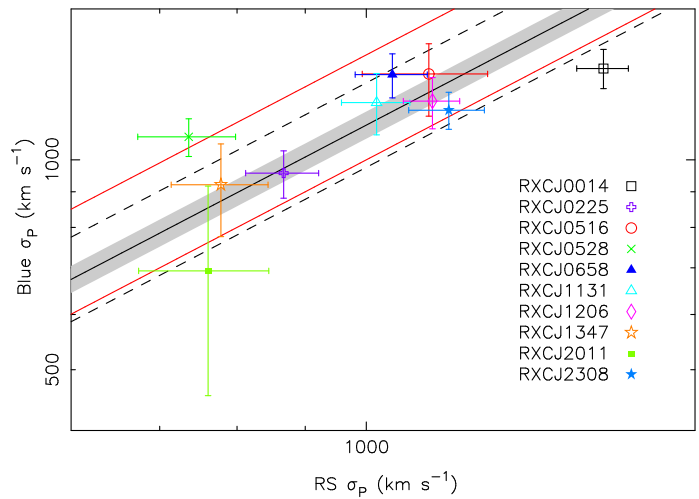}
\caption{Comparison of the velocity dispersions obtained using either the red-sequence (x-axis) or blue (y-axis) galaxies. Only galaxies within $R_{200}$ were used. See Fig. \ref{fig:JVc} for the legend. The two red lines trace $y=x$ and $y=\sqrt{2}x$, respectively.}
\label{fig:vdisp}
\end{figure}

All these considerations motivated us to investigate the impact of limiting the analysis to these galaxies. We first compared the velocity dispersions of the two populations (see Fig. \ref{fig:vdisp}). If late-type galaxies are gravitationally bound to the cluster but not yet virialised, their velocity dispersion should be $\lesssim\sqrt{2}$ larger than the value obtained for the red-sequence galaxies. With a mean ratio $\langle\sigma_{blue}/\sigma_{RS}\rangle=1.13\pm0.05$ (standard deviation $\sigma=0.15\pm0.04$), we confirm that blue galaxies have a larger velocity dispersion. It is interesting to note that RXCJ0014 is again a clear outlier (top-right point in Fig. \ref{fig:vdisp}). Its red-sequence galaxies have a velocity dispersion significantly larger than its blue population, which is a somewhat counter-intuitive result. A possible explanation is the presence of high-velocity substructures with a galaxy content dominated by red galaxies. In such a configuration, the bulk velocity of the merging sub-haloes outweighs the increased velocity dispersion of the blue population due to infalling galaxies.

\begin{figure}
\includegraphics[width=6.5cm, angle=-90]{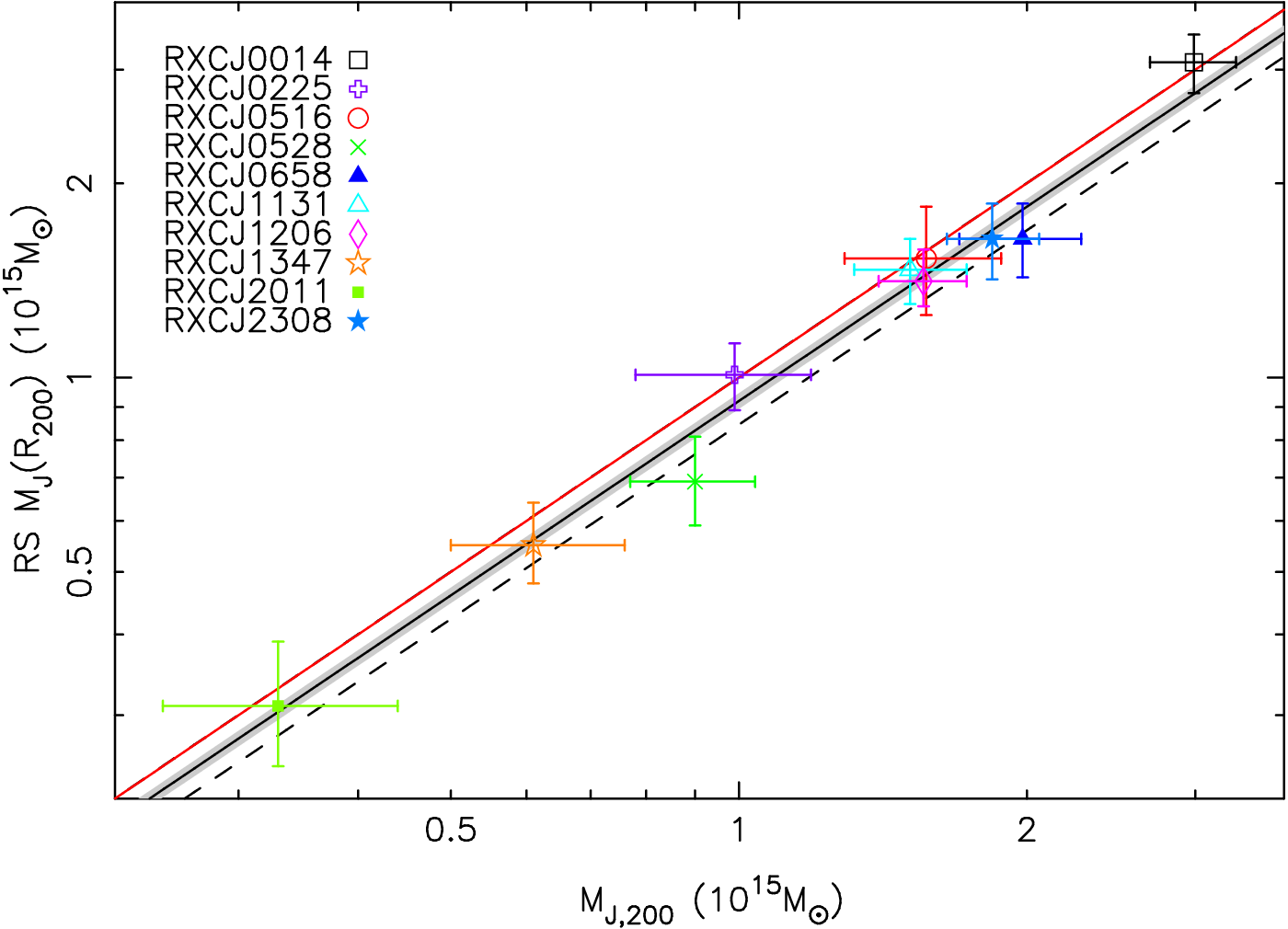}\\[4pt]
\includegraphics[width=6.5cm, angle=-90]{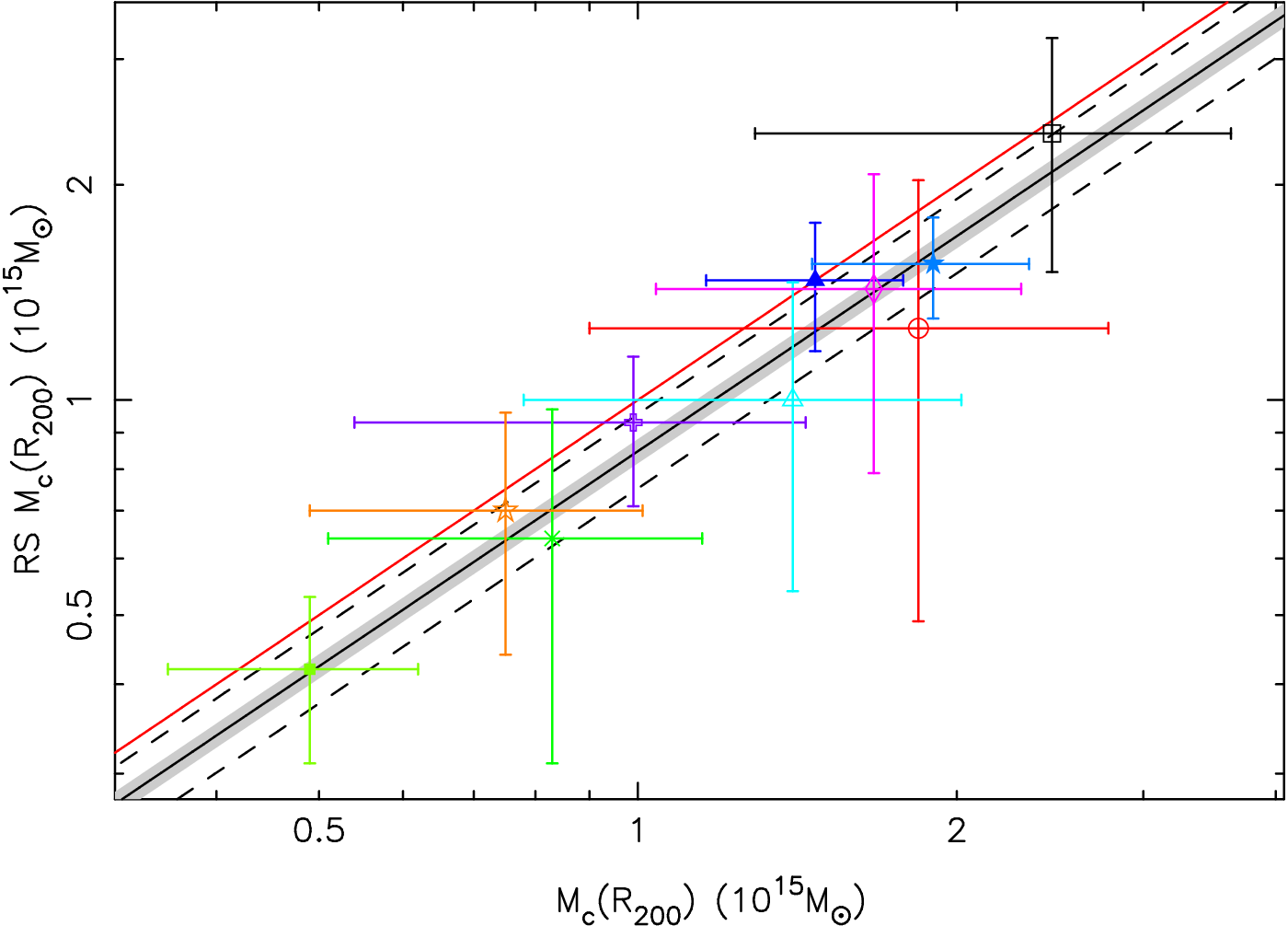}
\caption{Comparison of the Jeans (top panel) and caustic (bottom panel) masses estimated with (x-axis) or without (y-axis) the blue galaxies. See Fig. \ref{fig:JVc} for the legend.}
\label{fig:all_RS}
\end{figure}

Regarding the dynamical masses obtained with the red-sequence galaxies only, $M^{RS}$, we find a global offset with respect to the values derived previously when using both the red and blue populations (note that the new masses were estimated within the same radii $R_{200}$, given in the fourth column of Table \ref{table:mass}). For the Jeans estimator, we find an average mass ratio $\langle M^{RS}_J(R_{200})/M_{J,200}\rangle=0.92\pm0.02$ and small standard deviation $\sigma=0.08\pm0.02$ (see top panel of Fig. \ref{fig:all_RS}). We would like to point out that we re-estimated the characteristic radius of the galaxy density profile, to account for the specific spatial distribution of the red galaxies. For the virial theorem, the ratio is smaller: $\langle M_V^{RS}(R_{200})/M_V(R_{200})\rangle=0.86\pm0.04$, with a similar scatter. Interestingly, using the red-sequence galaxies leads to a better agreement between the two estimators, $\langle M_V^{RS}/M_J^{RS}\rangle=1.06\pm0.05$, whereas we had a ratio greater than 1 at the $3\sigma$ level. This result seems to confirm the comments made previously: a smaller contamination by interlopers leading to less underestimated harmonic radii and to a more adequate estimate of the SPT under the assumption of isotropic orbits.

With a mass ratio $\langle M_c^{RS}(R_{200})/M_c(R_{200})\rangle=0.85\pm0.03$, the caustic estimator presents the largest dependence on the galaxy population (see bottom panel of Fig. \ref{fig:all_RS}). This result might appear surprising since the caustic estimator is based on the escape velocity, which does not depend {\it a priori} on the galaxy population. Moreover, we used the same assumption regarding the filling factor, i.e. $\mathcal{F}_\beta=0.7$. While this approximation is justified for anisotropies $\beta\sim0.5$, we saw on Figure \ref{fig:Fbeta} that one should use a smaller value for an isotropic velocity field. Consequently, we should obtain even smaller caustic masses. This dependence on the galaxy population comes most likely from the combination of different effects. First of all, we remind here that the caustic amplitude is set by the average velocity dispersion within a certain radius. As shown above, the red-sequence galaxies have a smaller velocity dispersion; therefore, the density level in PPS satisfying the virial condition is smaller, leading to smaller masses. A second possibility could be the mislocating of the caustics due to a lack of galaxies observed at the escape velocity, which is better traced by the blue galaxies on their first orbit, prior to virialisation. Conversely, the presence of high-velocity interlopers in the blue population could also have biased high the caustic masses. We can mention here the work led by \cite{gifford13}, who analysed the impact of a colour selection on caustic mass estimates with semi-analytical simulations. They found a mass bias ranging from $\sim-5\%$ for purely red-sequence galaxies to $\sim+10\%$ when decreasing the fraction of red members to the average value of our sample $\sim0.6$. These values are consistent with the $15\%$ decrease in mass observed here.

To summarise, the red-sequence galaxies are characterised by smaller velocity dispersions. Using only these galaxies leads to dynamical masses that are $\sim10\%-15\%$ smaller. As a consequence, the average mass ratios between the dynamical and hydrostatic estimators are slightly smaller than the values obtained previously, but their scatter are essentially the same (see fourth part of Table \ref{table:comp}). A precise characterisation of the dynamical properties of the two broad populations of red and blue galaxies is beyond the scope of this paper. However, the present results seem to confirm the existence of a dynamical segregation that can potentially affect mass estimates.

\subsection{Impact of substructures}
\label{sec:newmass}

\begin{figure}
\includegraphics[width=6.5cm, angle=-90]{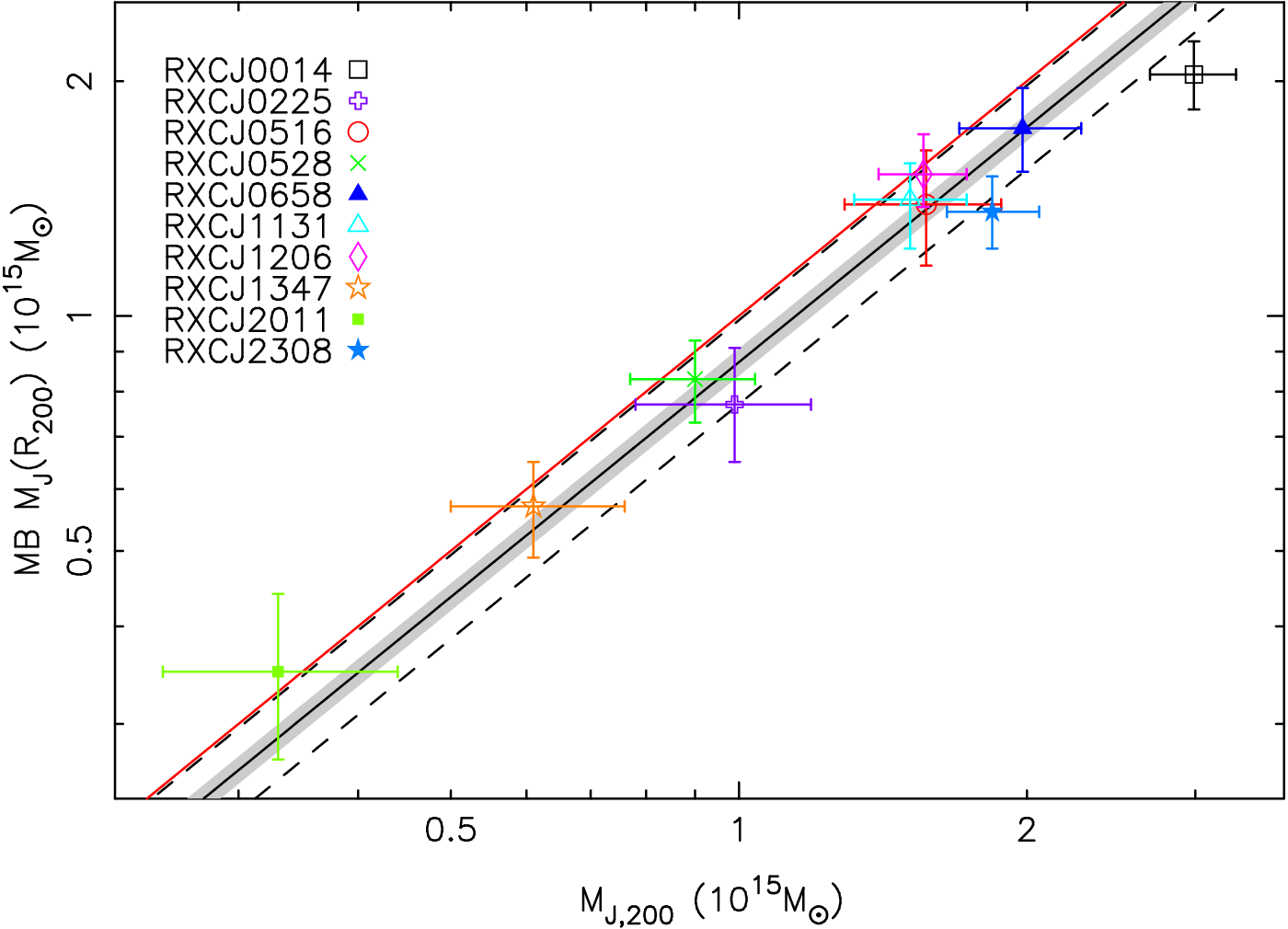}\\
\includegraphics[width=6.5cm, angle=-90]{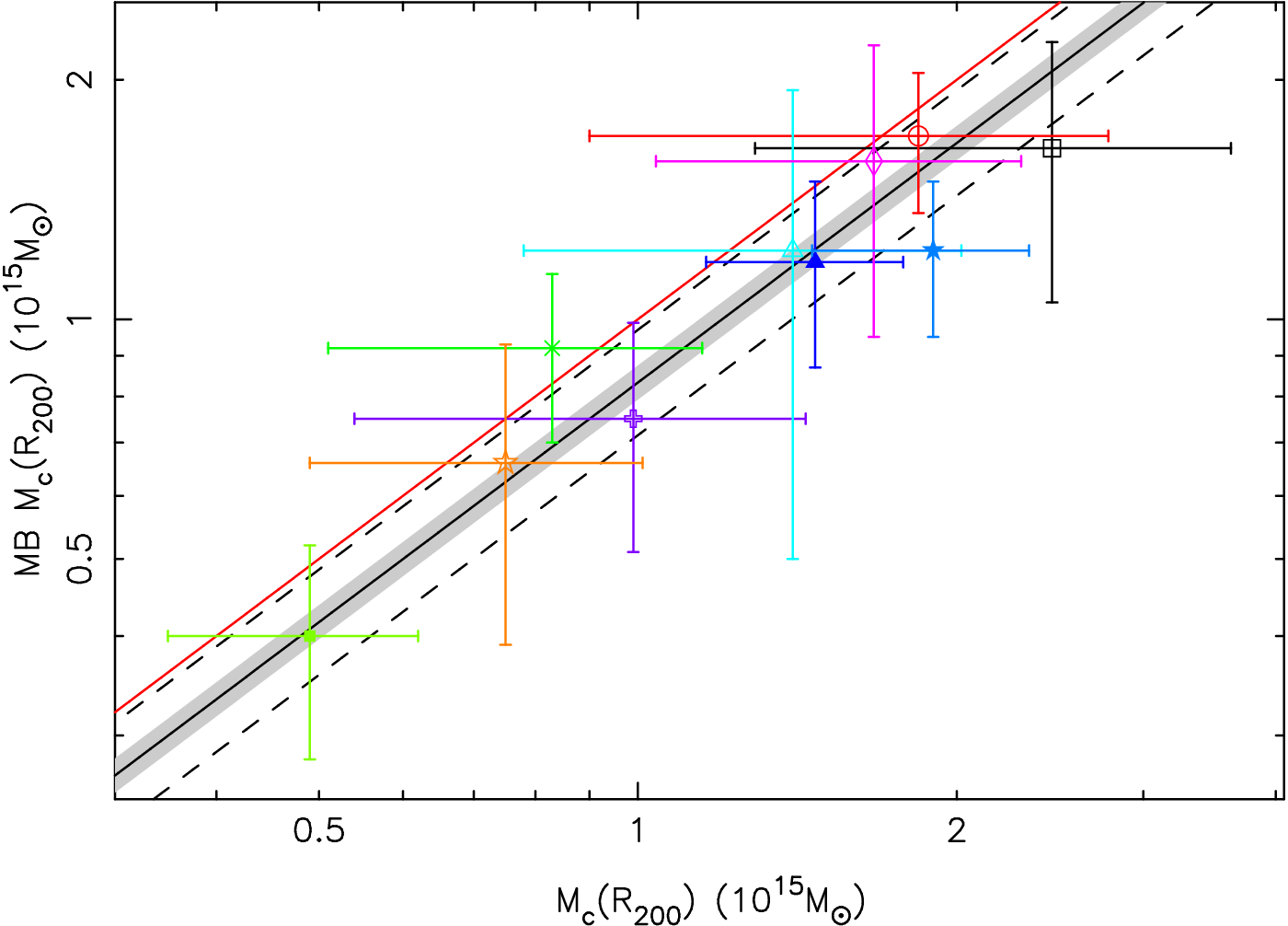}
\caption{Comparison of the Jeans (top panel) and caustic (bottom panel) masses estimated with (x-axis) or without (y-axis) the galaxies part of substructures. See Fig. \ref{fig:JVc} for the legend.}
\label{fig:all_MB}
\end{figure}

The correlation between the dynamical-to-hydrostatic mass ratio and the substructure indicators suggests that the dynamical estimates might be significantly biased by the presence of substructures. To further investigate this possibility, we now compare the masses obtained before and after excluding the galaxies that are associated with substructures (see Section \ref{sec:sub}). The results are presented in Figure \ref{fig:all_MB} and the new masses are listed in Table \ref{table:mass2}; the average ratios are given in the fifth part of Table \ref{table:comp}. For the three estimators, we obtain $\langle M^{MB}/M\rangle<1$ at the $\gtrsim3\sigma$ level, where $M^{MB}$ stands for the mass estimated after removing substructures. Their impact appears to be larger for the virial estimator, whose masses are $\sim20\%$ smaller.

We find that the mass ratios exhibit moderate anti-correlations with the substructure indicators: $\rho(M_J^{MB}/M_J,\Delta)=-0.56$ and $\rho(M_J^{MB}/M_J,f_{\mathrm{DS}})=-0.61$ for the Jeans estimator, $\rho(M_c^{MB}/M_c,\Delta)=-0.24$ and $\rho(M_c^{MB}/M_c,f_{\mathrm{DS}})=-0.68$ for the caustic masses, and $\rho(M_V^{MB}/M_V,\Delta)=-0.66$ and $\rho(M_V^{MB}/M_V,f_{\mathrm{DS}})=-0.42$ for the virial theorem (the corresponding p-values are given in Table \ref{table:corr}). The virial theorem appears to be more affected by the spatial distribution of galaxies, the harmonic radius depending on the inverse distance of galaxy pairs, whereas the Jeans and caustic estimators show a larger dependence on substructures in velocity space. These results indicate that a larger fraction of galaxies in substructures implies a larger impact on dynamical mass estimates. Interestingly, we also find a mass dependence of the relative substructure content. Using the Jeans masses as reference, we obtain correlation coefficients $\rho(M,\Delta)=0.77$, $\rho(M,f_{\mathrm{DS}})=0.71$, and $\rho(M,\Delta f)=0.79$. These values show that the more massive a cluster, the higher its level of substructure. This agrees well with the hierarchical model of cluster formation, according to which the largest objects are formed later by accretion and merging of smaller-scale systems, leaving them less time to reach a relaxed and homogenous sate.

\begin{table}
\centering 
\begin{threeparttable}
\caption{Correlations involving the substructure indicators.}
\label{table:corr}
\begin{tabular}{l c c c}
\hline\hline\noalign{\smallskip}
 & $\Delta$ & $f_{\mathrm{DS}}$ & $\Delta f$\\
\noalign{\smallskip}\hline\noalign{\smallskip}
$M_J/M_{HE}$ & 0.84(0.005) & 0.90(<0.001) & 0.95(<0.001)\\
$M_c/M_{HE}$ & 0.83(0.006) & 0.91(<0.001) & 0.95(<0.001)\\
$M_V/M_{HE}$ & 0.78(0.013) & 0.91(<0.001) & 0.93(<0.001)\\
\noalign{\smallskip}\hline\noalign{\smallskip}
$M_J^{MB}/M_J$ & -0.56(0.092) & -0.61(0.06) & -0.64(0.05)\\
$M_c^{MB}/M_c$ & -0.24(0.50) & -0.68(0.03) & -0.56(0.09)\\
$M_V^{MB}/M_V$ & -0.66(0.04) & -0.42(0.23) & -0.55(0.10)\\
\noalign{\smallskip}\hline\noalign{\smallskip}
$M_J$ & 0.77(0.009) & 0.71(0.021) & 0.79(0.007)\\
$M_c$ & 0.73(0.017) & 0.67(0.034) & 0.75(0.012)\\
$M_V$ & 0.73(0.017) & 0.70(0.024) & 0.77(0.009)\\
\noalign{\smallskip}\hline
\end{tabular}
    \begin{tablenotes}
      \small
      \item Pearson correlation coefficients and corresponding p-value for the null hypothesis (i.e. no correlation), estimated from a Student's t-distribution. $M_J$, $M_c$, $M_V$, and $M_{HE}$ denote the Jeans, caustic, virial theorem and hydrostatic mass estimators, respectively. The superscript $MB$ indicates masses estimated without the galaxies within substructures. 
    \end{tablenotes}
  \end{threeparttable}
\end{table}

To summarise, excluding galaxies that belong to substructures leads to dynamical masses $\sim15\%$ smaller. The impact of substructures increases with a cluster mass, as a consequence of two correlations: the larger the mass, the higher the level of substructure, and the higher this level, the larger the effect on the dynamical mass estimators. Therefore, a proper characterisation of substructures is mostly required when dealing with very massive clusters.

\begin{figure}
\includegraphics[width=6.5cm, angle=-90]{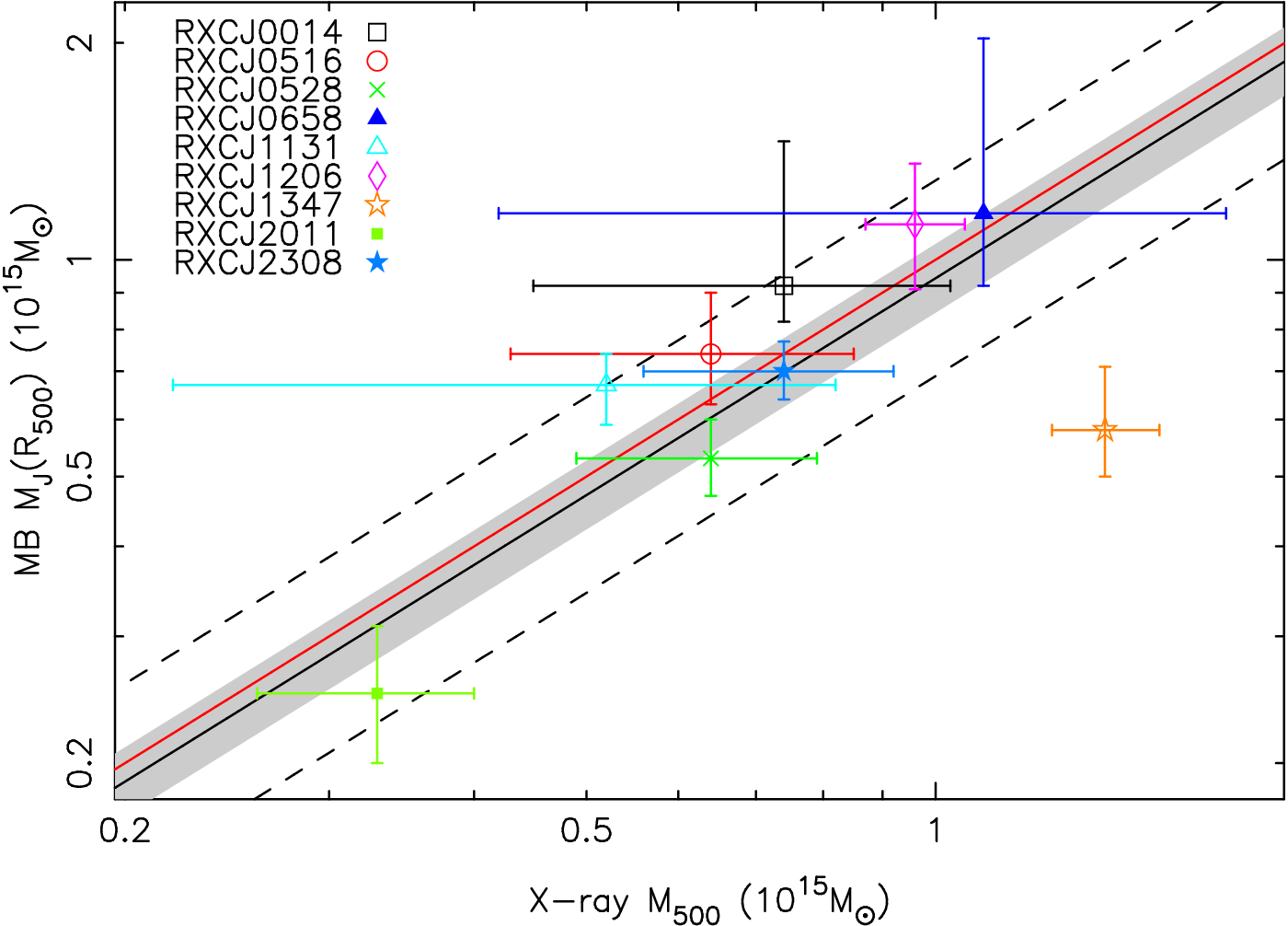}\\
\includegraphics[width=6.5cm, angle=-90]{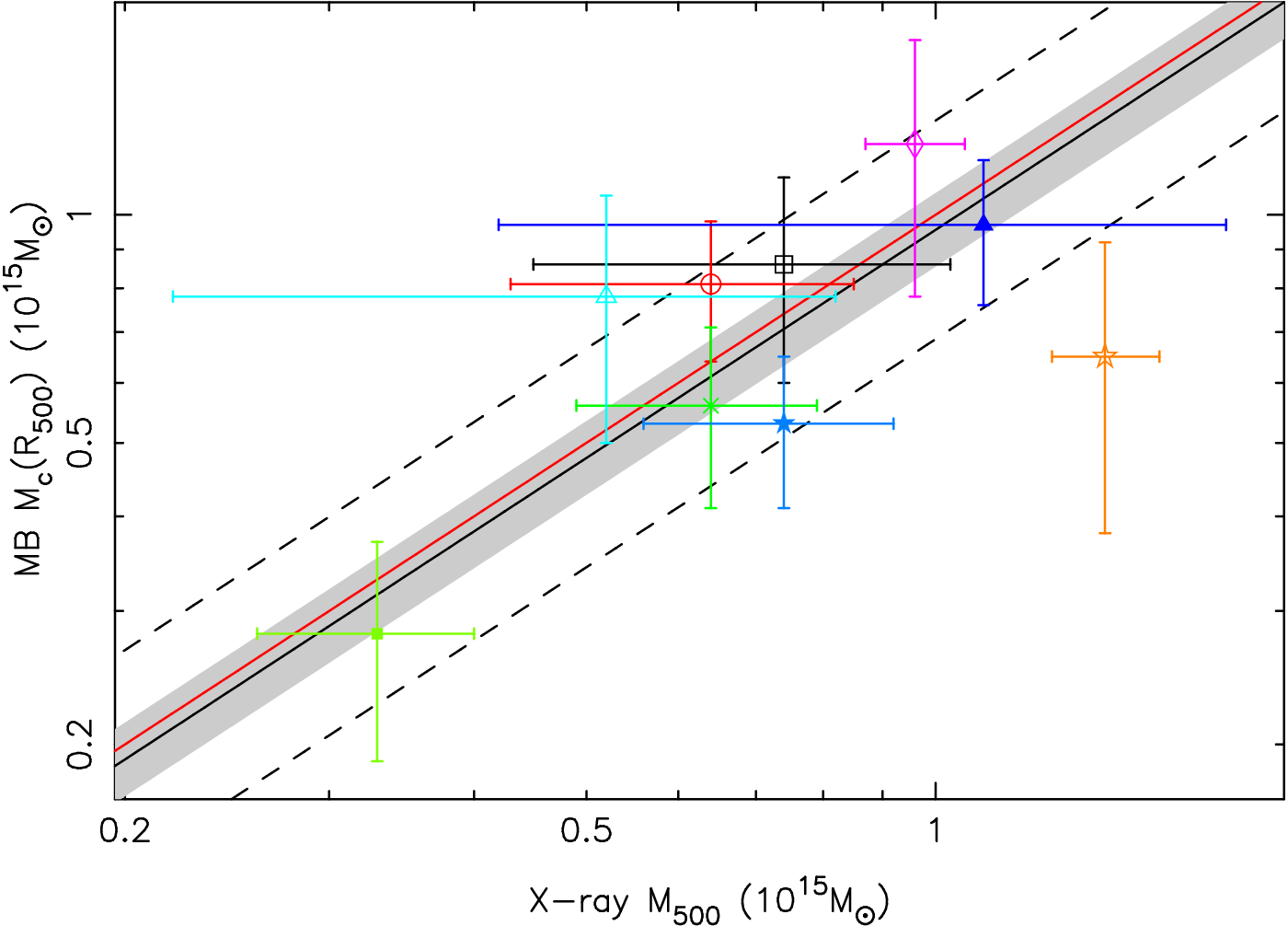}
\caption{Comparison of the hydrostatic and dynamical masses after the substructure analysis (top panel for the Jeans estimator, bottom panel for the caustic masses). See Fig. \ref{fig:JVc} for the legend.}
\label{fig:MB_MX}
\end{figure}

Having dealt with substructures, we can also check how the new dynamical masses compare with the hydrostatic values. The results are presented in Figure \ref{fig:MB_MX} for the Jeans and caustic masses (see also the last part of Table \ref{table:comp}). The main difference with the results obtained in Section \ref{sec:xray} concerns the average mass ratios, which now agree within $1\sigma$ with a vanishing fractional bias. We can also note that the standard deviations are decreased by roughly a factor 2, suggesting that our approach for treating substructures reduces efficiently their impact on dynamical mass estimates.

We have shown in Section \ref{sec:xray} that, prior to the substructure analysis, excluding RXCJ1347 from the sample leads to a mass bias that is significant at the $\gtrsim2\sigma$ level. Here, we find $\langle M_J^{MB}/M_{HE}\rangle=1.06\pm0.06$ ($\sigma=0.18$), $\langle M_c^{MB}/M_{HE}\rangle=1.07\pm0.09$ ($\sigma=0.25$), and $\langle M_V^{MB}/M_{HE}\rangle=1.17\pm0.11$ ($\sigma=0.30$) for the Jeans, caustic, and virial estimator, respectively. Thus, the standard deviations are decreased roughly by a factor 2 while the average ratios increase by $\sim+0.07$. However, they stay compatible with no bias at the $\sim1\sigma$ level (1.5 for the virial theorem estimator). In other words, excluding RXCJ1347 does not change our main conclusion: after the substructure analysis, we do not observe any significant bias between the hydrostatic and dynamical mass estimators.

We found in Section \ref{sec:RS} that the blue population tends to have a larger velocity dispersion than the red-sequence galaxies. \cite{girardi15} suggested that the dynamical state of a cluster could be responsible for this segregation. This is clearly the case for RXCJ0014, but with the opposite result, i.e. a larger velocity dispersion for the red-sequence galaxies. Now that we have isolated the galaxies belonging to the clusters' main body, we can check whether the two galaxy populations still present the same difference regarding their velocity dispersion. With an average ratio $\langle\sigma_{blue}/\sigma_{RS}\rangle=1.27\pm0.05$, we find that the segregation is even more pronounced; the corresponding standard deviation is $\sigma=0.16\pm0.03$, i.e. very similar to the value quoted in Sec. \ref{sec:RS}. Since we do not find a significant correlation between the ratio of velocity dispersions and either substructure indicators ($|\rho|<0.4$), our results indicate that the two populations have a different velocity distribution regardless of the dynamical state of their host. However, it is possible that our findings are specific to the cluster population analysed here, i.e. massive objects at moderate redshifts. A comparison with clusters at lower redshifts would provide more insights on a possible evolutionary trend of the dynamical segregation between early- and late-type galaxies (see e.g. \citealt{barsanti16}).

%%%%%%%%%%%%%%%%%%%%%%%%%%%%%%%%
%                          --------- mass after sub. table -----------
%%%%%%%%%%%%%%%%%%%%%%%%%%%%%%%%

\begin{table*}
\centering 
\begin{threeparttable}
\caption{Dynamical mass estimates after the substructure analysis.}
\label{table:mass2}
\begin{tabular}{l c c c c c c c}
\hline\hline\noalign{\smallskip}
Cluster & $M_{200}$ & $R_{200}$ & $M(R_{200}^{all})$ & $M_V(R_{200})$ & $M_V(R_{200}^{all})$ & $M_c(R_{200})$ & $M_c(R_{200}^{all})$\\
   & $(10^{15}\mathrm{M_\odot})$ & (Mpc) & $(10^{15}\mathrm{M_\odot})$ & $(10^{15}\mathrm{M_\odot})$ & $(10^{15}\mathrm{M_\odot})$ & $(10^{15}\mathrm{M_\odot})$ & $(10^{15}\mathrm{M_\odot})$\\
\noalign{\smallskip}\hline\noalign{\smallskip}
RXCJ0014 & $1.71_{-0.23}^{+0.26}$ & $2.22_{-0.10}^{+0.11}$ & $2.04_{-0.20}^{+0.21}$ & $1.84_{-0.25}^{+0.26}$ & $1.99_{-0.24}^{+0.25}$ & $1.48_{-0.50}^{+0.50}$ & $1.64_{-0.59}^{+0.59}$\\[3pt]
RXCJ0225 & $0.72_{-0.14}^{+0.17}$ & $1.72_{-0.12}^{+0.13}$ & $0.77_{-0.11}^{+0.14}$ & $1.01_{-0.17}^{+0.17}$ & $0.88_{-0.15}^{+0.17}$ & $0.71_{-0.23}^{+0.23}$ & $0.75_{-0.24}^{+0.24}$\\[3pt]
RXCJ0516 & $1.31_{-0.31}^{+0.34}$ & $2.04_{-0.17}^{+0.17}$ & $1.39_{-0.23}^{+0.24}$ & $1.25_{-0.27}^{+0.32}$ & $1.20_{-0.26}^{+0.31}$ & $1.57_{-0.31}^{+0.31}$ & $1.70_{-0.34}^{+0.34}$\\[3pt]
RXCJ0528 & $0.79_{-0.14}^{+0.15}$ & $1.73_{-0.11}^{+0.10}$ & $0.83_{-0.10}^{+0.10}$ & $0.89_{-0.13}^{+0.15}$ & $0.93_{-0.13}^{+0.14}$ & $0.87_{-0.21}^{+0.21}$ & $0.92_{-0.22}^{+0.22}$\\[3pt]
RXCJ0658 & $1.64_{-0.27}^{+0.32}$ & $2.20_{-0.13}^{+0.13}$ & $1.74_{-0.21}^{+0.22}$ & $1.59_{-0.21}^{+0.23}$ & $1.64_{-0.21}^{+0.24}$ & $1.15_{-0.28}^{+0.28}$ & $1.18_{-0.31}^{+0.31}$\\[3pt]
RXCJ1131 & $1.37_{-0.26}^{+0.24}$ & $2.07_{-0.14}^{+0.11}$ & $1.41_{-0.19}^{+0.16}$ & $1.29_{-0.19}^{+0.20}$ & $1.30_{-0.18}^{+0.22}$ & $1.20_{-0.70}^{+0.70}$ & $1.22_{-0.72}^{+0.72}$\\[3pt]
RXCJ1206 & $1.51_{-0.17}^{+0.25}$ & $2.03_{-0.08}^{+0.11}$ & $1.52_{-0.14}^{+0.19}$ & $1.85_{-0.18}^{+0.18}$ & $1.88_{-0.18}^{+0.20}$ & $1.57_{-0.62}^{+0.62}$ & $1.58_{-0.63}^{+0.63}$\\[3pt]
RXCJ1347 & $0.55_{-0.10}^{+0.13}$ & $1.44_{-0.10}^{+0.10}$ & $0.57_{-0.08}^{+0.08}$ & $0.53_{-0.09}^{+0.11}$ & $0.52_{-0.09}^{+0.09}$ & $0.65_{-0.27}^{+0.27}$ & $0.66_{-0.27}^{+0.27}$\\[3pt]
RXCJ2011 & $0.37_{-0.11}^{+0.14}$ & $1.35_{-0.15}^{+0.15}$ & $0.35_{-0.08}^{+0.09}$ & $0.37_{-0.10}^{+0.11}$ & $0.39_{-0.10}^{+0.11}$ & $0.42_{-0.13}^{+0.13}$ & $0.40_{-0.12}^{+0.12}$\\[3pt]
RXCJ2308 & $1.16_{-0.17}^{+0.20}$ & $1.96_{-0.10}^{+0.11}$ & $1.36_{-0.14}^{+0.15}$ & $1.30_{-0.19}^{+0.23}$ & $1.68_{-0.22}^{+0.26}$ & $0.99_{-0.21}^{+0.21}$ & $1.22_{-0.27}^{+0.27}$\\
\noalign{\smallskip}\hline
\end{tabular}
    \begin{tablenotes}
      \small
      \item Columns: (1) Cluster name. (2) Best-fit NFW mass derived from the Jeans analysis. As for the first analysis, the best-fit model combines the NFW mass profile with one of the three different anisotropy models. (3) Corresponding NFW radius. (4) Extrapolation of the Jeans best-fit NFW profile to the radius $R_{200}^{all}$ obtained before the substructure analysis (i.e. fourth column of Table \ref{table:mass}). (5,6) Mass derived from the virial theorem within an aperture of radius $R_{200}$ and $R_{200}^{all}$. (7,8) Caustic mass at the same two radii. The galaxy density in PPS, the density level $\kappa$, the velocity dispersion $\sigma_P(<R)$, and the caustic amplitude were re-estimated accordingly. Note: the masses given in column 4, 6, and 8 are the values used in Sect. \ref{sec:newmass} to investigate the impact of substructures.
    \end{tablenotes}
  \end{threeparttable}
\end{table*}

% mass comparison

\begin{table*}
\centering 
\begin{threeparttable}
\caption{Comparison of the different mass estimators.}
\label{table:comp}
\begin{tabular}{l c c c c c}
\hline\hline\noalign{\smallskip}
$M_X-M_Y$ & R & $\langle M_Y/M_X\rangle$ & $\sigma_{M_Y/M_X}$ & $\langle M_Y/M_X\rangle_g$ & $\sigma_\perp$\\
\noalign{\smallskip}\hline\noalign{\smallskip}
$M_J-M_c$ & $R_{200}$ & $1.04\pm0.06$ & $0.20\pm0.05$ & $1.02\pm0.06$ & $0.06\pm0.01$\\[3pt]
$M_J-M_V$ & $R_{200}$ & $1.15\pm0.05$ & $0.16\pm0.02$ & $1.14\pm0.05$ & $0.04\pm0.01$\\[3pt]
$M_c-M_V$ & $R_{200}$ & $1.14\pm0.08$ & $0.23\pm0.05$ & $1.11\pm0.08$ & $0.07\pm0.02$\\
\noalign{\smallskip}\hline\noalign{\smallskip}
$M_{HE}-M_J$ & $R_{500}$ & $1.22\pm0.18$ & $0.51\pm0.15$ & $1.12\pm0.17$ & $0.13\pm0.04$\\[3pt]
$M_{HE}-M_c$ & $R_{500}$ & $1.32\pm0.18$ & $0.51\pm0.13$ & $1.22\pm0.17$ & $0.13\pm0.03$\\[3pt]
$M_{HE}-M_V$ & $R_{500}$ & $1.51\pm0.26$ & $0.75\pm0.22$ & $1.33\pm0.25$ & $0.17\pm0.05$\\
\noalign{\smallskip}\hline\noalign{\smallskip}
$M_J-M^{RS}_J$ & $R_{200}$ & $0.92\pm0.02$ & $0.08\pm0.02$ & $0.92\pm0.03$ & $0.03\pm0.01$\\[3pt]
$M_c-M^{RS}_c$ & $R_{200}$ & $0.85\pm0.03$ & $0.10\pm0.02$ & $0.84\pm0.03$ & $0.04\pm0.01$\\[3pt]
$M_V-M^{RS}_V$ & $R_{200}$ & $0.86\pm0.04$ & $0.11\pm0.03$ & $0.85\pm0.04$ & $0.04\pm0.01$\\
\noalign{\smallskip}\hline\noalign{\smallskip}
$M_{HE}-M^{RS}_J$ & $R_{500}$ & $1.17\pm0.18$ & $0.53\pm0.14$ & $1.06\pm0.17$ & $0.14\pm0.04$\\[3pt]
$M_{HE}-M^{RS}_c$ & $R_{500}$ & $1.22\pm0.17$ & $0.49\pm0.14$ & $1.13\pm0.16$ & $0.13\pm0.03$\\[3pt]
$M_{HE}-M^{RS}_V$ & $R_{500}$ & $1.46\pm0.26$ & $0.76\pm0.18$ & $1.26\pm0.26$ & $0.18\pm0.04$\\
\noalign{\smallskip}\hline\noalign{\smallskip}
$M_J-M^{MB}_J$ & $R_{200}$ & $0.88\pm0.03$ & $0.11\pm0.02$ & $0.87\pm0.04$ & $0.04\pm0.01$\\[3pt]
$M_c-M^{MB}_c$ & $R_{200}$ & $0.84\pm0.04$ & $0.13\pm0.03$ & $0.83\pm0.04$ & $0.05\pm0.01$\\[3pt]
$M_V-M^{MB}_V$ & $R_{200}$ & $0.82\pm0.06$ & $0.17\pm0.04$ & $0.80\pm0.06$ & $0.07\pm0.01$\\
\noalign{\smallskip}\hline\noalign{\smallskip}
$M_{HE}-M^{MB}_J$ & $R_{500}$ & $0.98\pm0.09$ & $0.25\pm0.07$ & $0.94\pm0.10$ & $0.10\pm0.03$\\[3pt]
$M_{HE}-M^{MB}_c$ & $R_{500}$ & $1.00\pm0.10$ & $0.30\pm0.06$ & $0.96\pm0.11$ & $0.10\pm0.03$\\[3pt]
$M_{HE}-M^{MB}_V$ & $R_{500}$ & $1.08\pm0.13$ & $0.37\pm0.08$ & $1.01\pm0.14$ & $0.13\pm0.04$\\
\noalign{\smallskip}\hline
\end{tabular}
    \begin{tablenotes}
      \small
      \item Columns: (1) Pair of mass estimators. $M_J$, $M_c$, $M_V$, and $M_{HE}$ refers to the Jeans, caustic, virial theorem, and hydrostatic estimators, respectively. The superscript $RS$ and $MB$ indicate masses derived from the red-sequence galaxies and the clusters' main body, respectively. (2) Aperture within which the masses are estimated. The $R_{200}$ were obtained with the Jeans analysis (fourth column of Table \ref{table:mass}), whereas the $R_{500}$ come from the X-ray analysis. (3) Arithmetic mean of the mass ratio. (4) Standard deviation of the clusters around the mean ratio. (5) Geometrical mean of the mass ratio. (6) Logarithmic scatter (base 10) of the orthogonal residuals with respect to the geometrical mean ratio.
    \end{tablenotes}
  \end{threeparttable}
\end{table*}

\section{Conclusions}

In this paper we presented the dynamical analysis of ten X-ray luminous galaxy clusters at redshifts $z=0.2-0.45$ based on wide-field VIMOS spectroscopic data. Additional WFI photometric observations were used to compute photometric redshifts, in order to constrain the clusters' morphology, to fit their galaxy surface density profiles, and to locate substructures. We derived dynamical masses with three different approaches: the classical virial theorem estimator, a fully parametric forward fitting procedure to solve the Jeans equation, and the caustic method.

We introduced two substructure indicators to quantify the clusters' degree of relaxation. The first indicator is based on the galaxy surface density excess with respect to the best-fit 2D elliptical King model. The second indicator makes use of the Dressler and Shectman test to estimate the fraction of galaxies associated with substructures having a specific dynamics. We also described a methodology to properly identify the galaxies that belong to the main body of the clusters. It combines the visual identification of substructures in the galaxy surface density map with a minimisation-expectation algorithm to partition the galaxies into the different components found with the DS test.

After comparing the results of the three estimators, we constrained the well-known scaling relation between mass and velocity dispersion. Then we confronted our dynamical mass estimates with the hydrostatic values derived from the X-ray data. Finally, we investigated the effect of a colour-based selection of cluster members and the impact of substructures on the dynamical measurements. Our results can be summarised as follows.

\begin{itemize}

\item The Jeans and caustic estimators give very similar results, $\langle M_c/M_J\rangle=1.04\pm0.06$, with a standard deviation $\sigma=0.20\pm0.05$. Masses from the virial theorem are $\sim15\%$ larger, probably because of underestimated SPTs and/or overestimated harmonic radii.\\

\item Our calibration of the scaling relation between mass and velocity dispersion agrees perfectly with the theoretical prediction and the results of numerical simulations for dark matter particles. In particular, the relation calibrated with the Jeans masses is indistinguishable from that of \cite{biviano06}; the caustic masses lead to a slightly smaller logarithmic slope, but the agreement is nonetheless very good over the mass range probed by our sample. Such results indicate the absence of a strong dynamical segregation between galaxies and dark matter.\\

\item On average, the dynamical masses are larger than the hydrostatic values: $\langle M_J/M_{HE}\rangle=1.22\pm0.18$, $\langle M_c/M_{HE}\rangle=1.32\pm0.18$, and $\langle M_V/M_{HE}\rangle=1.51\pm0.26$ for the Jeans, caustic, and virial theorem estimators, respectively.\\

\item The large scatters of the $\langle M_{dyn}/M_{HE}\rangle$ ratios ($\sim50\%$ for the Jeans and caustic estimators, $\sim75\%$ for the virial theorem) are mainly driven by two clusters, RXCJ0014 and RXCJ1347. The former clearly has an overestimated dynamical mass due to substructures (highest value of our indicators), whereas the latter most likely has an overestimated hydrostatic measurement. Excluding RXCJ1347 from the sample leads to a positive fractional mass bias that is significant at the $\gtrsim2\sigma$ level.\\

\item We found strong correlations between the $\langle M_{dyn}/M_{HE}\rangle$ ratios and the substructure indicators, in particular with $f_{\mathrm{DS}}$ (substructures in velocity space). This highlights the limitations of the dynamical mass estimators for clusters observed in a highly disturbed state.\\

\item With an average ratio $\langle\sigma_{blue}/\sigma_{RS}\rangle=1.13\pm0.05$ (scatter of $\sim15\%$), we confirmed previous claims that the blue star-forming galaxies are characterised by a larger velocity dispersion than the passive red-sequence galaxies. The trend is stronger when excluding the galaxies belonging to substructures: $\langle\sigma_{blue}/\sigma_{RS}\rangle=1.27\pm0.05$. Moreover, the absence of a correlation between this ratio and the substructure indicators suggests an intrinsic dynamical segregation between the two galaxy populations regardless of the dynamical state of their cluster host.\\

\item The dynamical masses obtained with the red-sequence galaxies are $10\%-15\%$ smaller than the value derived with the whole population. The impact of the colour selection is the largest for the caustic estimator, most likely due to a lack of red galaxies observed at the escape velocity.\\

\item The removal of substructures leads to dynamical masses that are $\sim15\%$ percent smaller; the largest impact was found for RXCJ0014, whose mass is reduced by a factor of $\sim1.5$. We found that more massive clusters have a larger fraction of substructures and that the more substructures, the larger the impact on dynamical mass estimates. These correlations imply that a proper treatment of substructures is mainly required for the most massive galaxy clusters.\\

\item The dynamical masses perfectly agree with the hydrostatic values after the removal of substructures. Namely, we obtained average ratios $\langle M_J/M_{HE}\rangle=0.98\pm0.09$, $\langle M_c/M_{HE}\rangle=1.00\pm0.10$, and $\langle M_V/M_{HE}\rangle=1.08\pm0.13$ for the Jeans, caustic, and virial theorem estimators, respectively. The standard deviations around the means are roughly divided by two, indicating that our procedure properly corrects the impact of substructure on dynamical masses. Excluding RXCJ1347 from the sample no longer gives a significant fractional mass bias between the hydrostatic and dynamical estimators.\\

\end{itemize}

The takeaway message of this work is that substructures can significantly affect dynamical mass measurements of massive clusters, leading to a spurious mass bias when comparing to the hydrostatic estimates. However, a careful analysis based on large spectroscopic data sets allows to efficiently reduce their impact. Therefore, dynamical mass estimators are an interesting alternative to lensing measurements for the mass calibration of clusters' scaling relations. Our work was focused on the high-mass end of the cluster population, with results that seem to rule out the hydrostatic mass bias commonly found with lensing studies. Conducting a similar analysis on a larger sample including lower-mass systems would allow for a better characterisation of the possible systematics of the X-ray, dynamical, and lensing mass estimators. 

\begin{acknowledgements}

The authors thank Mischa Schirmer for precious advice regarding the reduction of WFI images with THELI. We would like to acknowledge support from the Deutsche Forschungsgemeinschaft through the Transregio Program TR33 and through the Munich Excellence Cluster ``Structure and Evolution of the Universe'', as well as from Deutsches Zentrum f\"ur Luft- und Raumfahrt through grant No. 50OR1601.

\end{acknowledgements}

%
%_____________________________________________________________

\bibliography{../references}

\appendix

\section{Notes on individual clusters}

\subsection*{RXCJ0014.3-3023}
This cluster presents the highest level of substructure in our sample. Its exceptional nature was recently confirmed by \cite{jauzac16} with the latest data from the HST-Frontier Field campaign. Our dynamical results are very similar to those obtained by \cite{owers11} with the same spectroscopic data set. In particular, we find that the cluster core is made of two high-velocity substructures, whose spatial overlap produces the main peak in the galaxy surface density map. As for the mass estimate of the whole system, we can mention the weak-lensing results of \cite{medezinski16}, who reported $M_{200}=2.06\pm0.42\times10^{15}\mathrm{M_\odot}$ when using a single mass component. Their multi-halo model leads to a smaller total mass of $M=1.76\pm0.23\times10^{15}\mathrm{M_\odot}$. Using their $R_{200}$ to rescale our results, we find that the Jeans analysis gives a mass $M(r<2.35\,\mathrm{Mpc})=1.81\pm0.18\times10^{15}\mathrm{M_\odot}$ after the removal of substructures, which agrees well with their findings. The caustic analysis returns a smaller mass $M=1.48\pm0.50\times10^{15}\mathrm{M_\odot}$ that agrees nonetheless with the other estimates within their $1\sigma$ uncertainties. This is because the removal of the two central high-velocity substructures leaves the PPS empty below $\sim0.5$ Mpc (see Fig. \ref{fig:PPS_all}). As a consequence, the caustic amplitude is largely underestimated within this region, thus giving a smaller mass when integrated to larger radii. Prior removing the substructures, the caustic mass within the same aperture is $M=2.36\pm1.10\times10^{15}\mathrm{M_\odot}$, which also agrees with the lensing estimates. Our work presents the first dynamical mass estimates for this cluster.

%The number and locations of the different substructures, as well as a precise modelling of the merger scenario, has not yet reached a consensus (e.g. \citealt{owers11,merten11,medezinski16,jauzac16}).

\subsection*{RXCJ0225.9-4154}
A detailed analysis of this cluster was presented in \cite{foex17}. On the NE side of the cluster core lies an infalling galaxy clump. On the SW part, two additional clumps are found with a small rest-frame velocity, indicating a future merger close to the plane of the sky. The largest of these two clumps has a mass similar to that of the cluster core, hence the dynamical analysis most likely underestimate the total mass of the system by a factor $\sim2$. Several other smaller galaxy clumps are found along the NE-SW axis, extending over at least 8 Mpc. The X-ray emission of the cluster core shows a secondary peak, probably due to a recent merger.

\subsection*{RXCJ0516.6-5430}
This cluster possesses a large amount of substructures in its galaxy surface density map (second largest $\Delta$). It has a bimodal core containing three BCGs and several other galaxy clumps are distributed along the N-S axis (see Fig. \ref{fig:KMM_1}). These remarks suggest that RXCJ0516 is observed at an early stage of its formation history. This is also confirmed by the presence of high-velocity galaxies in the substructures north and south from the core, and by the complex X-ray morphology that is also elongated on the N-S axis (see Fig. B.3 of \citealt{weissmann13}). \cite{mcinnes09} made a weak-lensing analysis of this cluster, as part of the Southern Cosmology Survey. They obtained a mass $M_{200}=0.56\pm0.25\times10^{14}\mathrm{M_\odot}$. In the corresponding radius, $R_{200}\approx1.5$ Mpc, the Jeans analysis gives a mass $M(r<1.5\,\mathrm{Mpc})=1.06\pm0.13\times10^{15}\mathrm{M_\odot}$ after the substructure analysis; the caustic mass at the same radius is $M=1.05\pm0.22\times10^{15}\mathrm{M_\odot}$. Both estimators return a mass roughly twice larger, even after excluding the high-velocity substructures. The same authors provide a mass estimate based on the cluster optical luminosity, $M(L_{200})=1.7\times10^{15}\mathrm{M_\odot}$, which is in good agreement with our $M_{200}$ derived from the Jeans analysis. The consistency of the dynamical estimates, their relatively good agreement with the hydrostatic value, and the high richness and optical luminosity indicate that their lensing mass is most likely underestimated. Our work presents the first dynamical mass estimates for this cluster.

\subsection*{RXCJ0528.9-3927}
A detailed analysis of this cluster was presented in \cite{foex17}. It is located in a rather poor environment, and has a simple dynamical structure (no substructure from the DS test, smallest $f_{\mathrm{DS}}$ value). A second BCG 200 kpc north from the central galaxy and a couple of small substructures in the galaxy surface density map suggests an accretion history along the N-S axis. The X-ray emission peak is slightly offset along the same N-S axis from the centroid estimated at larger scale, which also indicates that the cluster has not yet reached a fully relaxed state.

\subsection*{RXCJ0658.5-5556}
The Bullet cluster is one of the most studied galaxy clusters owing to the peculiar configuration of a merger taking place nearly in the plane of the sky (e.g. \citealt{markevitch04}). The first dynamical analysis of this object was performed by \cite{barrena02}. They applied the virial theorem estimator (accounting for the SPT) on 71 galaxies located within an aperture of radius $\sim1.5$ Mpc. They extrapolated the resulting mass to the radius $R_{200}=\sqrt{3}\sigma_\upsilon/(10H_z)\approx2.65$ Mpc to obtain $M_{200}=1.24\times10^{15}\mathrm{M_\odot}$. After the removal of substructures, our Jeans analysis gives a mass $M(r<2.65\,\mathrm{Mpc})=1.91\pm0.25\times10^{15}\mathrm{M_\odot}$, while the caustic estimator returns $M=1.26\pm0.37\times10^{15}\mathrm{M_\odot}$. The latter agrees well with their estimate, whereas the Jeans mass is somewhat larger. Recently \cite{melchior15} conducted a weak-lensing analysis of the Bullet cluster, finding a mass $M_{200}=1.3\pm0.6\times10^{14}\mathrm{M_\odot}$. At the corresponding radius $R_{200}\approx2$ Mpc, we find $M=1.49\pm0.18\times10^{15}\mathrm{M_\odot}$ and $M=1.11\pm0.26\times10^{15}\mathrm{M_\odot}$ for the Jeans and caustic estimator, respectively. The Jeans analysis gives again a larger mass, however, both values agree perfectly with the weak-lensing estimate. Interestingly, the hydrostatic mass agrees with the dynamical mass too, even though the cluster is known for being dynamically young (the merger takes place on the plane of the sky, hence it does not affect the dynamical measurements).

Regarding the structure of the cluster, it is worth making a few comments. On the one hand, the Bullet did not appear in the DS test. We added a KMM partition anyway and found a rest-frame velocity $\upsilon\approx550\,\mathrm{km\,s^{-1}}$, which is slightly smaller than the value quoted by \cite{barrena02}. On the other hand, the DS test found a group of $\sim10$ high-velocity galaxies ($\upsilon\approx-1200\,\mathrm{km\,s^{-1}}$) between the Bullet and the main clump (blue circles on Fig. \ref{fig:KMM_2}); this high-velocity substructure is clearly visible in the PPS diagram (Fig. \ref{fig:PPS_all}) and is responsible for the bimodal velocity distribution (Fig. \ref{fig:nz_all}). The cluster itself is at the centre of a rich larger-scale environment with a clear filamentary-like extension towards south and an additional large galaxy clump northwest from the Bullet (see Fig. \ref{fig:KMM_2}). The dynamics from the southern galaxies suggest that the filament is located behind the cluster (the KMM partition associated with the blue triangles has a rest-frame velocity $\upsilon\approx-1000\,\mathrm{km\,s^{-1}}$). Finally, we can note that \cite{melchior15} found a similar NW-SE orientation in the galaxy distribution at even larger scales (see their Fig. 8).  

\subsection*{RXCJ1131.9-1955}
A detailed analysis of this cluster was presented in \cite{ziparo12}. Our analysis confirms that the cluster possesses a complex structure that is characteristic of a dynamically young object. Several galaxy overdensities are found within its virial radius. Two of them present a specific dynamics that can be associated to two filamentary structures located north (positive rest-frame velocity) and south (negative rest-frame velocity) from the core. The latter filament can be traced further south with two additional KMM partitions (see Fig. \ref{fig:KMM_2}). \cite{ziparo12} used the scaling relation of \cite{biviano06} to estimate the mass of the cluster's main body from its velocity dispersion. They obtained $M_{200}=1.1\pm0.2\times10^{15}\mathrm{M_\odot}$ and a corresponding $R_{200}\approx1.9$ Mpc. Within the same aperture, the Jeans analysis and the caustic estimator give, after the substructures removal, a mass $M(r<1.9\,\mathrm{Mpc})=1.26\pm0.15\times10^{15}\mathrm{M_\odot}$ and $M=1.16\pm0.64\times10^{15}\mathrm{M_\odot}$, respectively. Both values agree with the prediction of the $M(\sigma_\upsilon)$ scaling law.

\subsection*{RXCJ1206.2-0848}
Our analysis of RXCJ1206 is based on the same data set that was used for the detailed dynamical analyses of \cite{biviano13}, \cite{lemze13}, and \cite{girardi15}. Therefore, this cluster provides a consistency check for our results. In particular, \cite{biviano13} reported a mass $M(r<1.96\,\mathrm{Mpc})=1.37\pm0.18\times10^{15}\mathrm{M_\odot}$ from the Jeans analysis, and $M(r<2.08\,\mathrm{Mpc})=1.63\pm0.58\times10^{15}\mathrm{M_\odot}$ from the caustic method. Once rescaled to the same radii, our implementation of the Jeans approach gives $M=1.51\pm0.14\times10^{15}\mathrm{M_\odot}$, while the caustic mass profile leads to $M=1.70\pm0.64\times10^{15}\mathrm{M_\odot}$, i.e. a perfect agreement for both estimators. Different lensing analyses of the cluster have found very similar results, i.e. $M_{200}\approx1.5\times10^{15}\mathrm{M_\odot}$ (see e.g. the latest work by \citealt{sereno17}).

\subsection*{RXCJ1347.5-1144}
This cluster, which is the brightest in the REFLEX sample \citep{bohringer04}, has generated a great deal of discussion regarding its mass estimate using either X-ray, SZ, lensing, or dynamical data sets (see e.g. \citealt{cohen02,gitti07,bradac08}). Most of the debate originates from the first dynamical analysis by \cite{cohen02}, who reported a velocity dispersion $\sigma_P=910\pm130\,\mathrm{km s^{-1}}$, whereas lensing analyses consistently find an equivalent $\sigma_P\sim1300-1500\,\mathrm{km s^{-1}}$ (e.g. \citealt{fischer97,bradac08,lu10,foex12,hoekstra15}). The merger of a substructure located close to the core has been invoked to explain the large discrepancies between the different mass estimates. Such a substructure can boost the strong-lensing signal, leading to an overestimated mass when extrapolated to large radii, while shocks associated with the merger can significantly enhance the gas temperature and X-ray luminosity, leading to a biased high hydrostatic mass. On the other hand, if the merger is taking place on the plane of the sky, a dynamical analysis would fail at estimating the total mass of the system. The lack of substructure found by the DS test and the NE-SW elongation of the galaxy distribution (see Fig. \ref{fig:KMM_2}) support the merger configuration described by \cite{kreisch16}.

Our dynamical results tend to agree with the original work of \cite{cohen02} since they obtained a velocity dispersion $\sigma_P=820\pm110\,\mathrm{km s^{-1}}$ after a $3\sigma$ clipping selection of the cluster members. Using this value and accounting for the SPT correction at a radius of 1 Mpc, their mass estimate becomes $M\approx4\times 10^{14}\mathrm{M_\odot}$, which is perfectly consistent with our results (interpolated to the same radius) $M=3.9\pm0.5\times 10^{14}\mathrm{M_\odot}$ and $M=5.2\pm1.6\times10^{14}\mathrm{M_\odot}$ for the Jeans analysis and caustic estimators, respectively. More recently, \cite{lu10} reported a dynamical mass $M_{200}=1.16\pm0.3\times10^{15}\mathrm{M_\odot}$ (at a corresponding $R_{200}\approx1.85$ Mpc) under the assumption of a singular isothermal density profile. They also obtained a caustic mass $M\sim1\times10^{15}\mathrm{M_\odot}$ within the same aperture. Our analysis gives $M=0.75\pm0.1\times10^{15}\mathrm{M_\odot}$ and $M=0.91\pm0.3\times10^{15}\mathrm{M_\odot}$ for the Jeans and caustic estimators, respectively. The former is somewhat smaller, as expected given the velocity dispersion $\sigma_P=1163\pm97\,\mathrm{km s^{-1}}$ reported by these authors. They also derived a weak-lensing mass $M=1.47\pm0.45\times10^{15}\mathrm{M_\odot}$, which is again larger than the dynamical measurements. Several galaxy clumps found in the vicinity of the cluster, in particular in the SW region, could contribute to the shear signal, thus explaining the larger weak-lensing masses.  

\subsection*{RXCJ2011.3-5725}
This cluster is embedded in a rich star field. It was observed only in the B and V bands, making the photometric redshifts much less accurate, and thus preventing a precise characterisation of the cluster morphology. The DS test finds a small high-velocity substructure right in the cluster centre, so we removed the corresponding KMM partition before re-estimating the dynamical masses. Since the results of the DS test could be the consequence of low statistics, we also verified that our conclusions in Sec. \ref{sec:newmass} are essentially unchanged if we do not remove this KMM partition prior to the mass estimation. Our work presents the first dynamical results for this cluster.

\subsection*{RXCJ2308.3-0211}
A detailed analysis of this cluster was presented in \cite{foex17}. It is part of a supercluster, with several galaxy over-densities distributed along two main axes, N-S and E-W. The cluster core shows signs of a recent merger with significant X-ray residuals and possibly two high-velocity substructures following orbits close to the line of sight. \cite{newman13} obtained a combined strong+weak lensing mass $M_{200}=1.32\pm0.12\times10^{15}\mathrm{M_\odot}$ for a radius $R_{200}\approx2.05$ Mpc. After removing the substructures, the Jeans analysis gives a mass $M(r<2.05\,\mathrm{Mpc})=1.22\pm0.13\times10^{15}\mathrm{M_\odot}$ that agrees perfectly with their result. The caustic estimator gives $M=1.06\pm0.23\times10^{15}\mathrm{M_\odot}$. As for RXCJ0014, this lower value is the consequence of the PPS being empty in the central region after the removal of the two high-velocity substructures. RXCJ0014 and RXCJ2308 are two good examples for which our approach to treat substructures is not optimal for the caustic estimator. We note that prior to the substructure analysis, we obtain a caustic mass $M(r<2.05\,\mathrm{Mpc})=1.73\pm0.41\times10^{15}\mathrm{M_\odot}$ that agrees with the lensing value.

\section{Substructure indicators}

\begin{figure}
\center
\includegraphics[width=6.5cm, angle=-90]{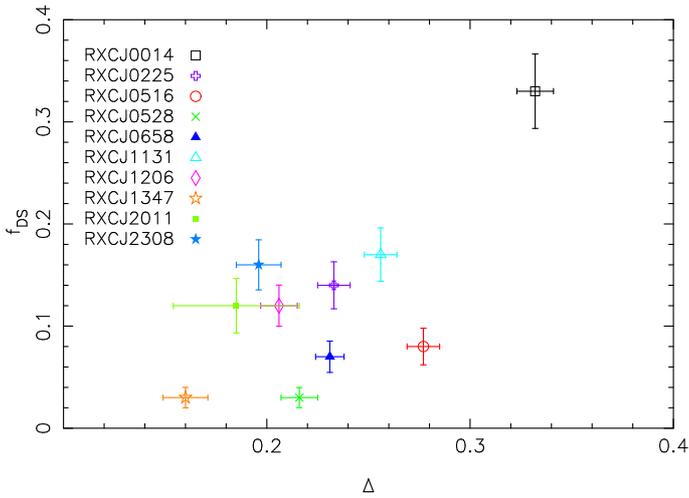}
\caption{Comparison between the photometric ($\Delta$) and dynamical ($f_{\mathrm{DS}}$) substructure indicators.}
\label{fig:PDsub} 
\end{figure}

\begin{figure}
\center
\includegraphics[width=6.5cm, angle=-90]{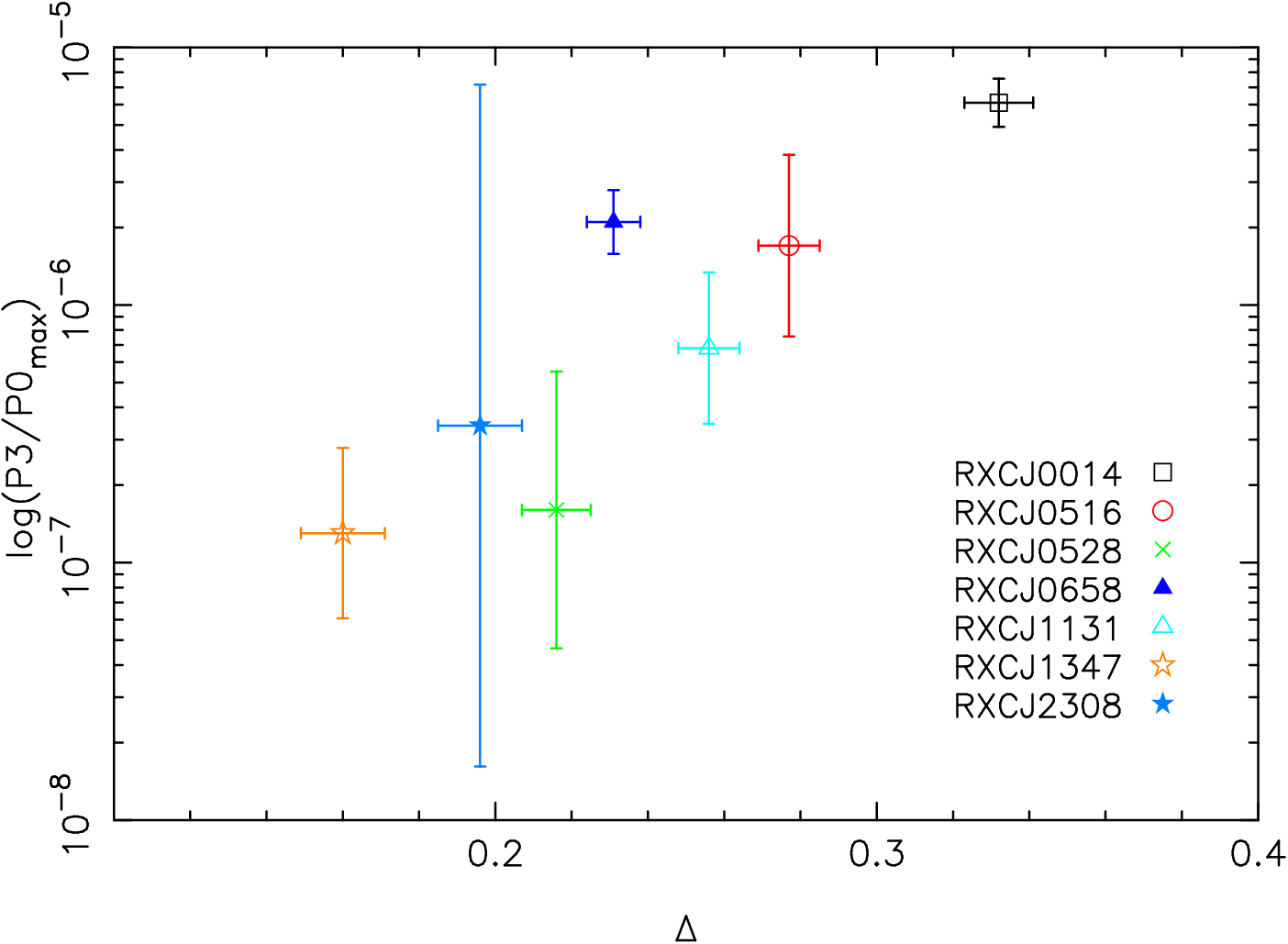}\\
\includegraphics[width=6.5cm, angle=-90]{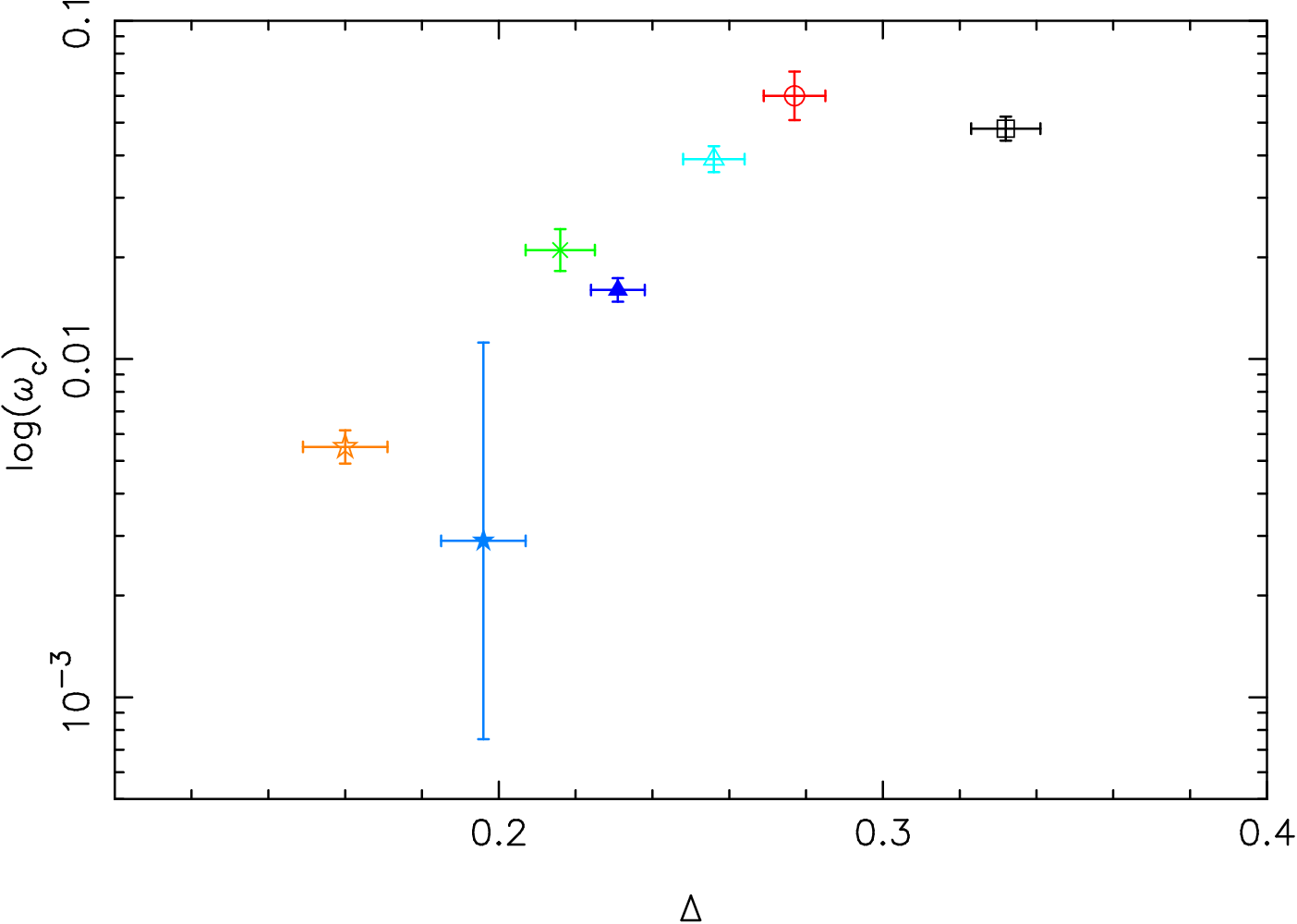}
\caption{Comparison between the photometric substructure indicator and the X-ray power ratio (top panel) and centre shift (bottom panel).}
\label{fig:Xsub} 
\end{figure}

The Figure \ref{fig:PDsub} shows the photometric and dynamical substructure indicators defined in Sections \ref{sec:delta} and \ref{sec:DS}. In Figure \ref{fig:Xsub} we present the comparison between the photometric indicator and the X-ray morphology estimators; the latter were taken from \cite{weissmann13}.

\section{Intermediate results}

We provide in the following some intermediate results of the photometric and dynamical analyses. In particular, Figs. \ref{fig:KMM_1}-\ref{fig:KMM_3} present the spatial distribution of the spectroscopically confirmed members, the galaxy surface density contours, and the different KMM partitions associated with substructures.

\begin{figure*}
\includegraphics[width=0.23\textwidth, angle=-90]{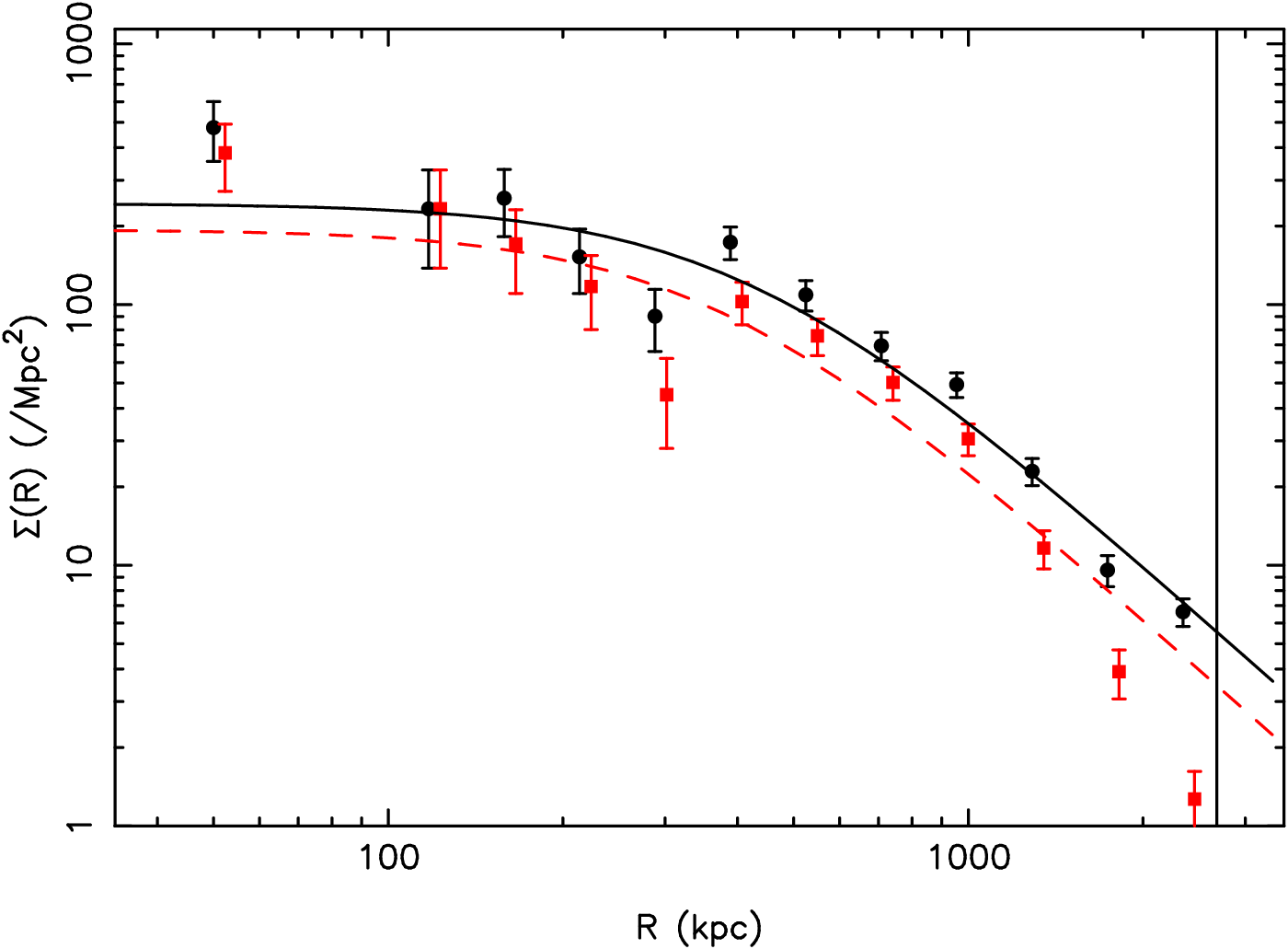}\hspace{0.02\textwidth}%
\includegraphics[width=0.23\textwidth, angle=-90]{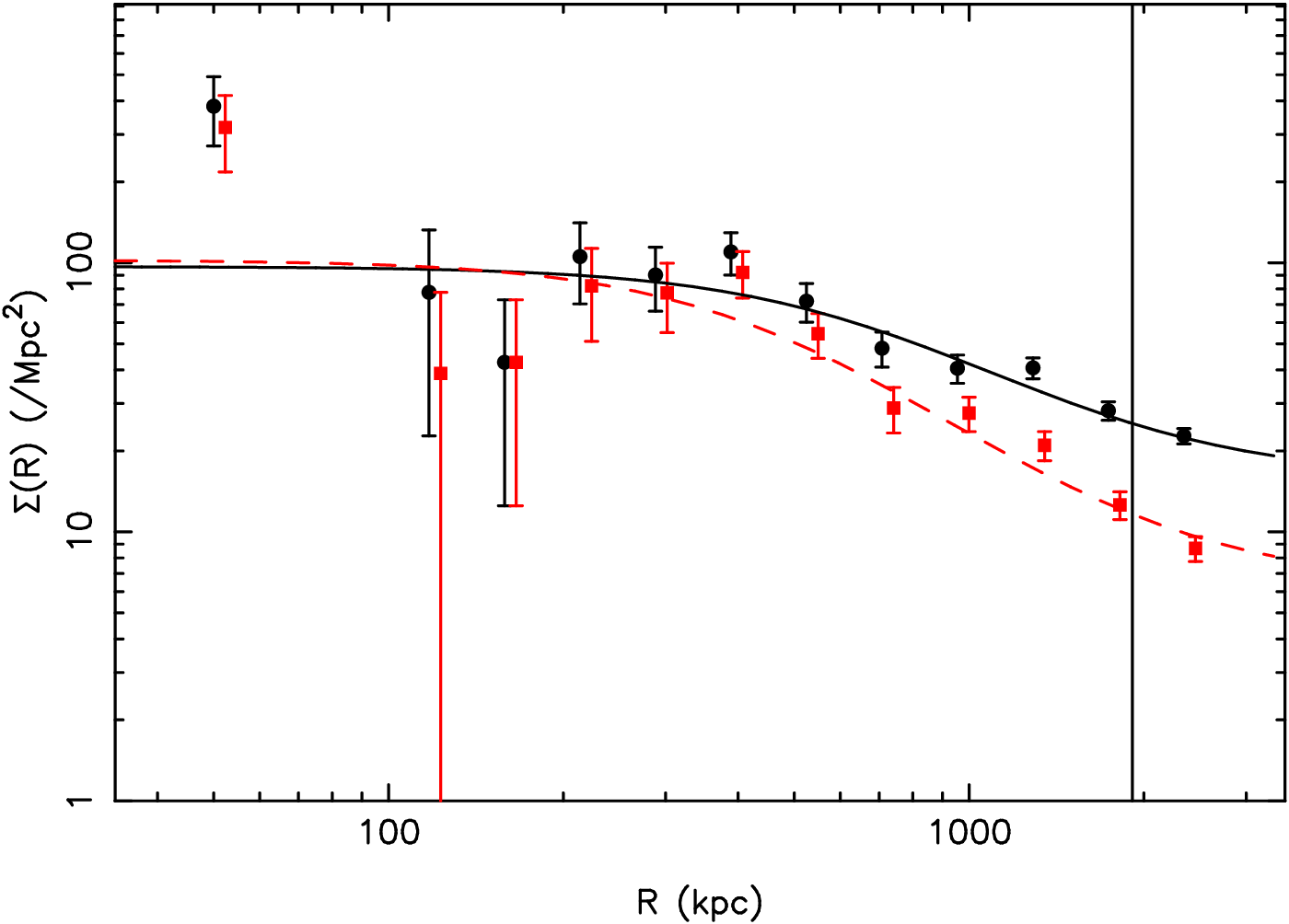}\hspace{0.02\textwidth}%
\includegraphics[width=0.23\textwidth, angle=-90]{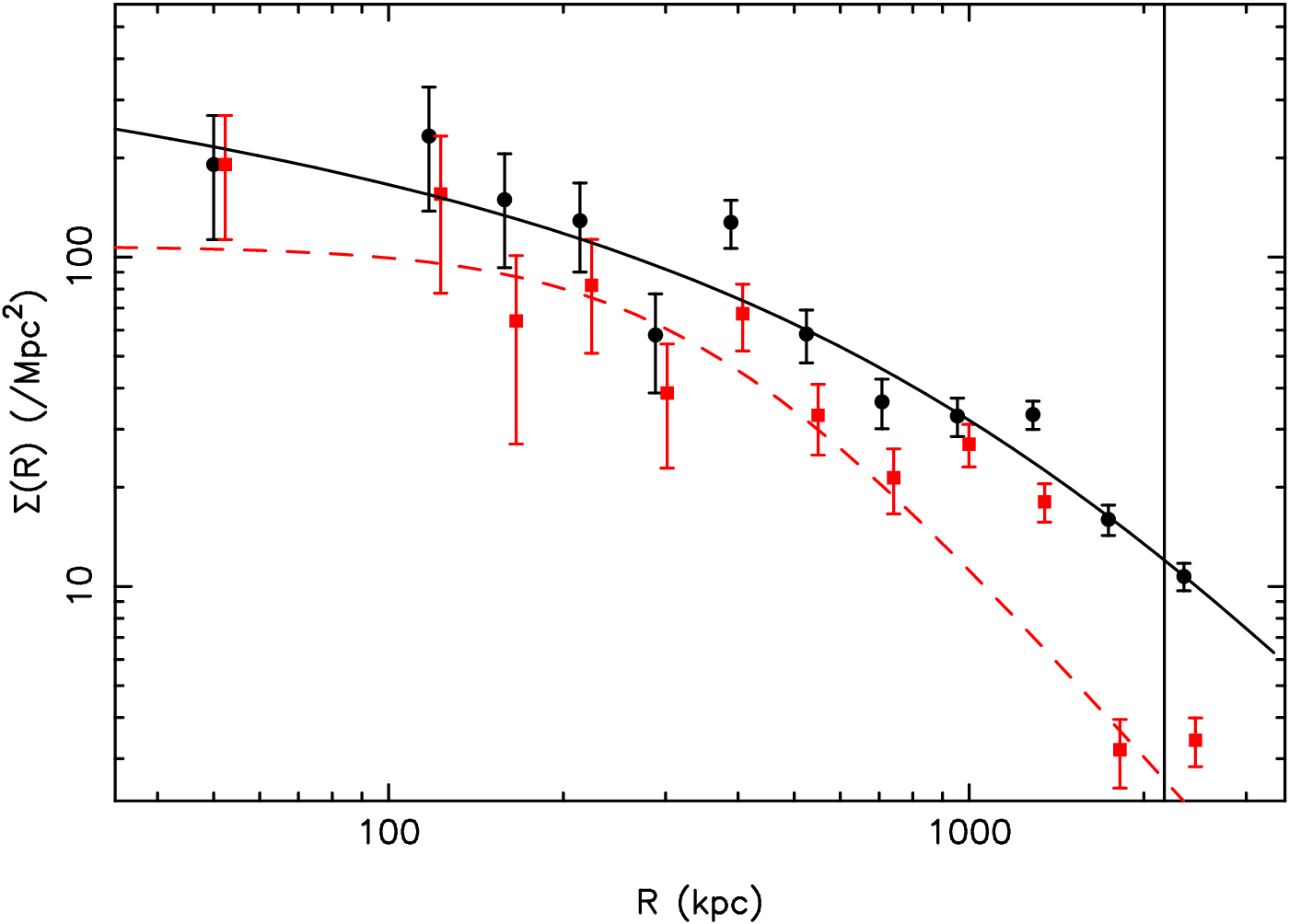}\\
\includegraphics[width=0.24\textwidth, angle=-90]{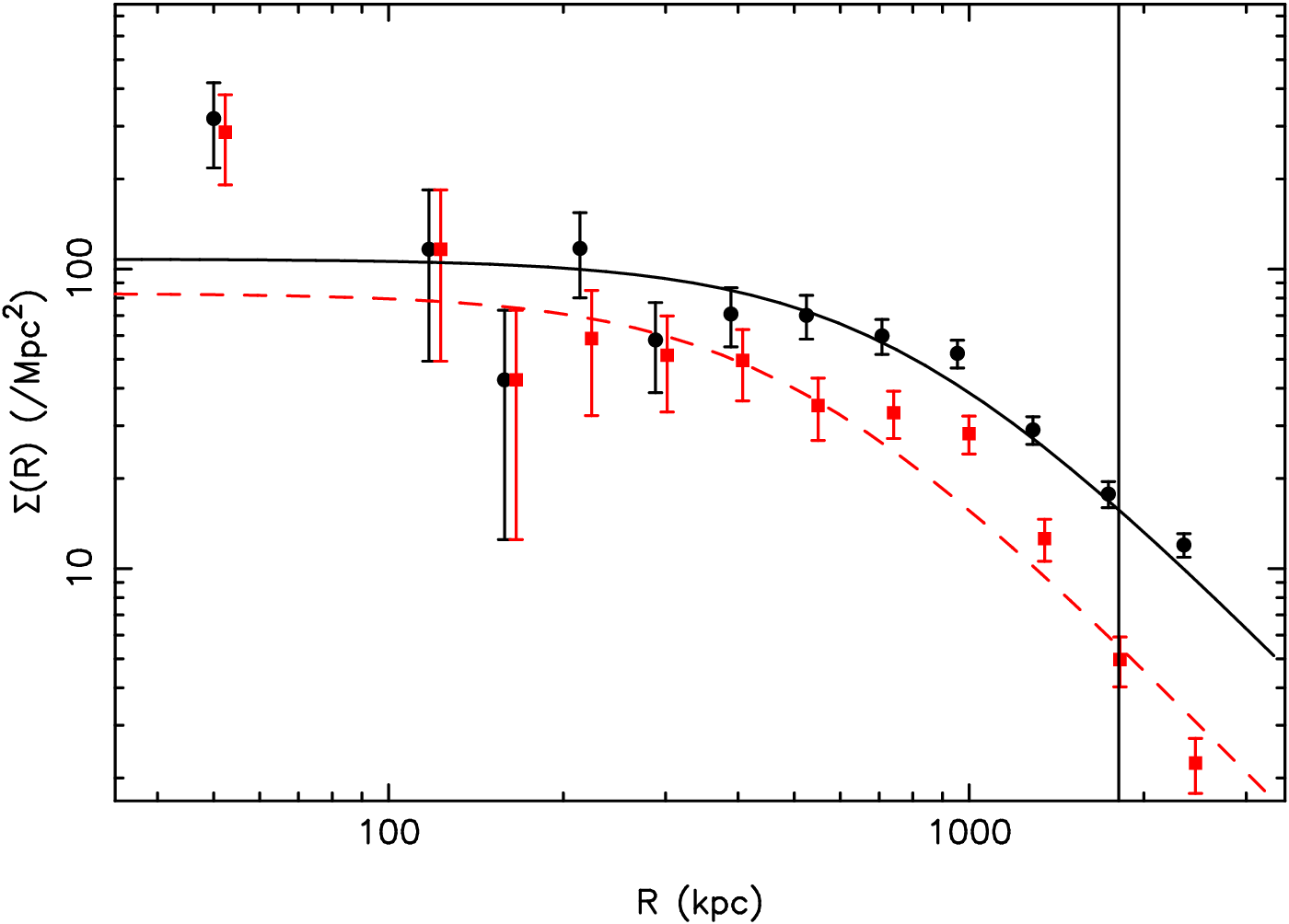}\hspace{0.02\textwidth}%
\includegraphics[width=0.23\textwidth, angle=-90]{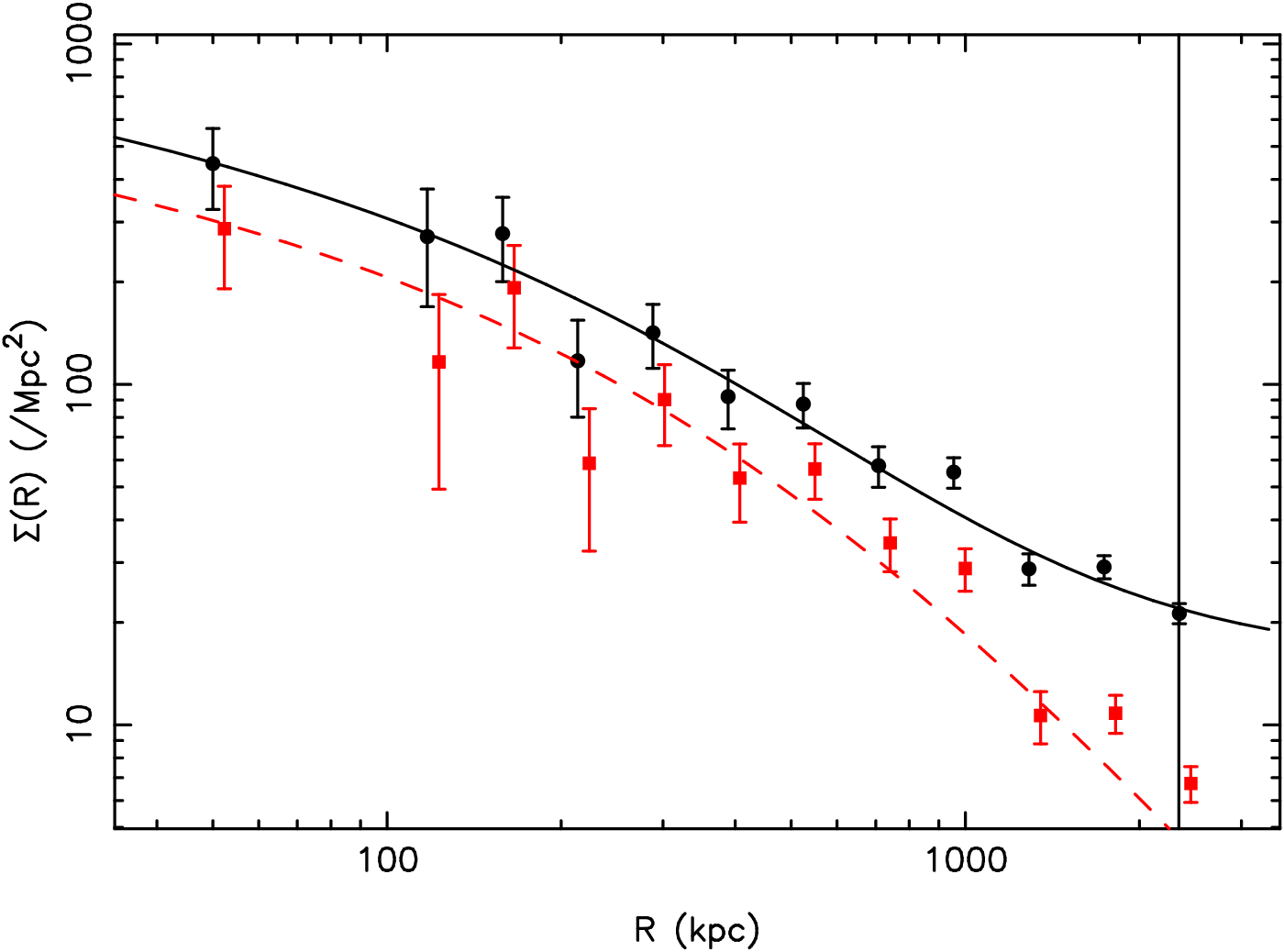}\hspace{0.02\textwidth}%
\includegraphics[width=0.23\textwidth, angle=-90]{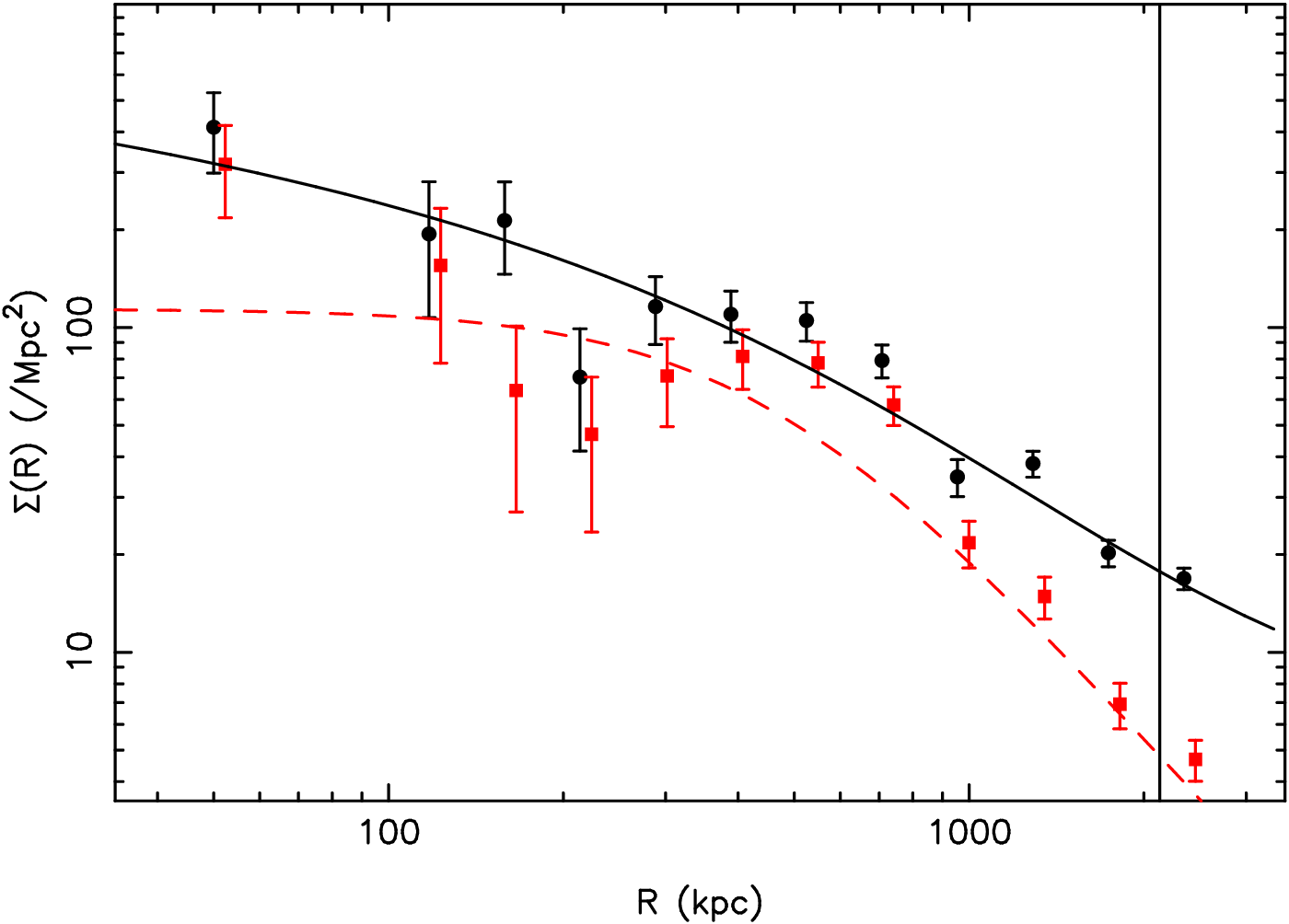}\\
\includegraphics[width=0.23\textwidth, angle=-90]{RXCJ1206.eps}\hspace{0.02\textwidth}%
\includegraphics[width=0.23\textwidth, angle=-90]{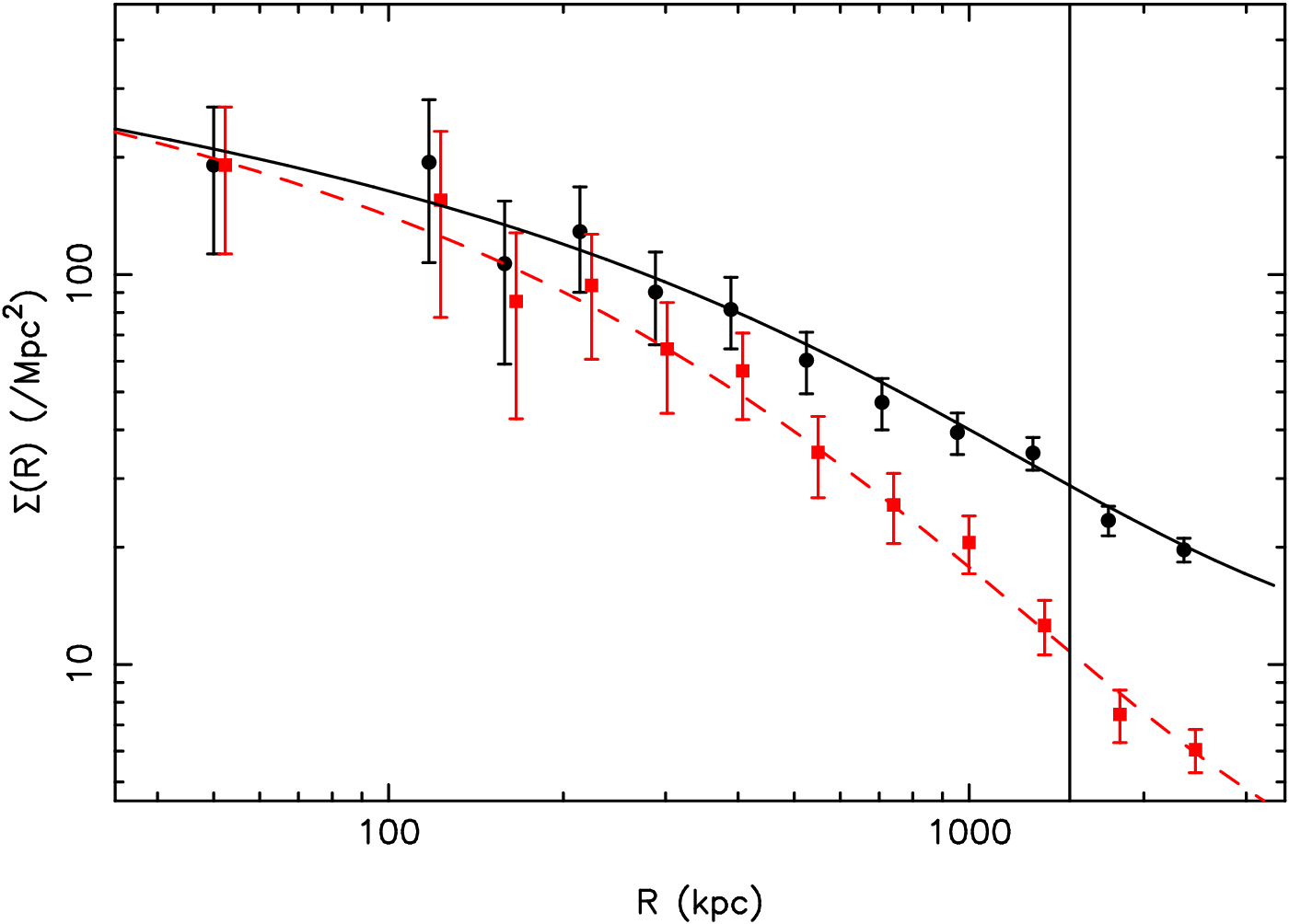}\hspace{0.02\textwidth}%
\includegraphics[width=0.23\textwidth, angle=-90]{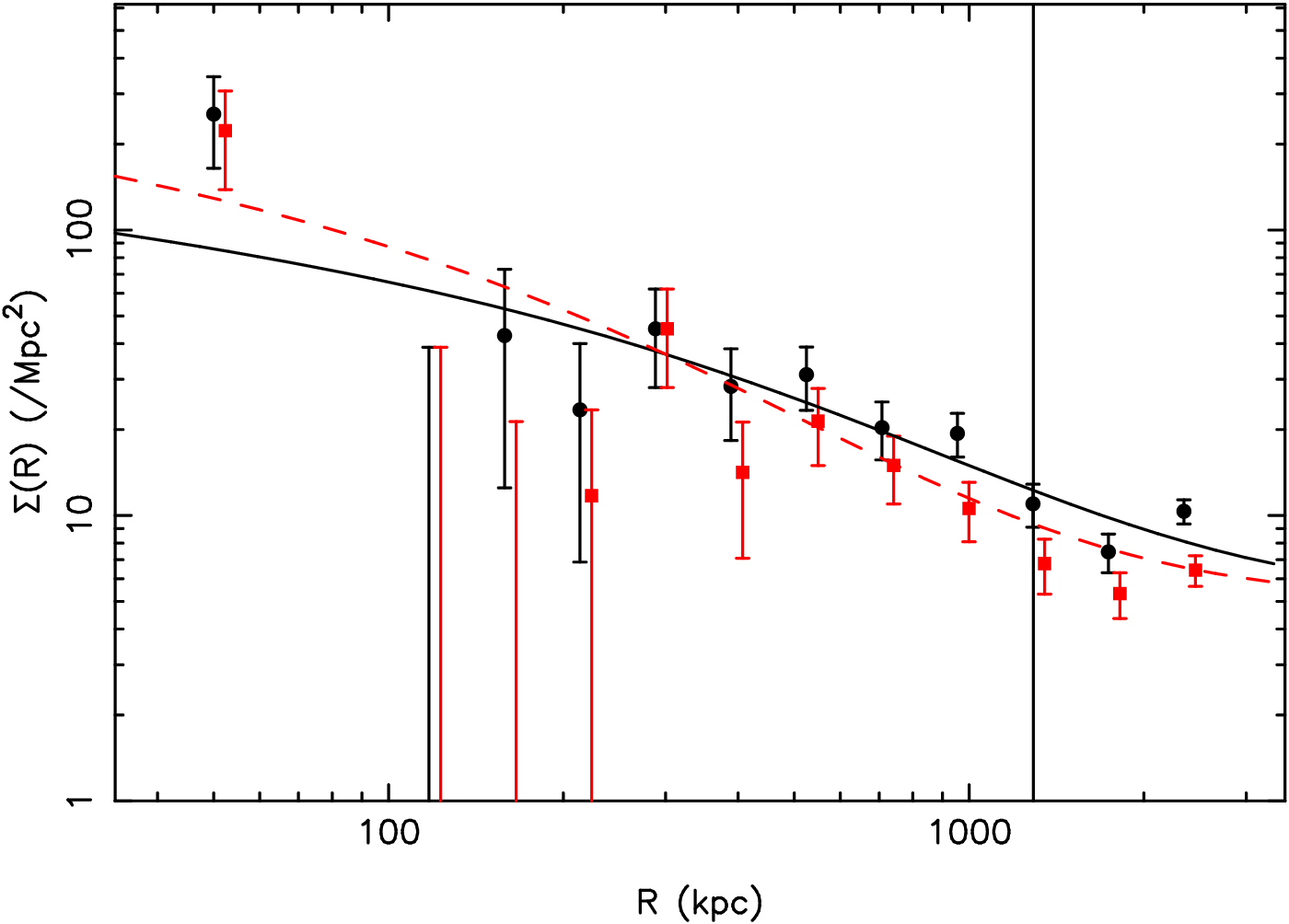}\\
\includegraphics[width=0.23\textwidth, angle=-90]{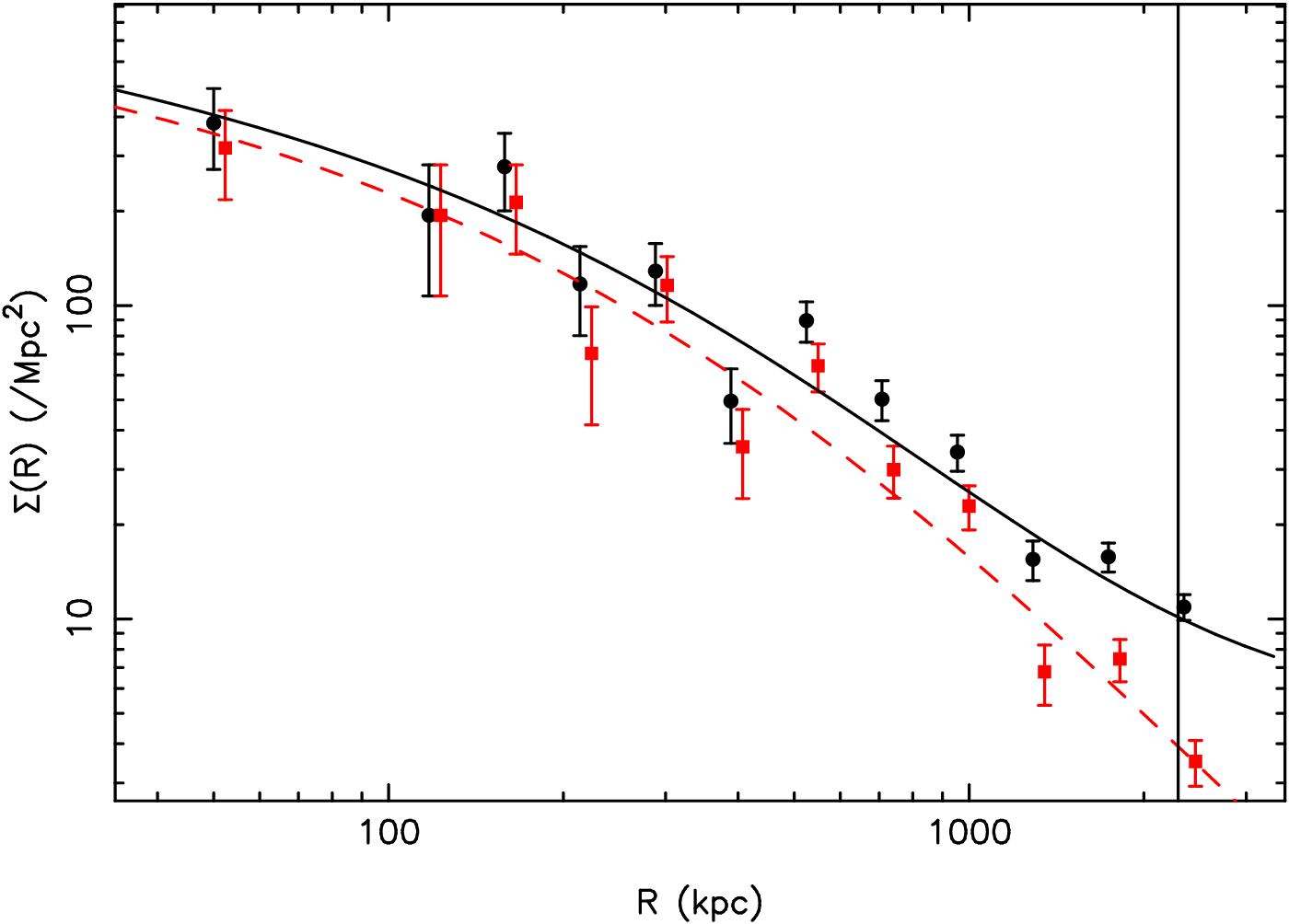}
\caption{Galaxy surface density profile (black points) and the best-fit model (black curve) combining either a King or NFW profile with a constant background residual. The red-dashed line and red points correspond to the population of red-sequence galaxies. The best-fit characteristic radii are used during the Jeans analysis, and to estimate the SPT correction for the virial theorem. The vertical line shows $R_{200}$, as estimated from the Jeans equation, prior the substructure analysis. From left to right, top to bottom: RXCJ0014, RXCJ0225, RXCJ0516, RXCJ0528, RXCJ0658, RXCJ1131, RXCJ1206, RXCJ1347, RXCJ2011, and RXCJ2308.}\label{fig:sigma_prof_all}
\end{figure*}

\begin{figure*}
\includegraphics[width=0.23\textwidth, angle=-90]{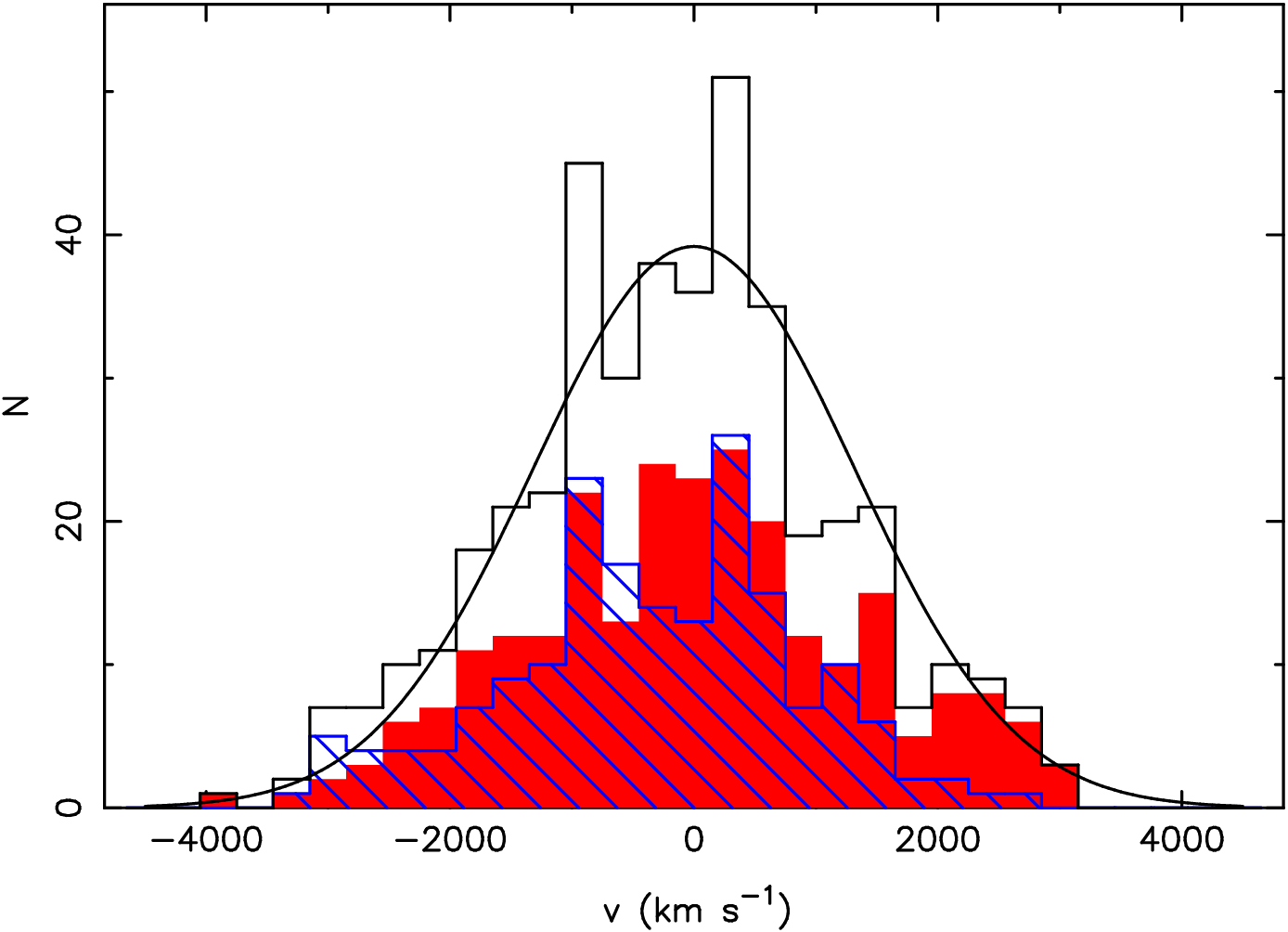}\hspace{0.02\textwidth}%
\includegraphics[width=0.23\textwidth, angle=-90]{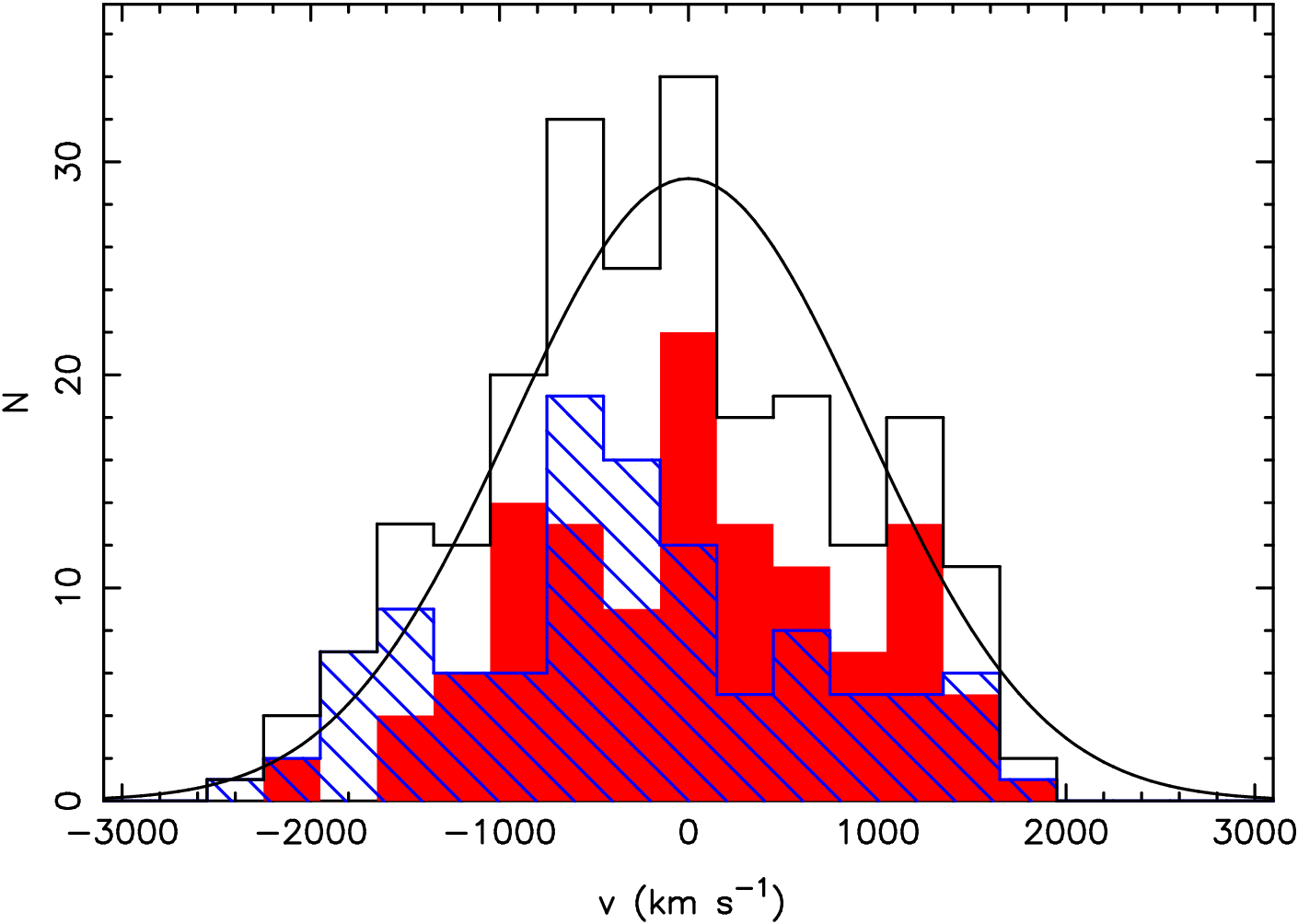}\hspace{0.02\textwidth}%
\includegraphics[width=0.23\textwidth, angle=-90]{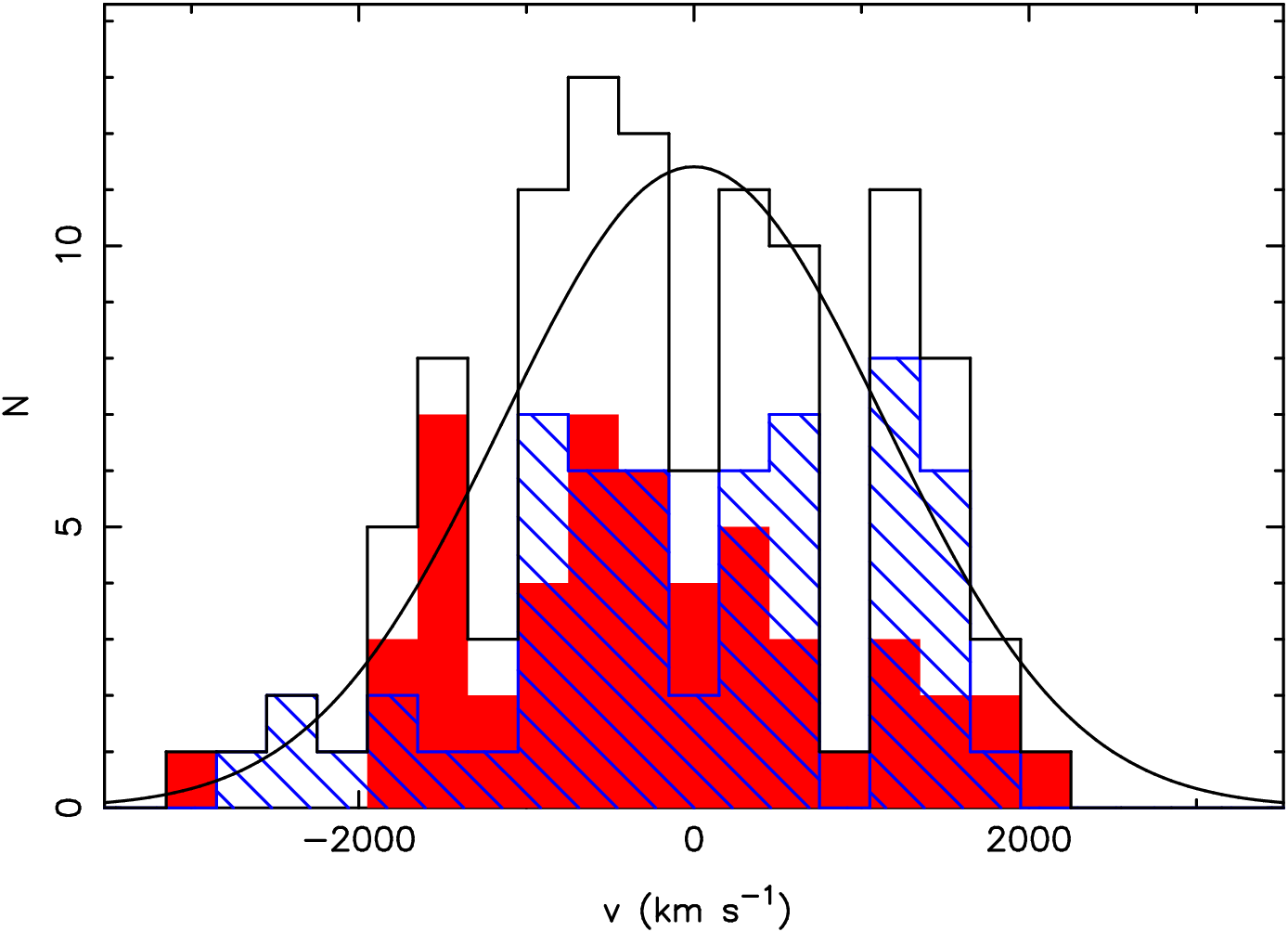}\\
\includegraphics[width=0.23\textwidth, angle=-90]{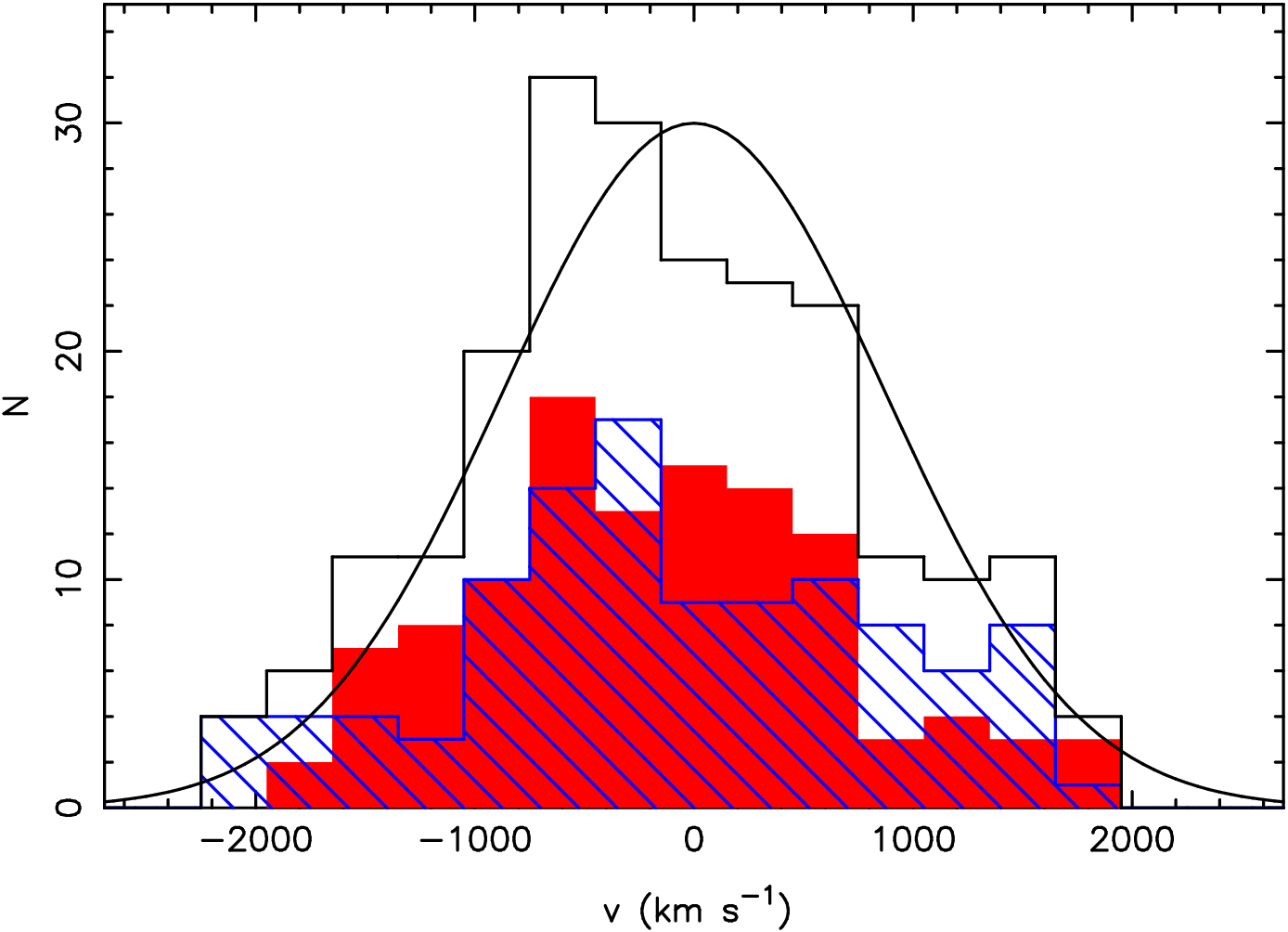}\hspace{0.02\textwidth}%
\includegraphics[width=0.23\textwidth, angle=-90]{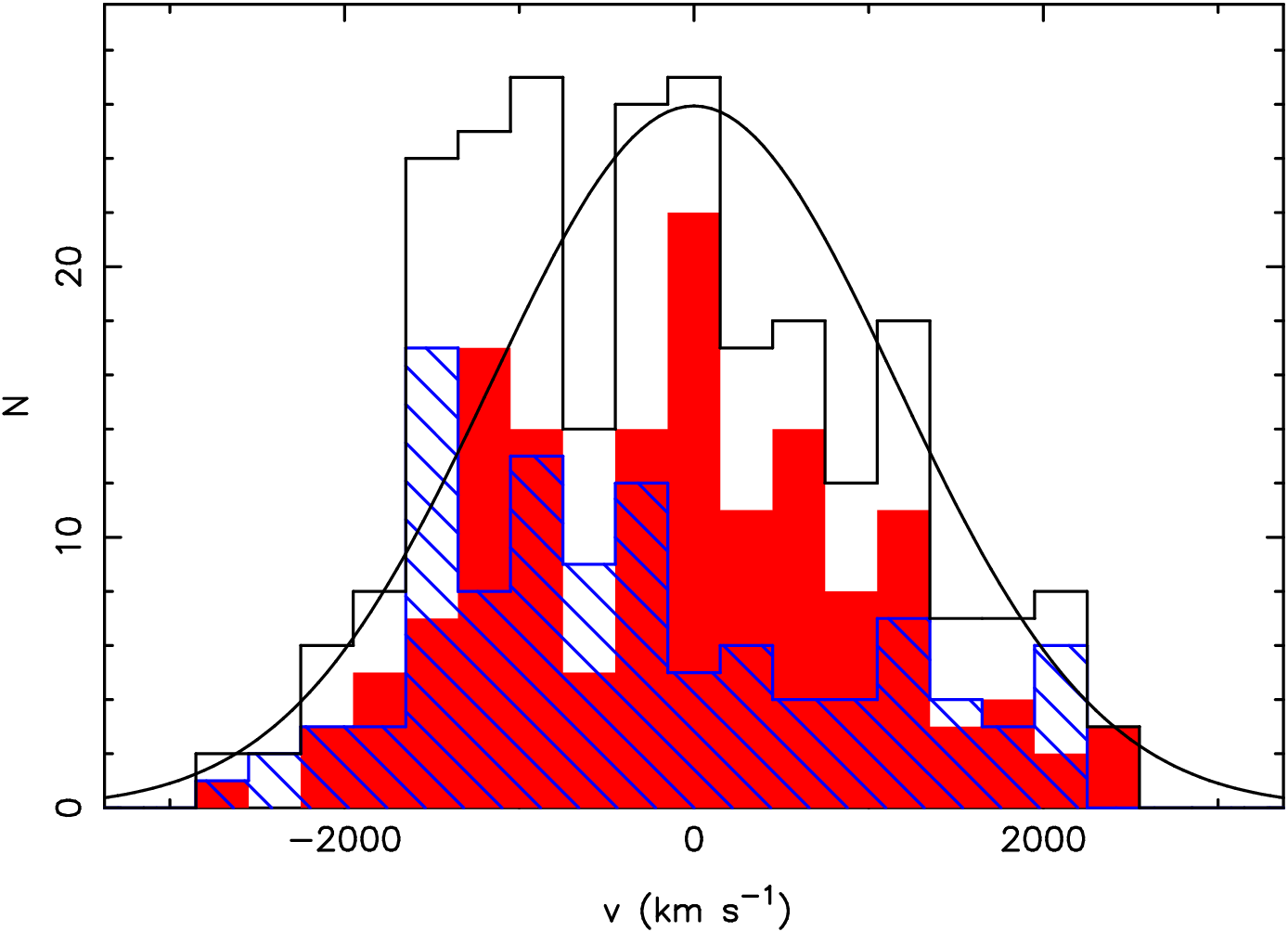}\hspace{0.02\textwidth}%
\includegraphics[width=0.23\textwidth, angle=-90]{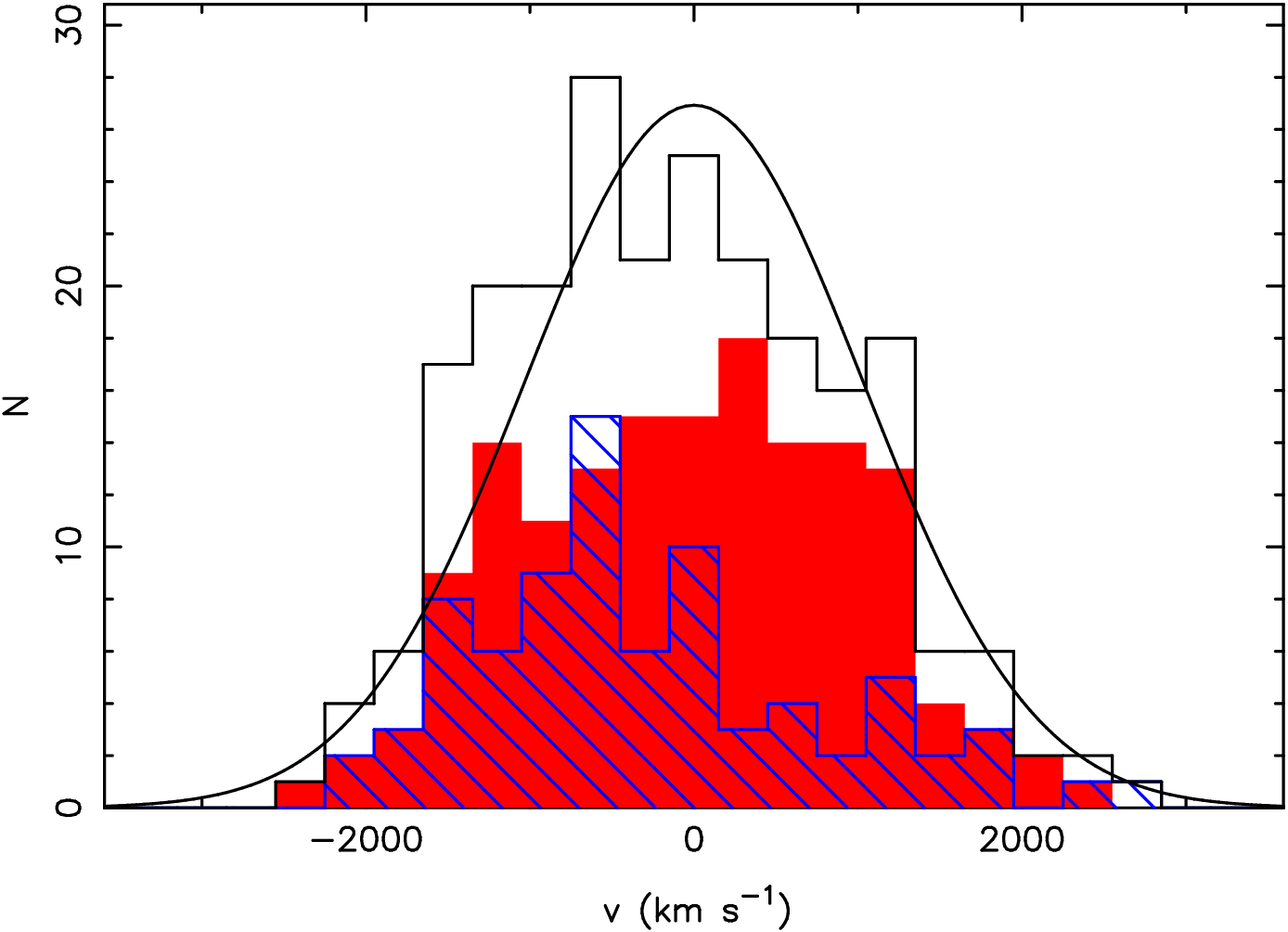}\\
\includegraphics[width=0.23\textwidth, angle=-90]{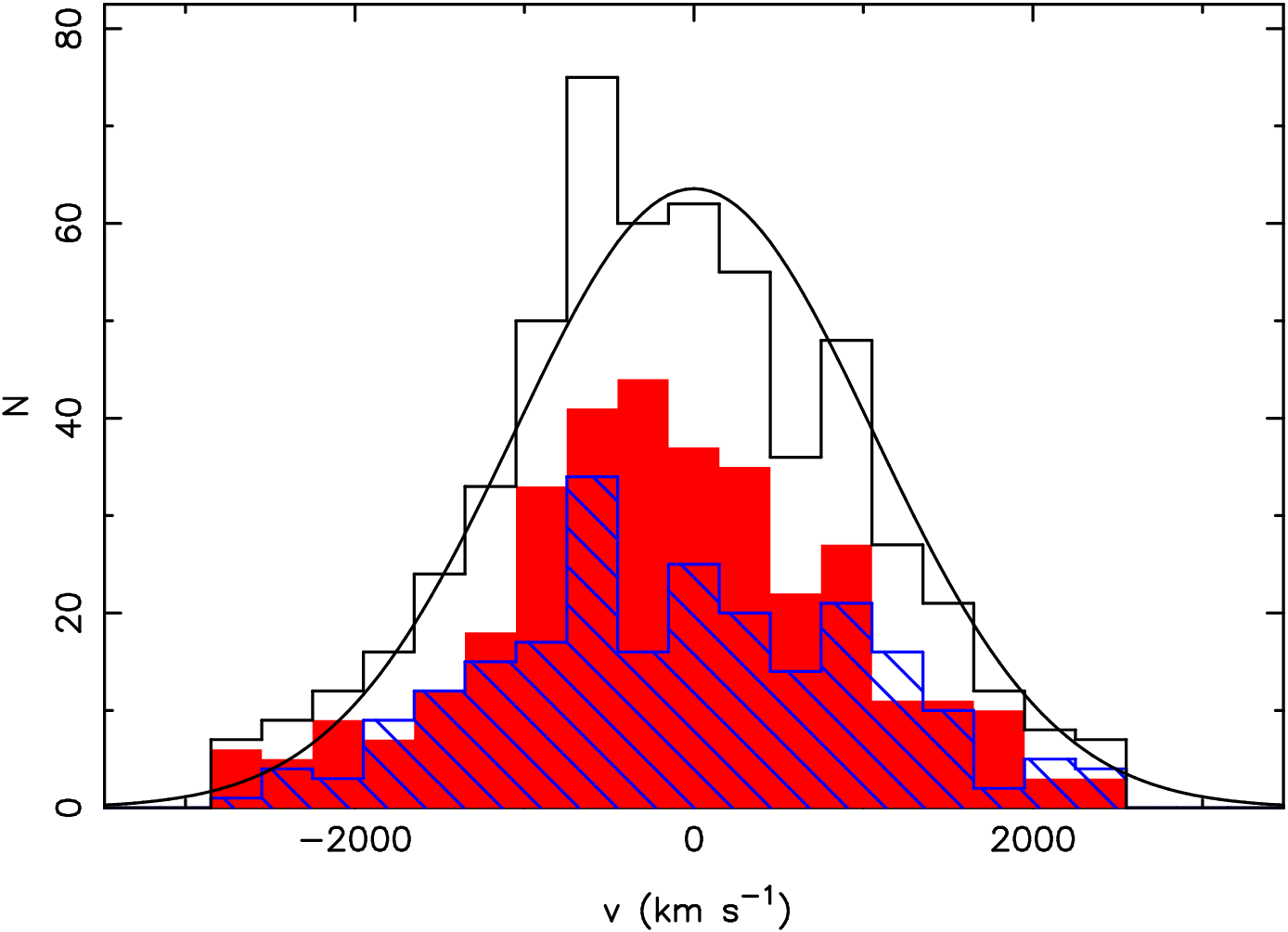}\hspace{0.02\textwidth}%
\includegraphics[width=0.23\textwidth, angle=-90]{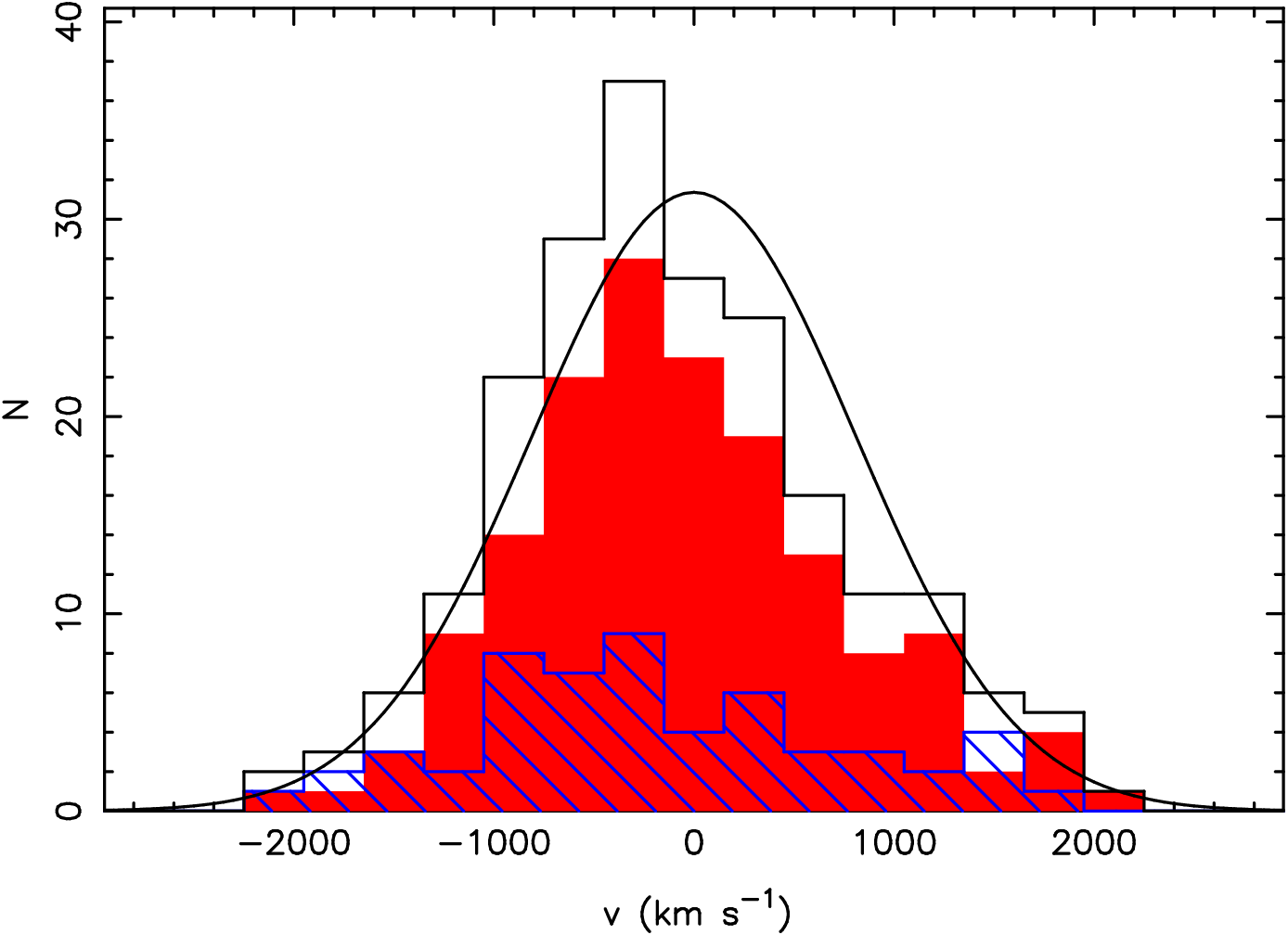}\hspace{0.02\textwidth}%
\includegraphics[width=0.23\textwidth, angle=-90]{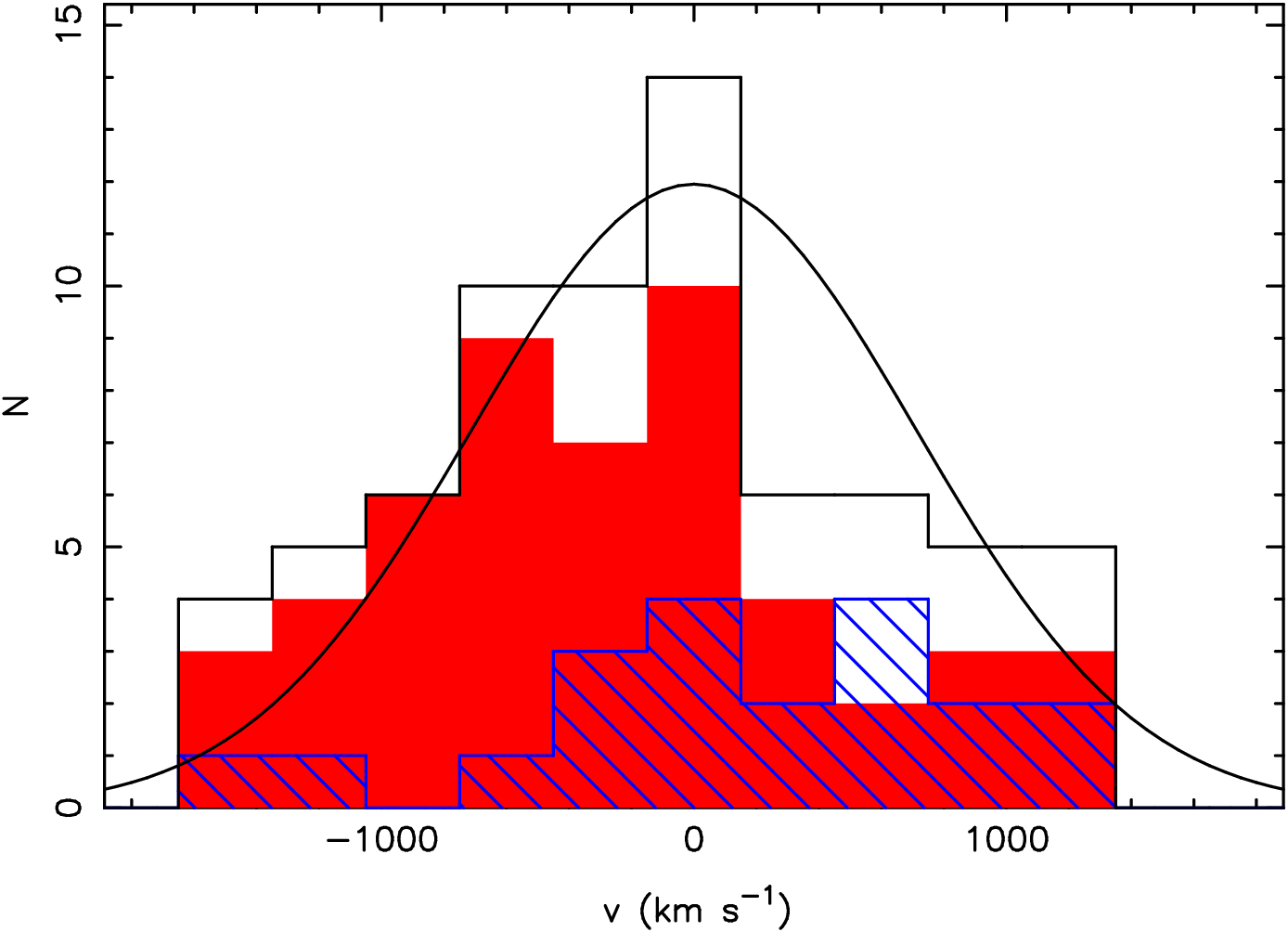}\\
\includegraphics[width=0.23\textwidth, angle=-90]{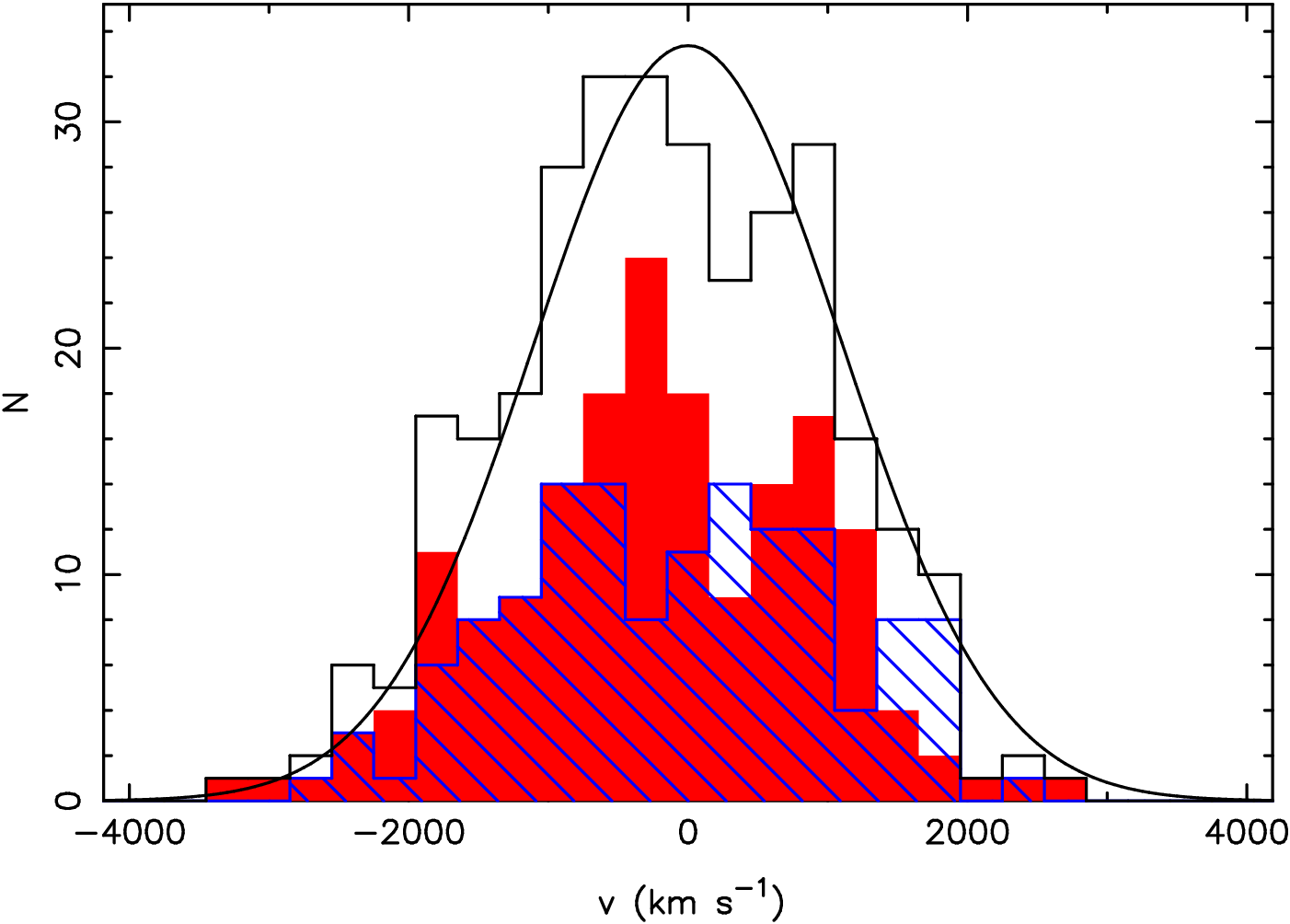}
\caption{Rest-frame velocity distribution of all the spectroscopically confirmed cluster members (including those outside $R_{200}$). The red-shaded histogram is for the red-sequence galaxies, while the blue-hatched one is for the blue population. The velocity dispersion of the total population is represented by the Gaussian curve, which is centred on 0. From left to right, top to bottom: RXCJ0014, RXCJ0225, RXCJ0516, RXCJ0528, RXCJ0658, RXCJ1131, RXCJ1206, RXCJ1347, RXCJ2011, and RXCJ2308.}\label{fig:nz_all}
\end{figure*}

\begin{figure*}
\includegraphics[width=0.23\textwidth, angle=-90]{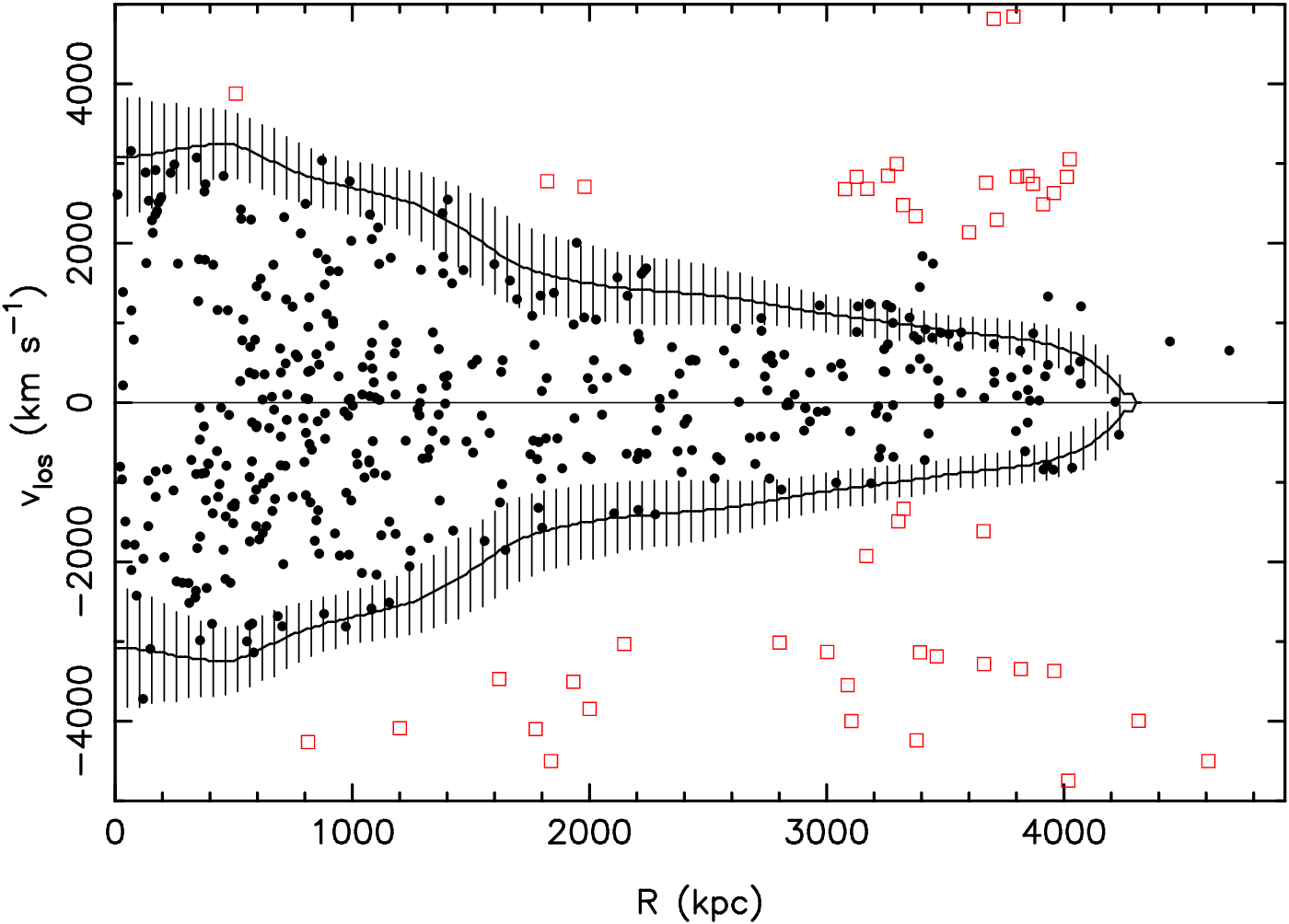}\hspace{0.02\textwidth}%
\includegraphics[width=0.23\textwidth, angle=-90]{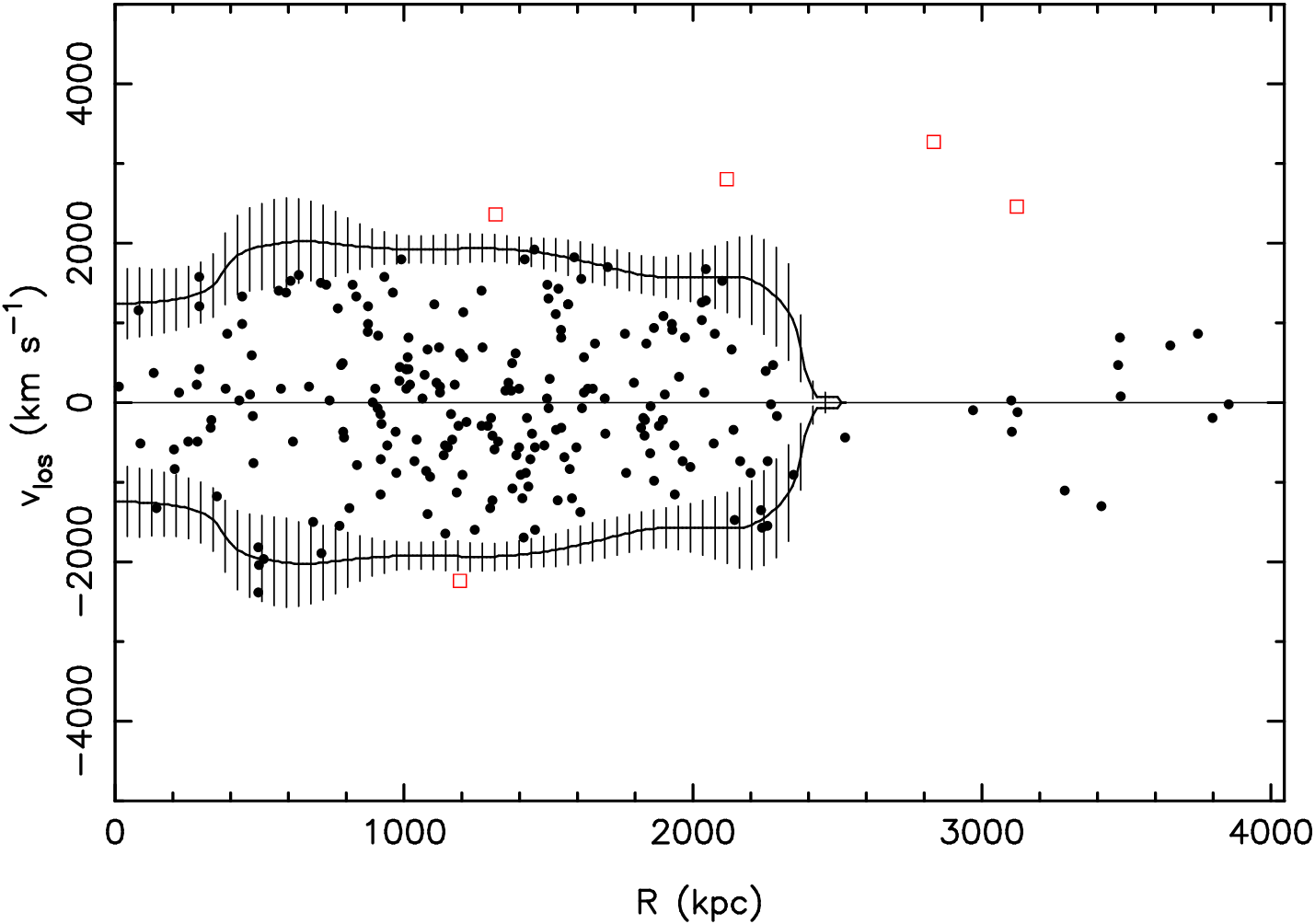}\hspace{0.02\textwidth}%
\includegraphics[width=0.23\textwidth, angle=-90]{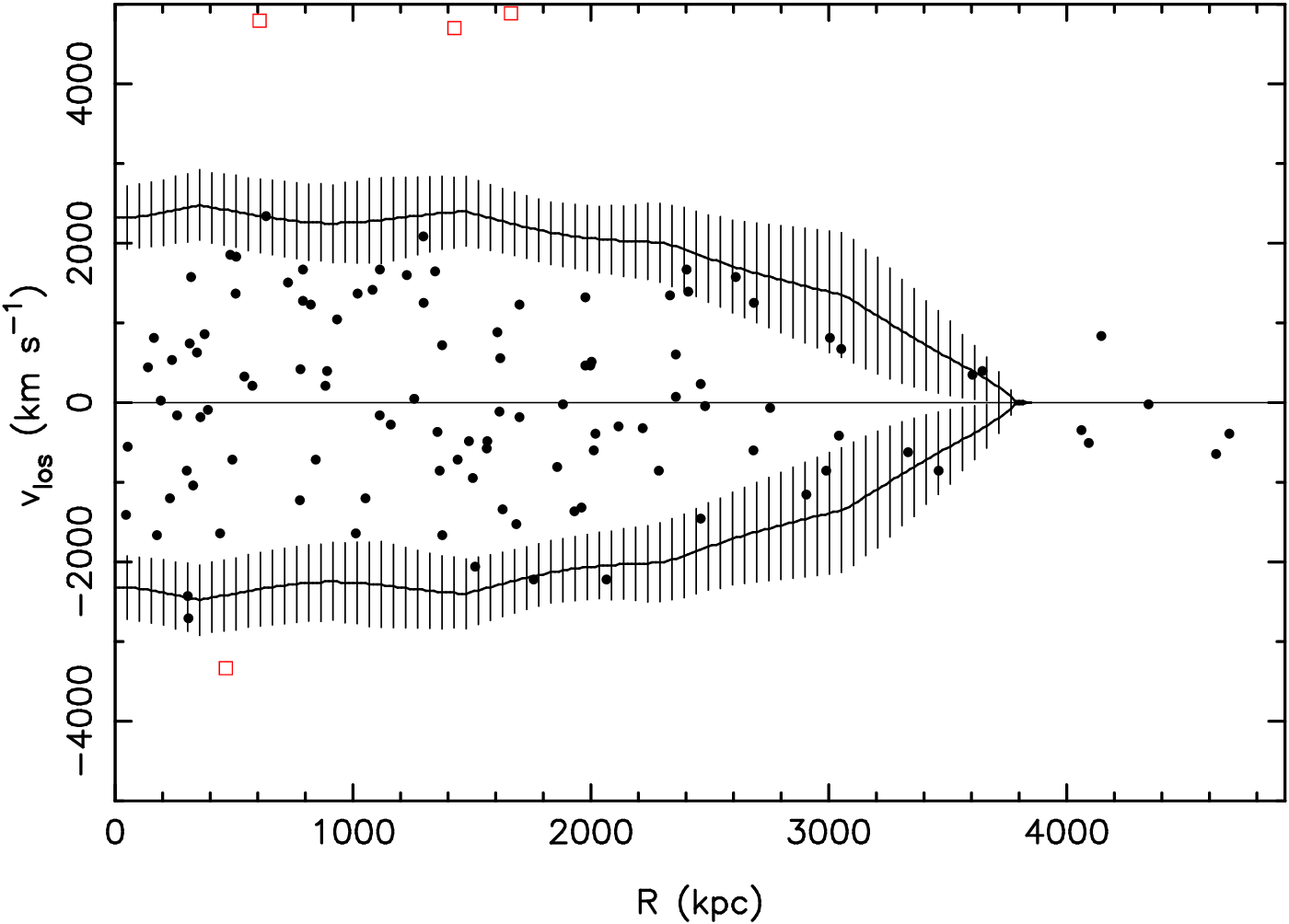}\\
\includegraphics[width=0.23\textwidth, angle=-90]{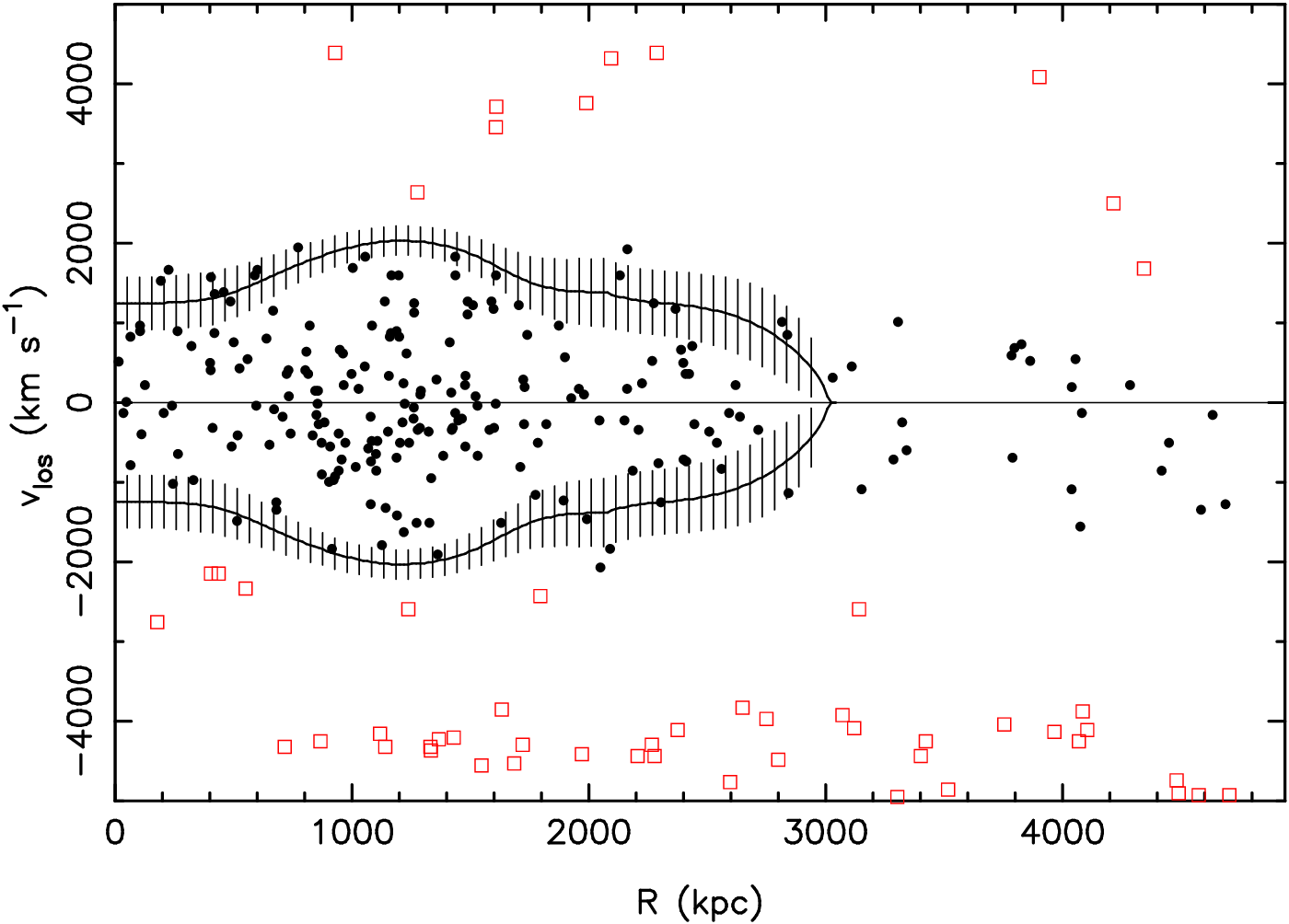}\hspace{0.02\textwidth}%
\includegraphics[width=0.23\textwidth, angle=-90]{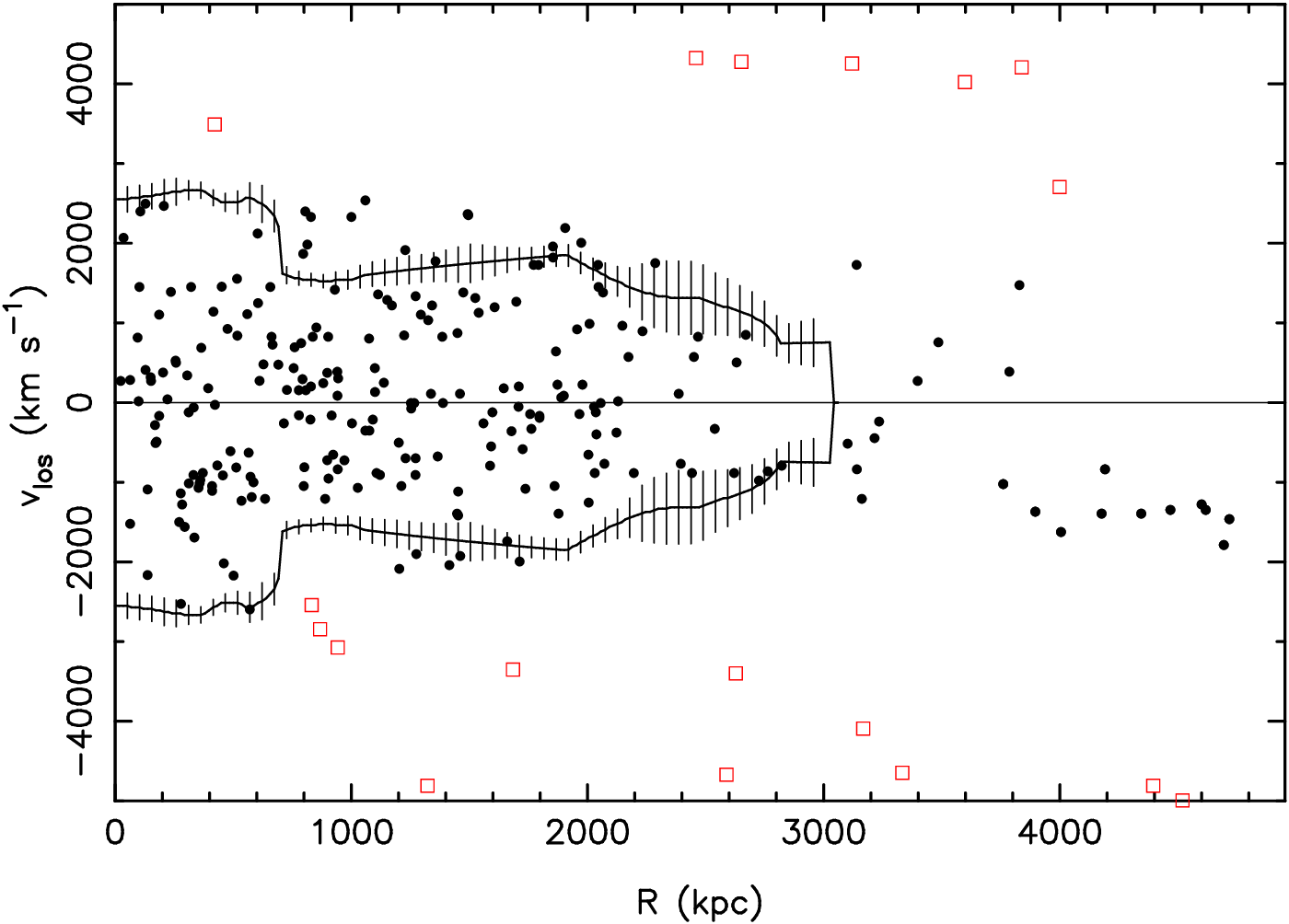}\hspace{0.02\textwidth}%
\includegraphics[width=0.23\textwidth, angle=-90]{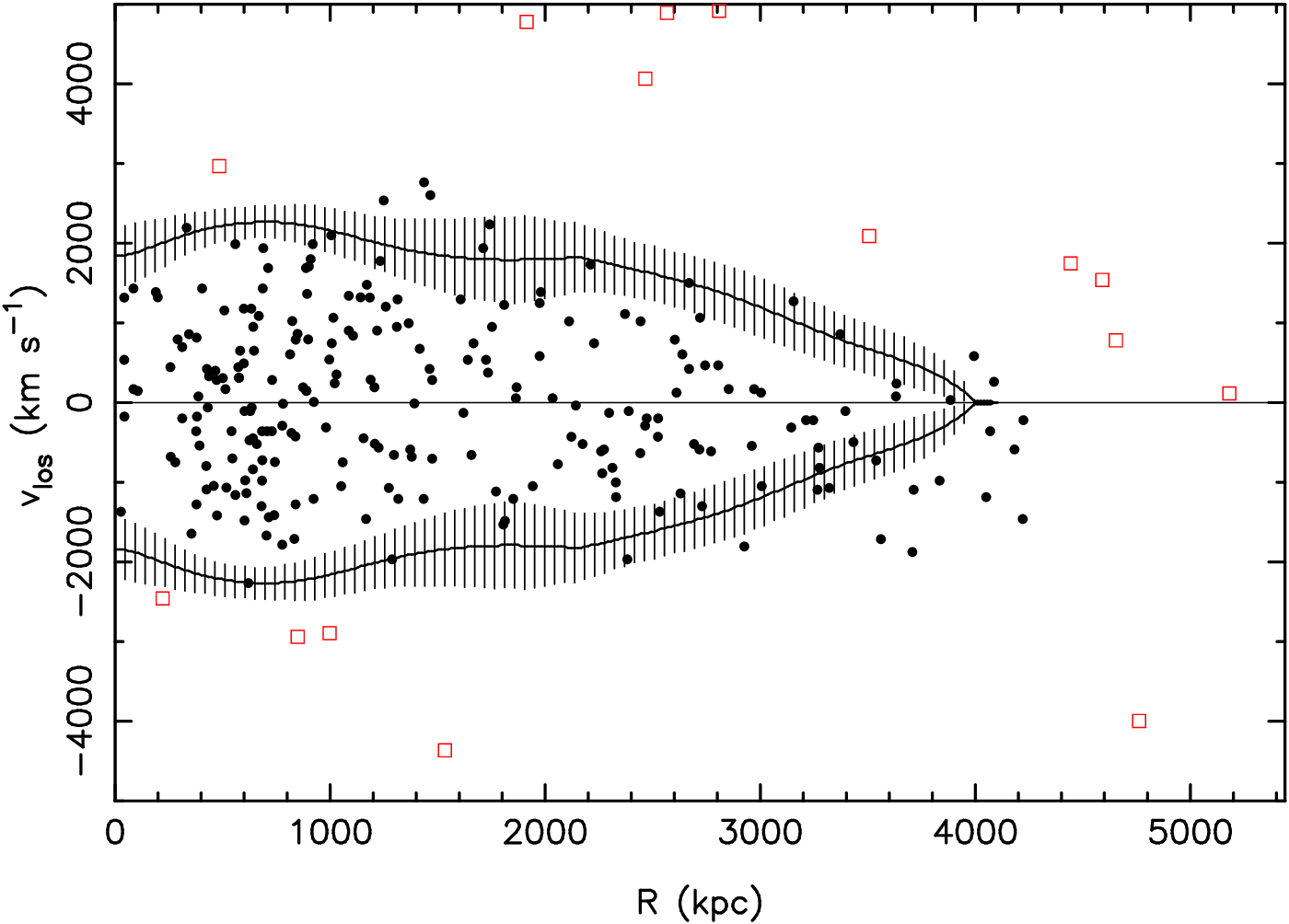}\\
\includegraphics[width=0.23\textwidth, angle=-90]{RXCJ1206_caus.eps}\hspace{0.02\textwidth}%
\includegraphics[width=0.23\textwidth, angle=-90]{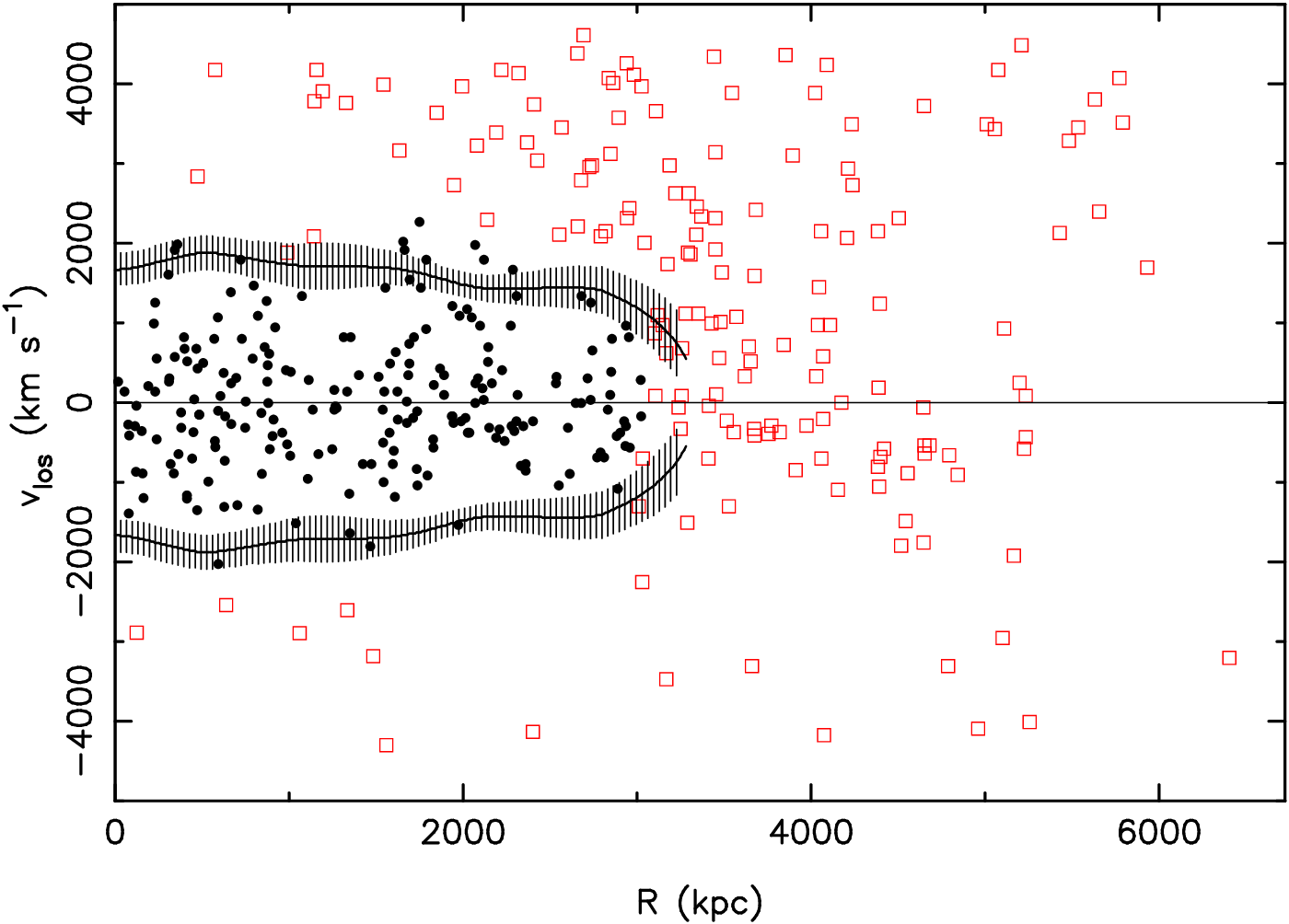}\hspace{0.02\textwidth}%
\includegraphics[width=0.23\textwidth, angle=-90]{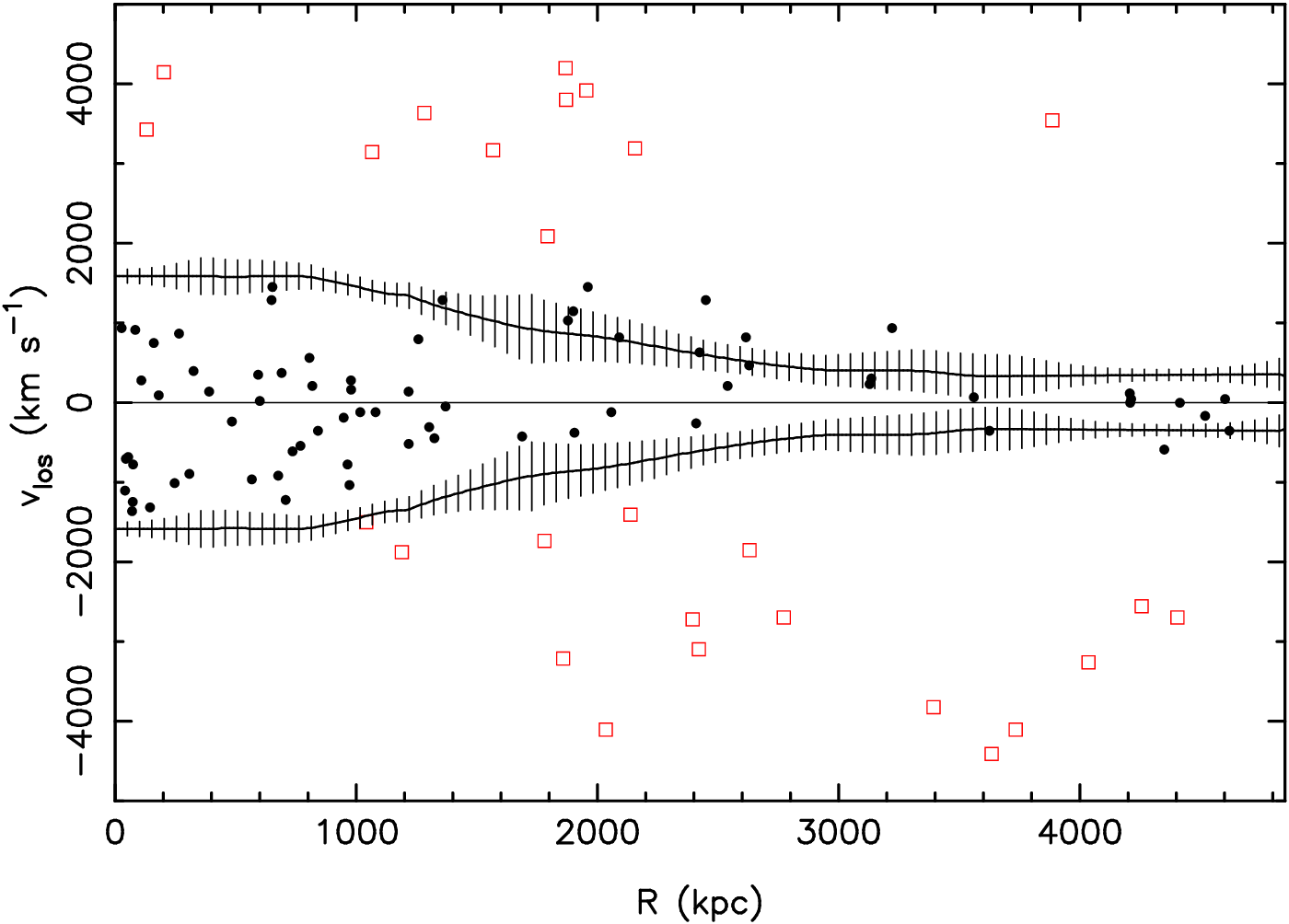}\\
\includegraphics[width=0.23\textwidth, angle=-90]{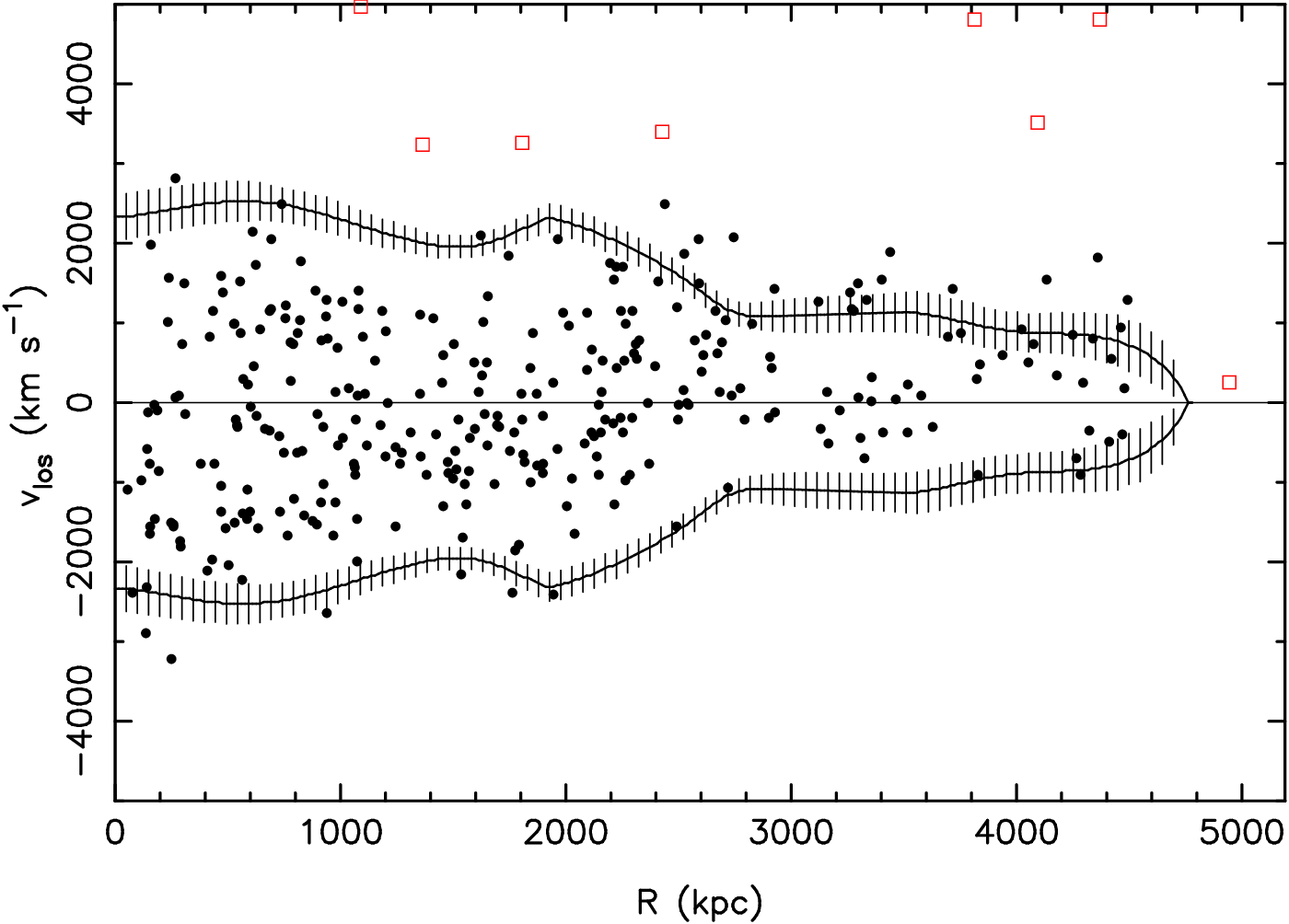}
\caption{Galaxy distribution in PPS. Black points are galaxies retained as cluster members by our iterative $3\sigma$ clipping procedure, whereas red squares are galaxies that were discarded. The black curves represent the caustic amplitude (the vertical lines show their $1\sigma$ uncertainty). We arbitrarily limited the member selection to within 3 Mpc for RXCJ1347; the cluster main body is hardly distinguishable from interlopers beyond this radius. From left to right, top to bottom: RXCJ0014, RXCJ0225, RXCJ0516, RXCJ0528, RXCJ0658, RXCJ1131, RXCJ1206, RXCJ1347, RXCJ2011, and RXCJ2308.}\label{fig:PPS_all}
\end{figure*}

\begin{figure*}
\includegraphics[width=0.23\textwidth, angle=-90]{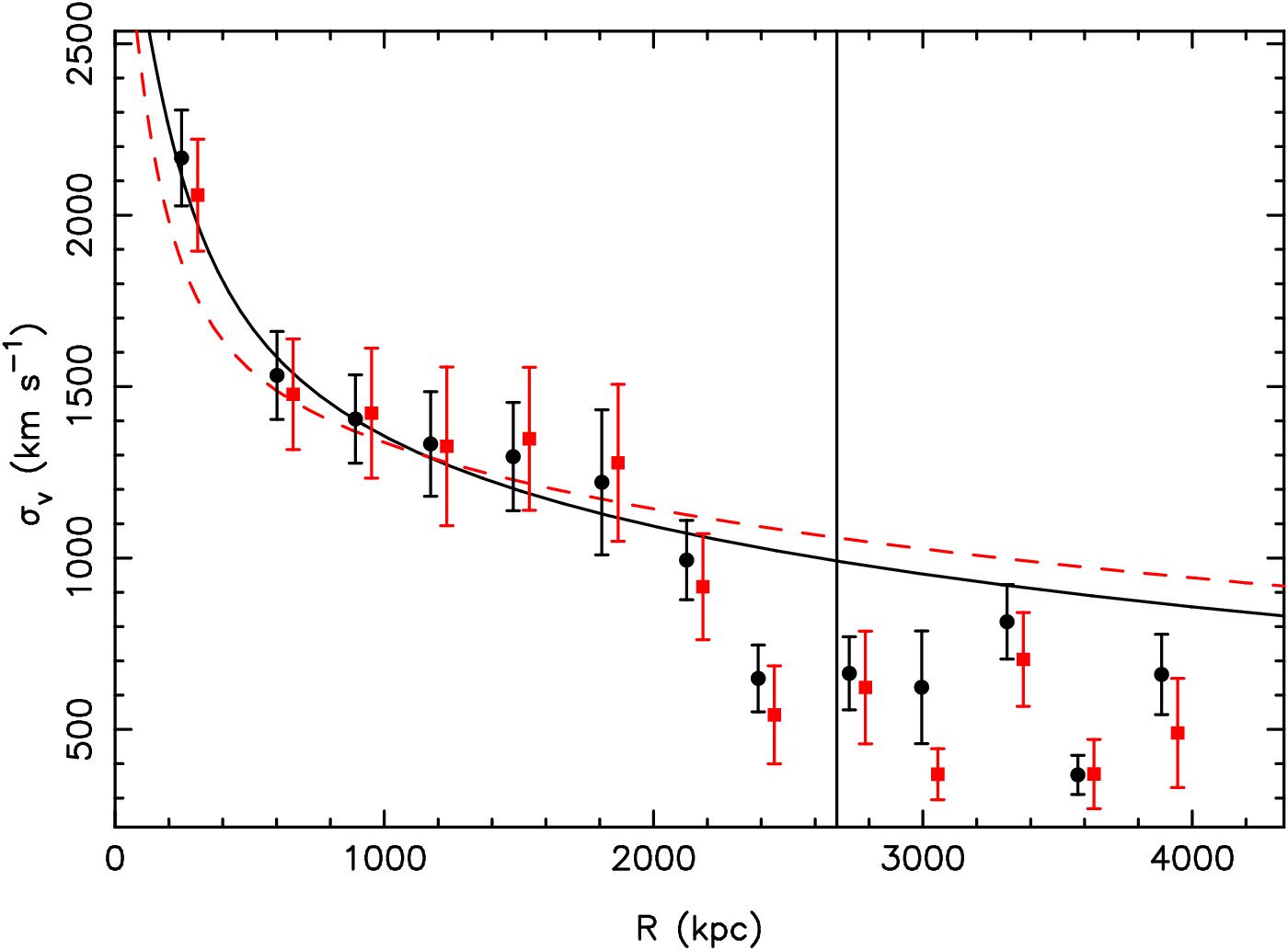}\hspace{0.02\textwidth}%
\includegraphics[width=0.23\textwidth, angle=-90]{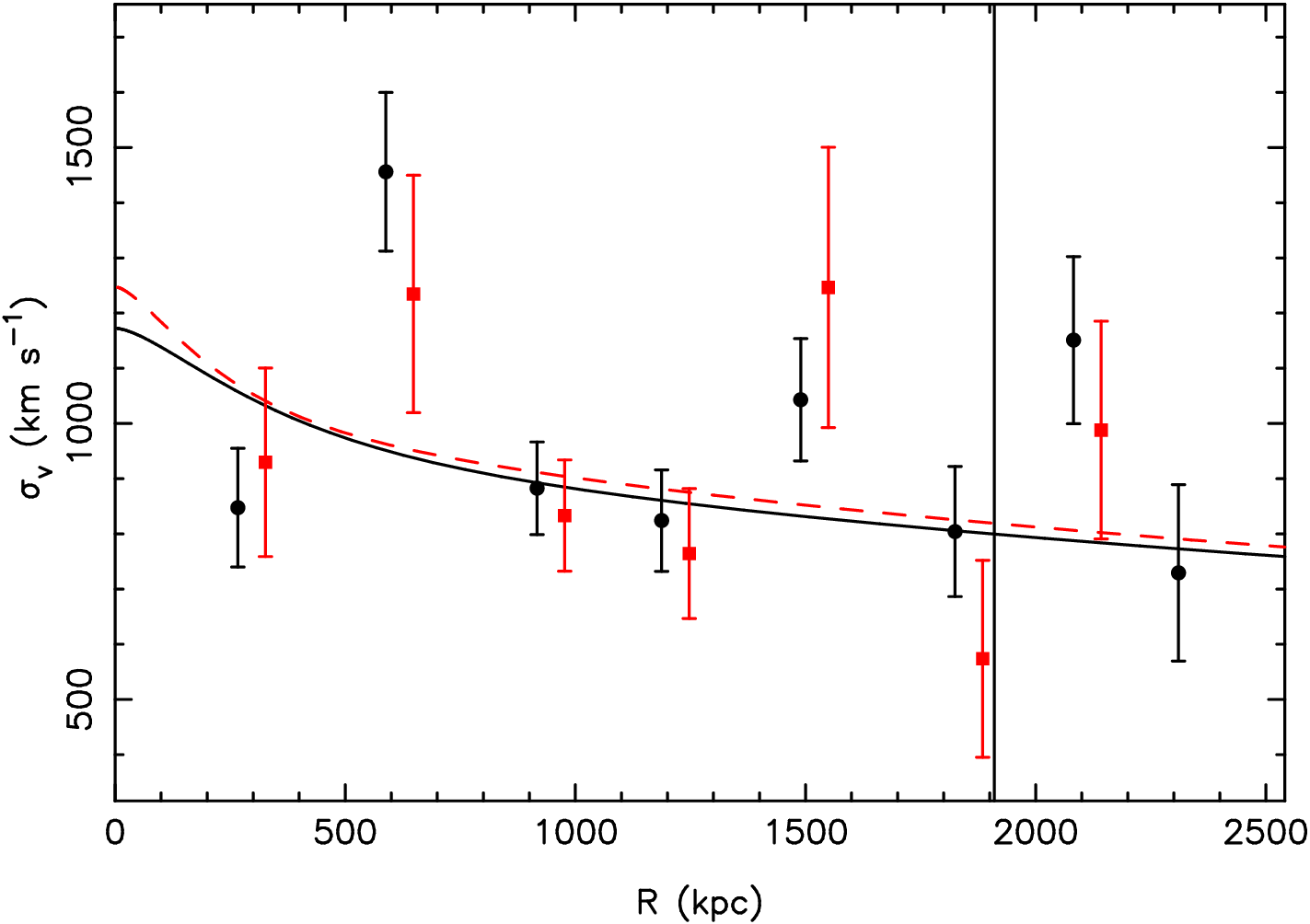}\hspace{0.02\textwidth}%
\includegraphics[width=0.23\textwidth, angle=-90]{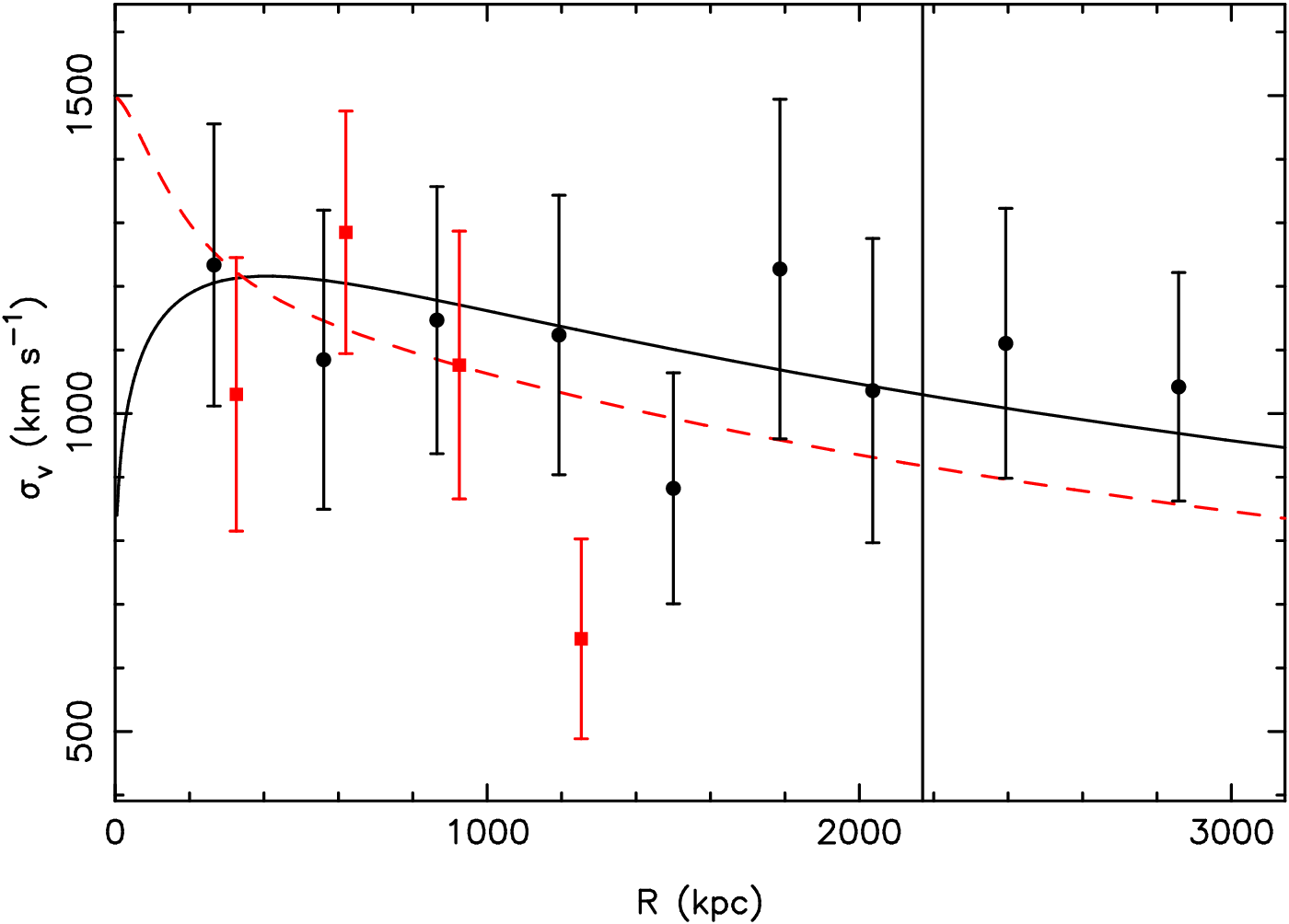}\\
\includegraphics[width=0.23\textwidth, angle=-90]{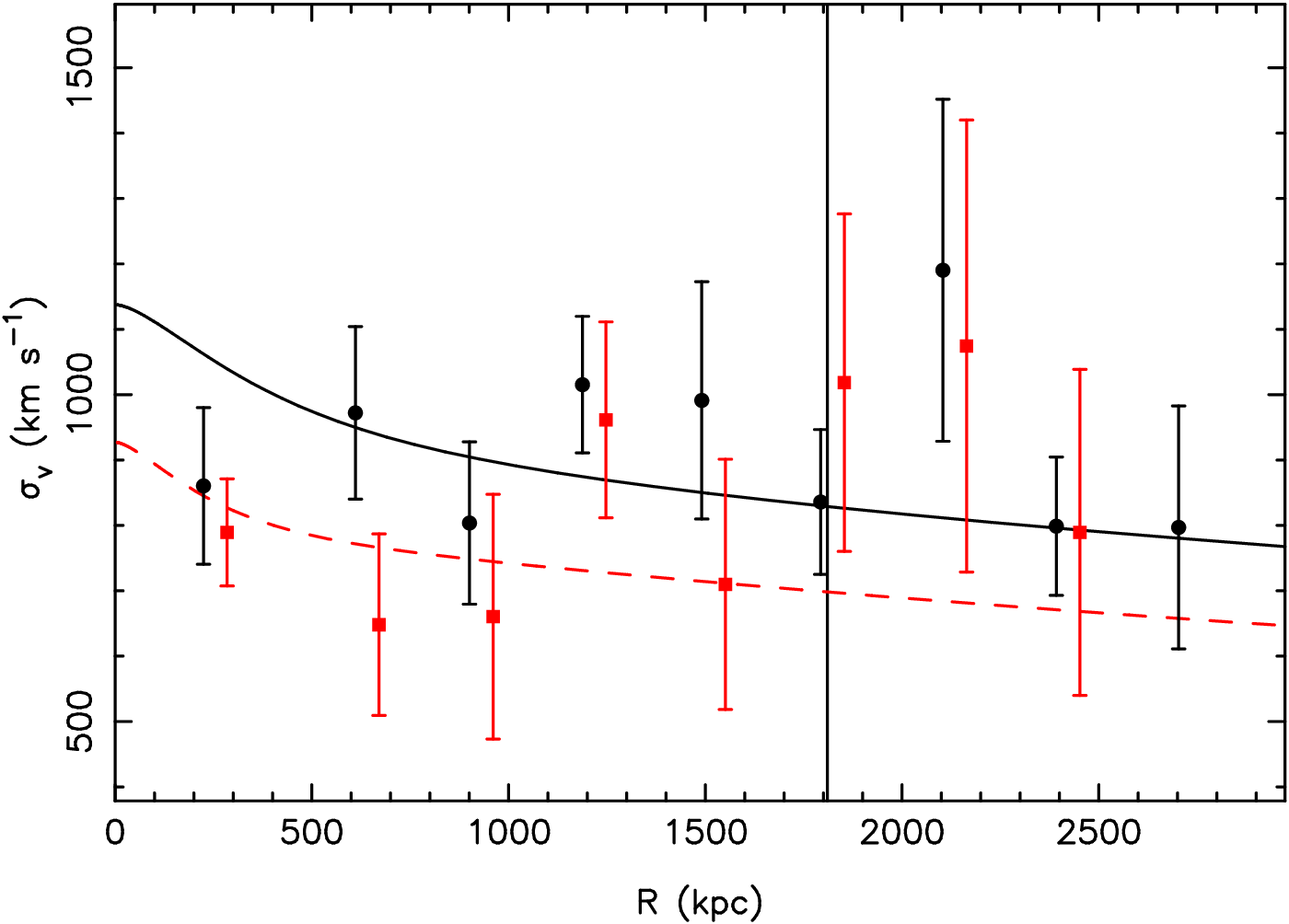}\hspace{0.02\textwidth}%
\includegraphics[width=0.23\textwidth, angle=-90]{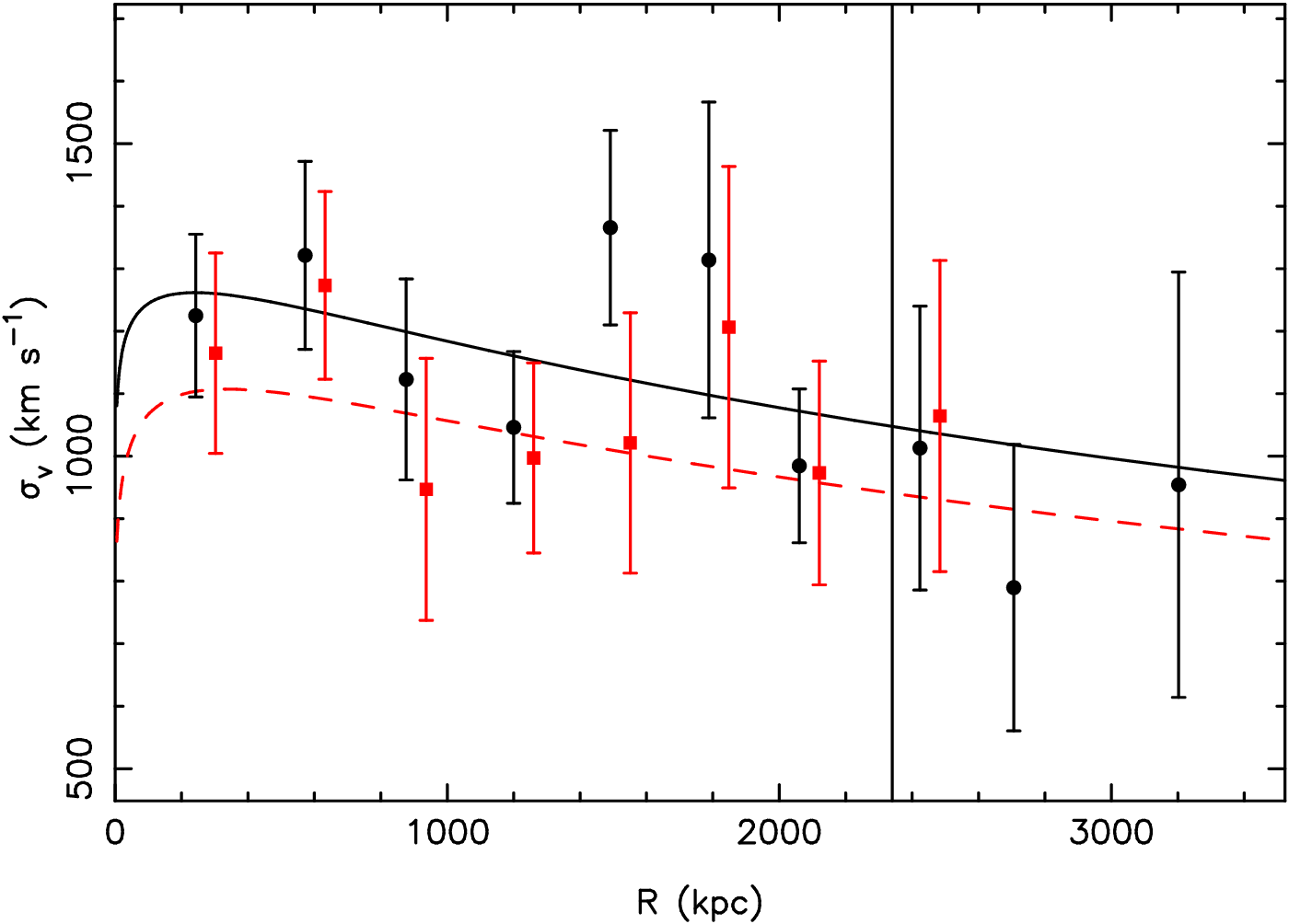}\hspace{0.02\textwidth}%
\includegraphics[width=0.23\textwidth, angle=-90]{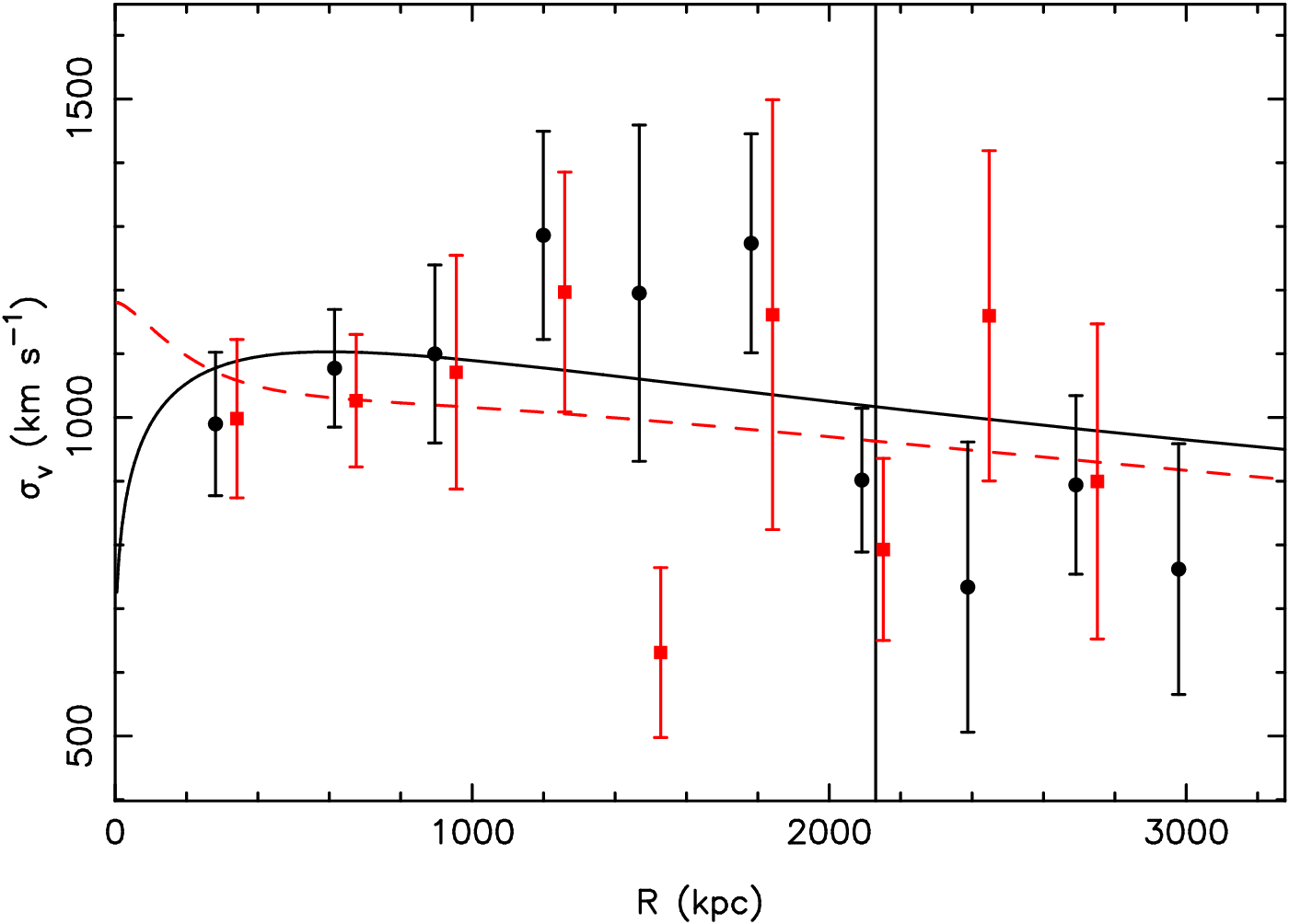}\\
\includegraphics[width=0.23\textwidth, angle=-90]{RXCJ1206_vdisp.eps}\hspace{0.02\textwidth}%
\includegraphics[width=0.23\textwidth, angle=-90]{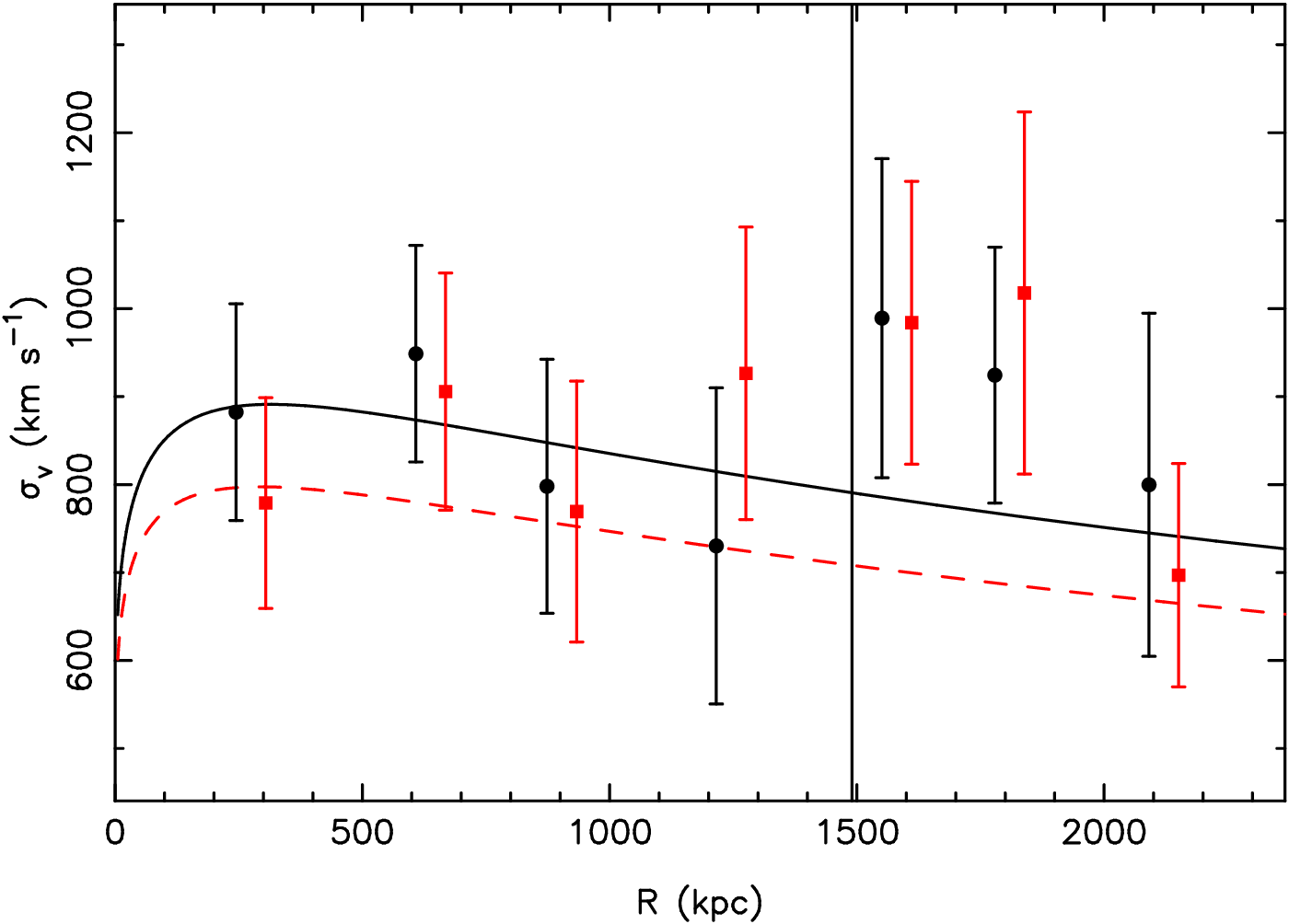}\hspace{0.02\textwidth}%
\includegraphics[width=0.23\textwidth, angle=-90]{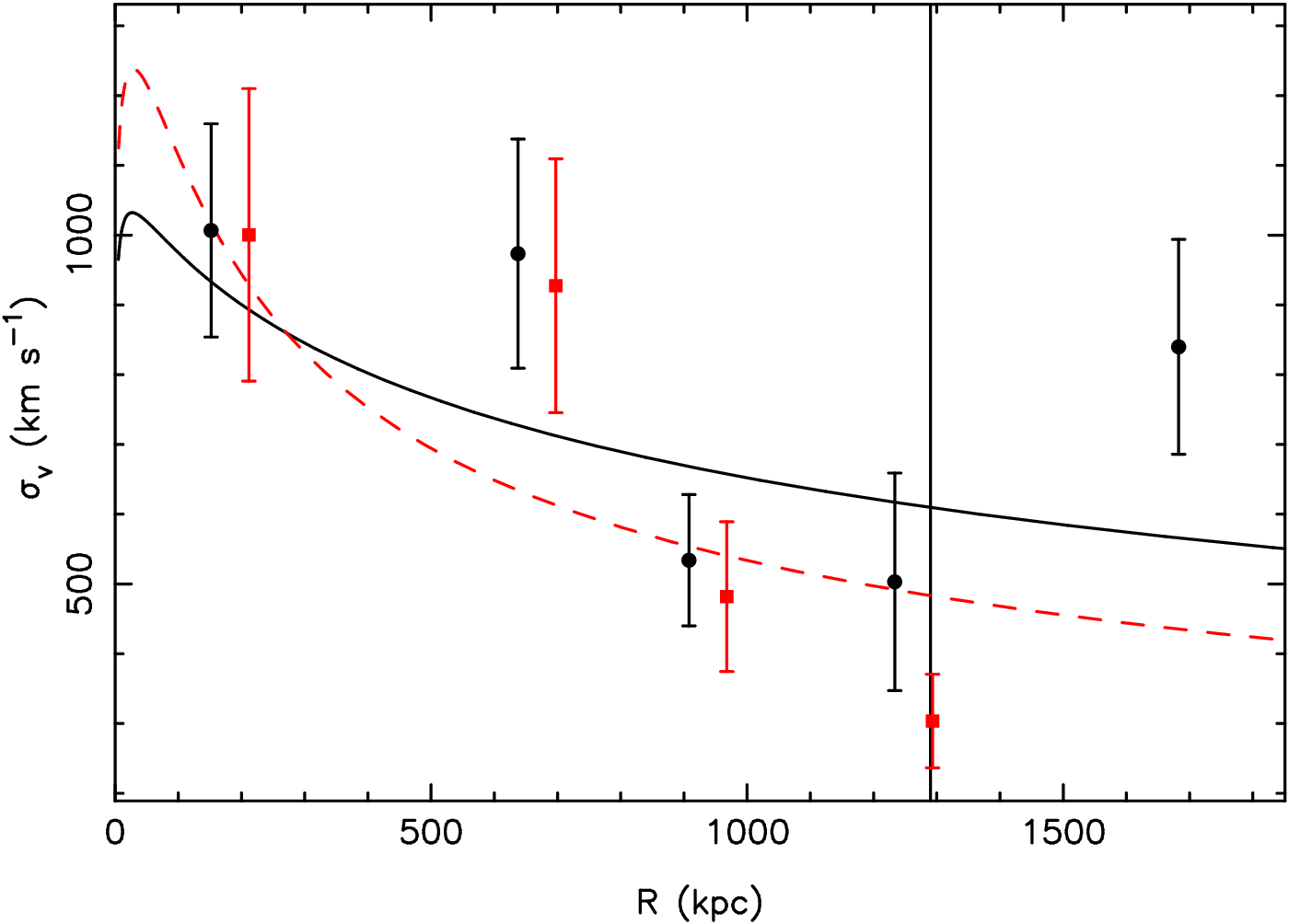}\\
\includegraphics[width=0.23\textwidth, angle=-90]{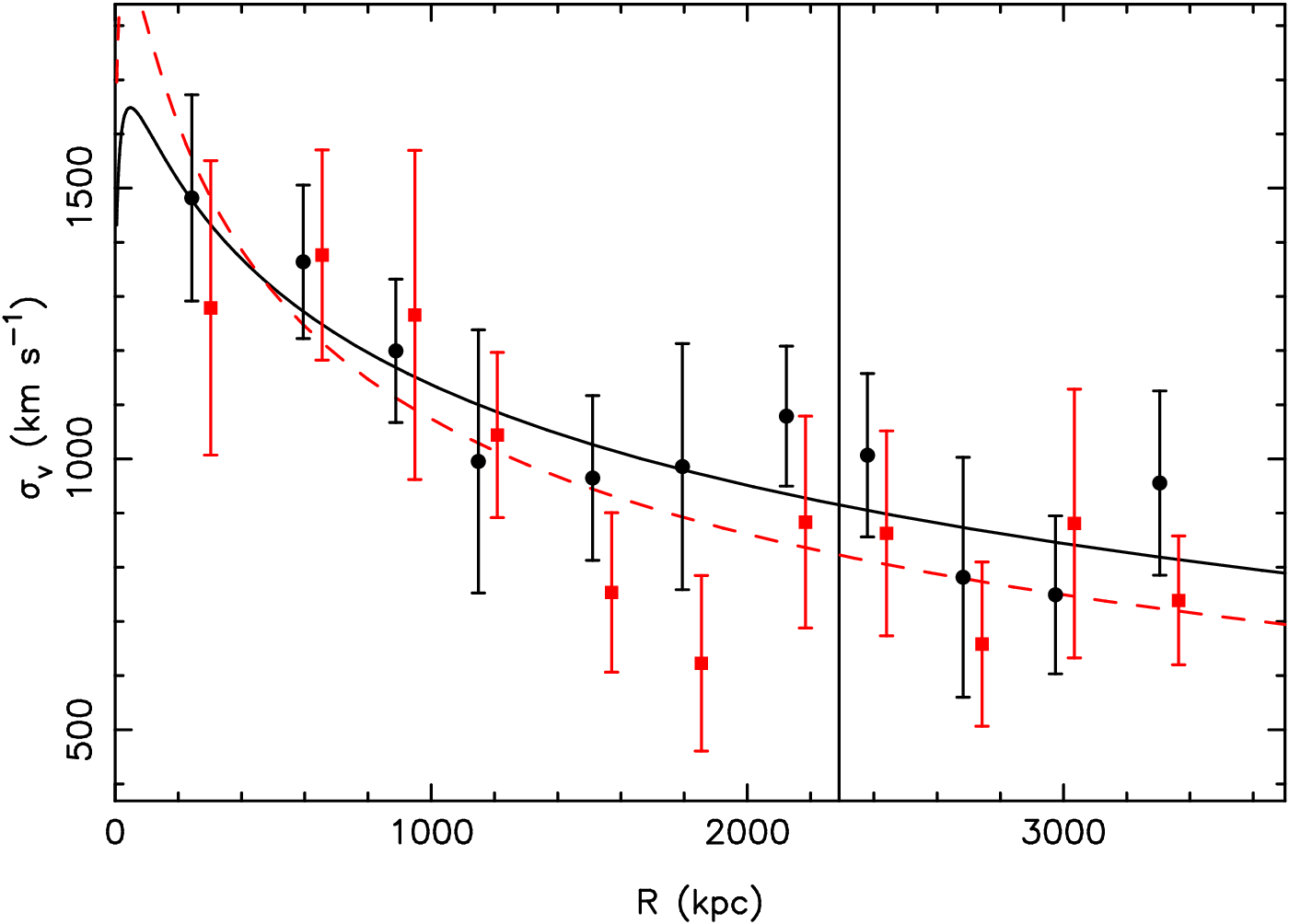}
\caption{Observed profile of the projected velocity dispersion. The black curve is the predicted profile from the best model of the Jeans analysis; it is not the best-fit to the observed profile. The red points and red-dashed curve are for the red-sequence galaxies. The vertical line shows $R_{200}$, as estimated from the Jeans equation, prior the substructure analysis. From left to right, top to bottom: RXCJ0014, RXCJ0225, RXCJ0516, RXCJ0528, RXCJ0658, RXCJ1131, RXCJ1206, RXCJ1347, RXCJ2011, and RXCJ2308.}\label{fig:vdisp_all}
\end{figure*}

\begin{figure*}
\includegraphics[width=0.23\textwidth, angle=-90]{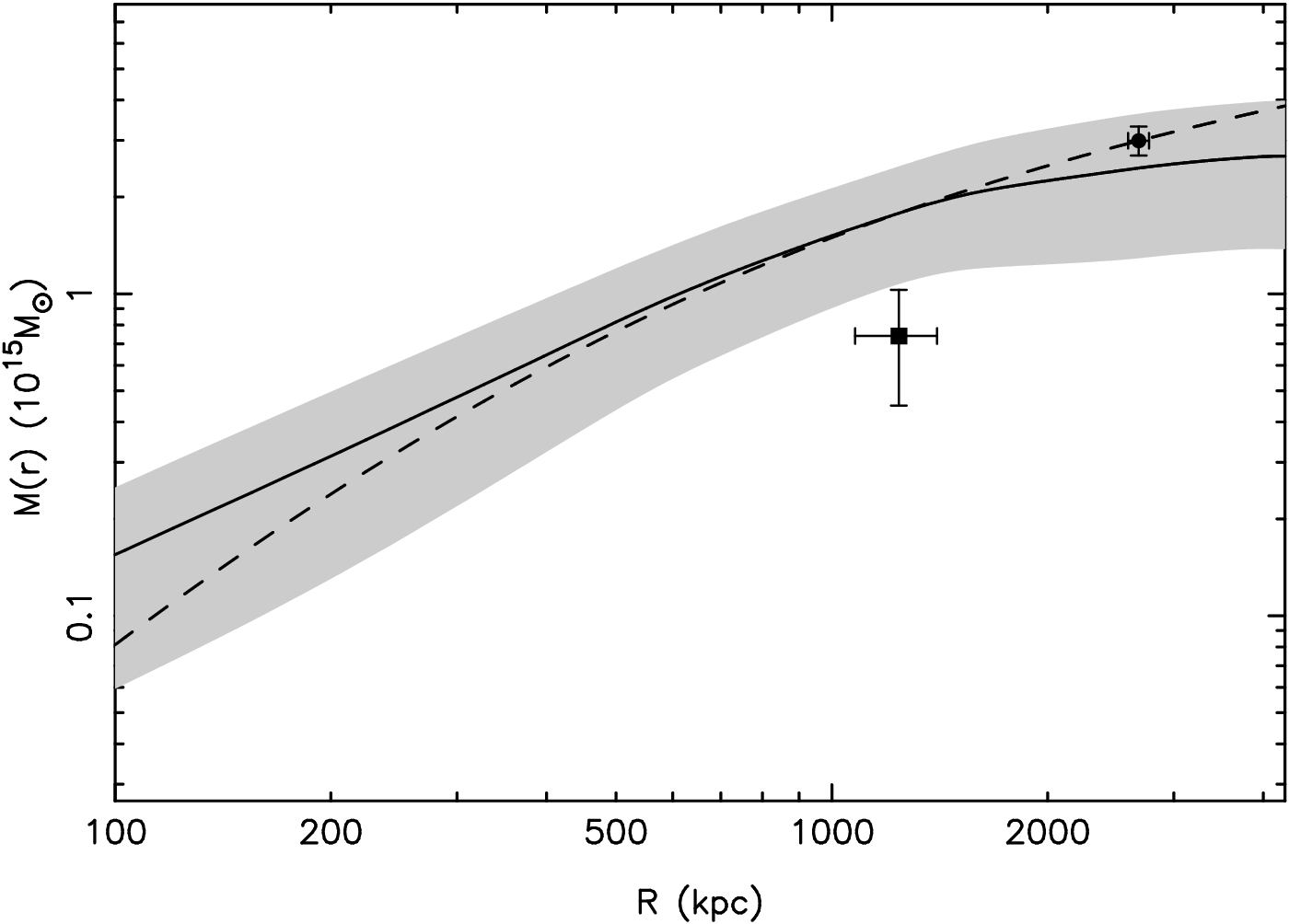}\hspace{0.02\textwidth}%
\includegraphics[width=0.23\textwidth, angle=-90]{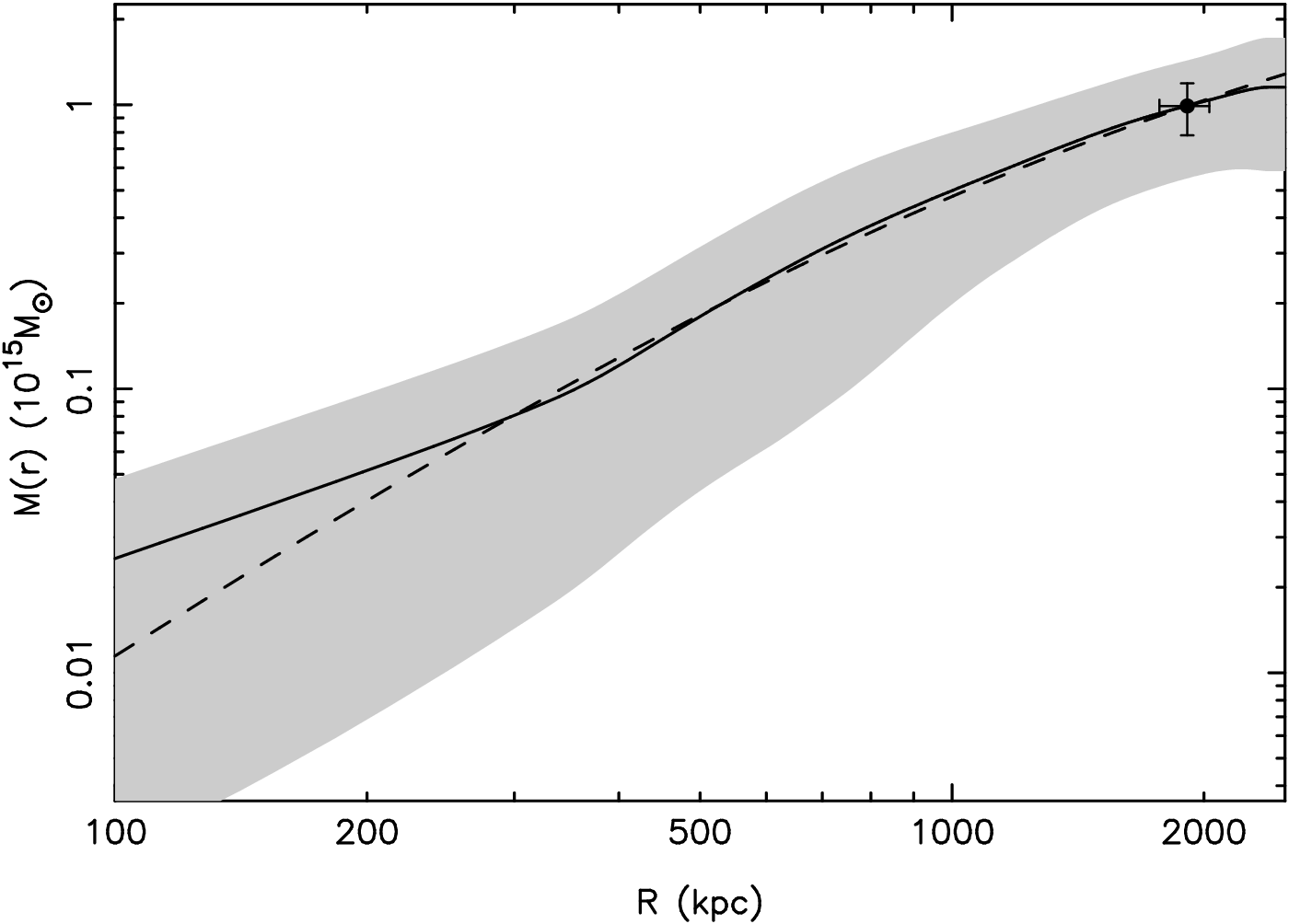}\hspace{0.02\textwidth}%
\includegraphics[width=0.23\textwidth, angle=-90]{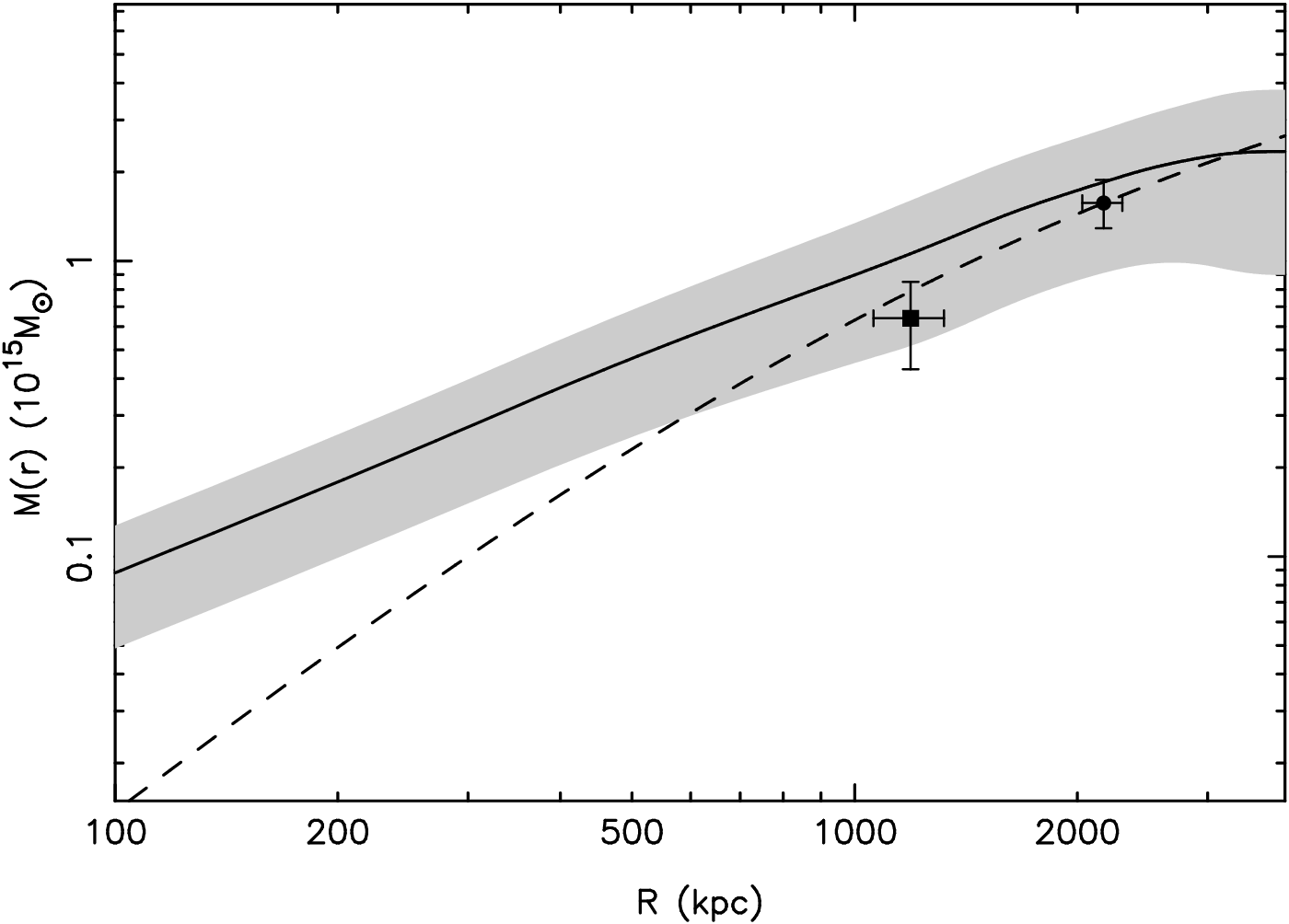}\\
\includegraphics[width=0.23\textwidth, angle=-90]{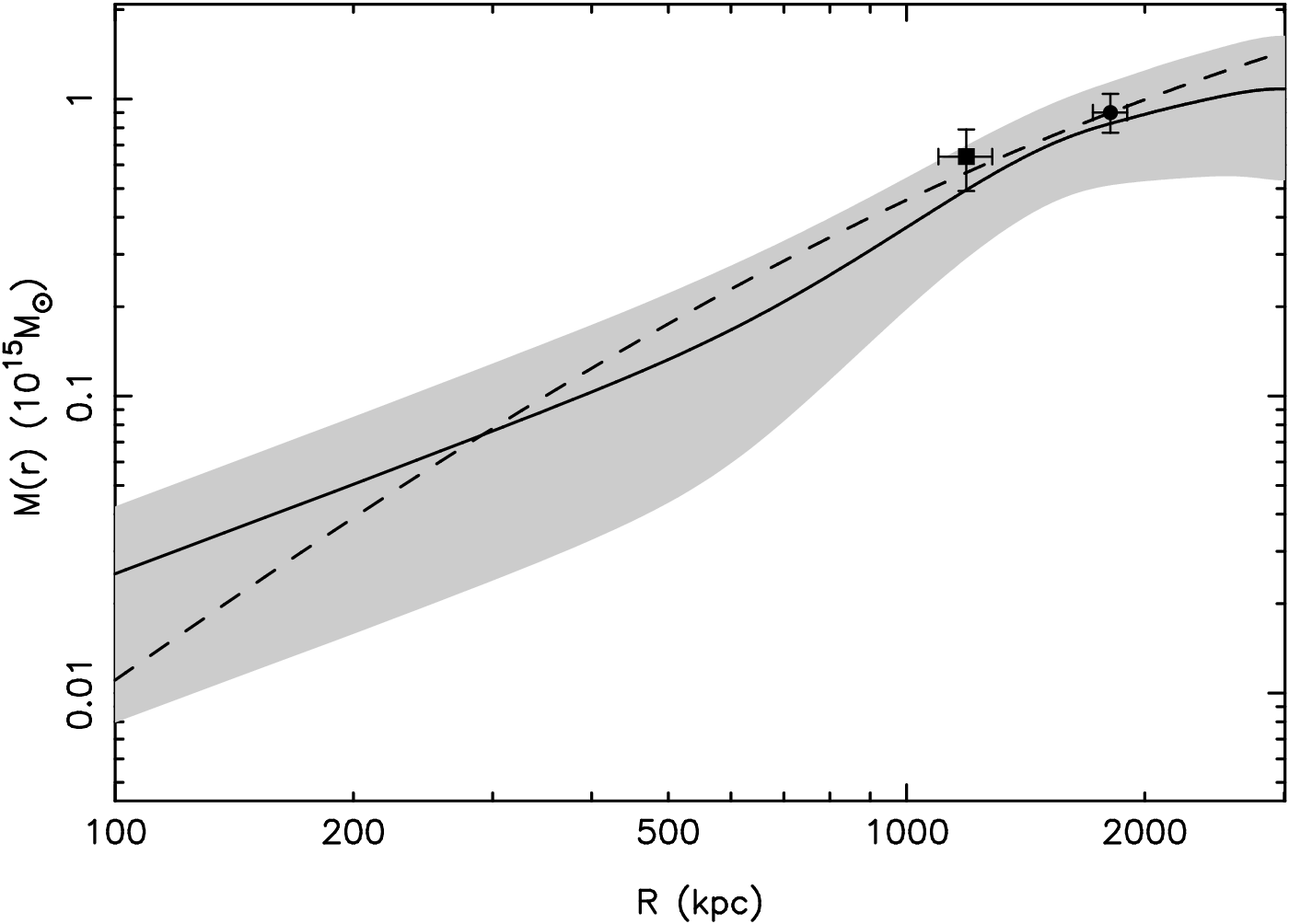}\hspace{0.02\textwidth}%
\includegraphics[width=0.23\textwidth, angle=-90]{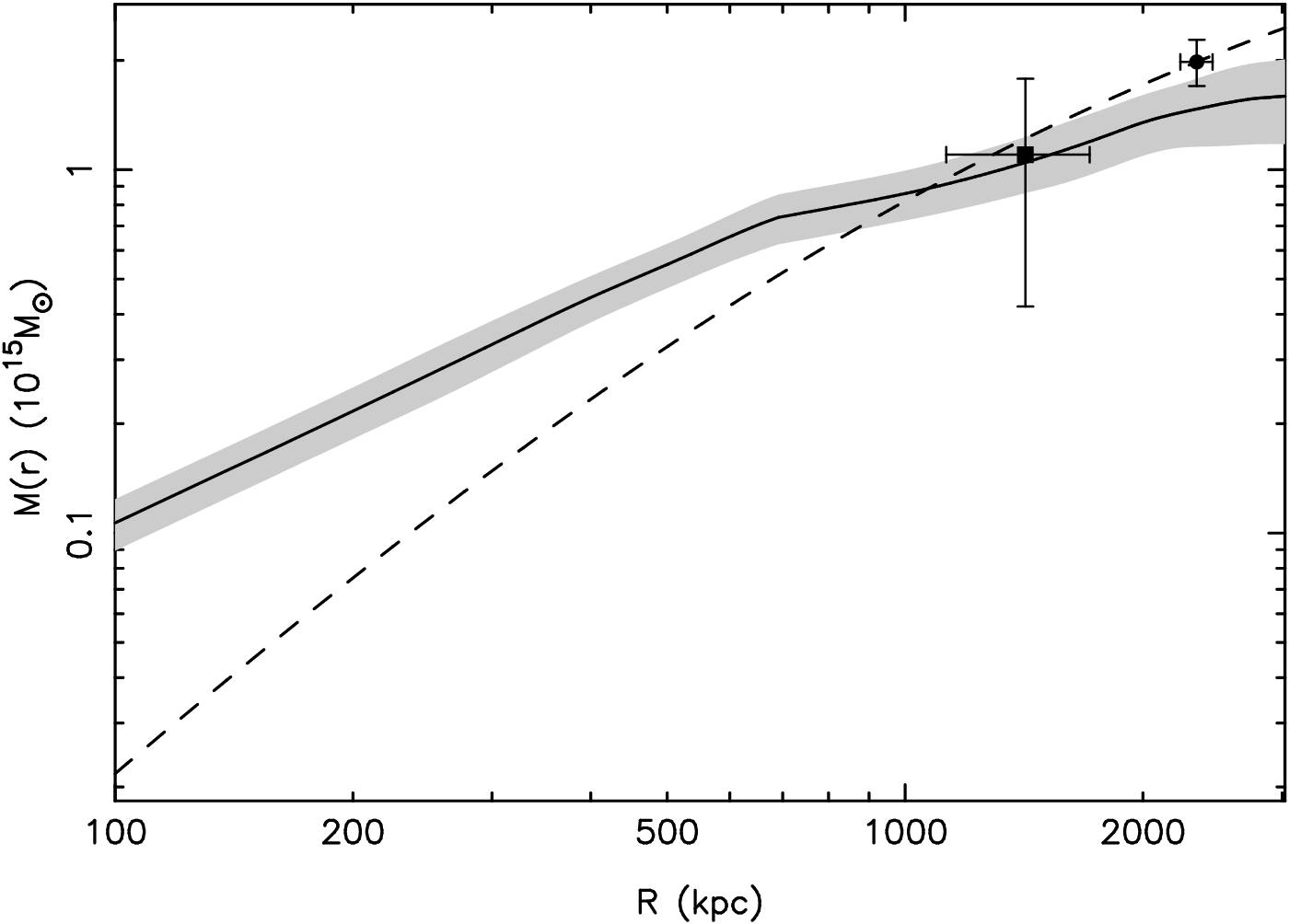}\hspace{0.02\textwidth}%
\includegraphics[width=0.23\textwidth, angle=-90]{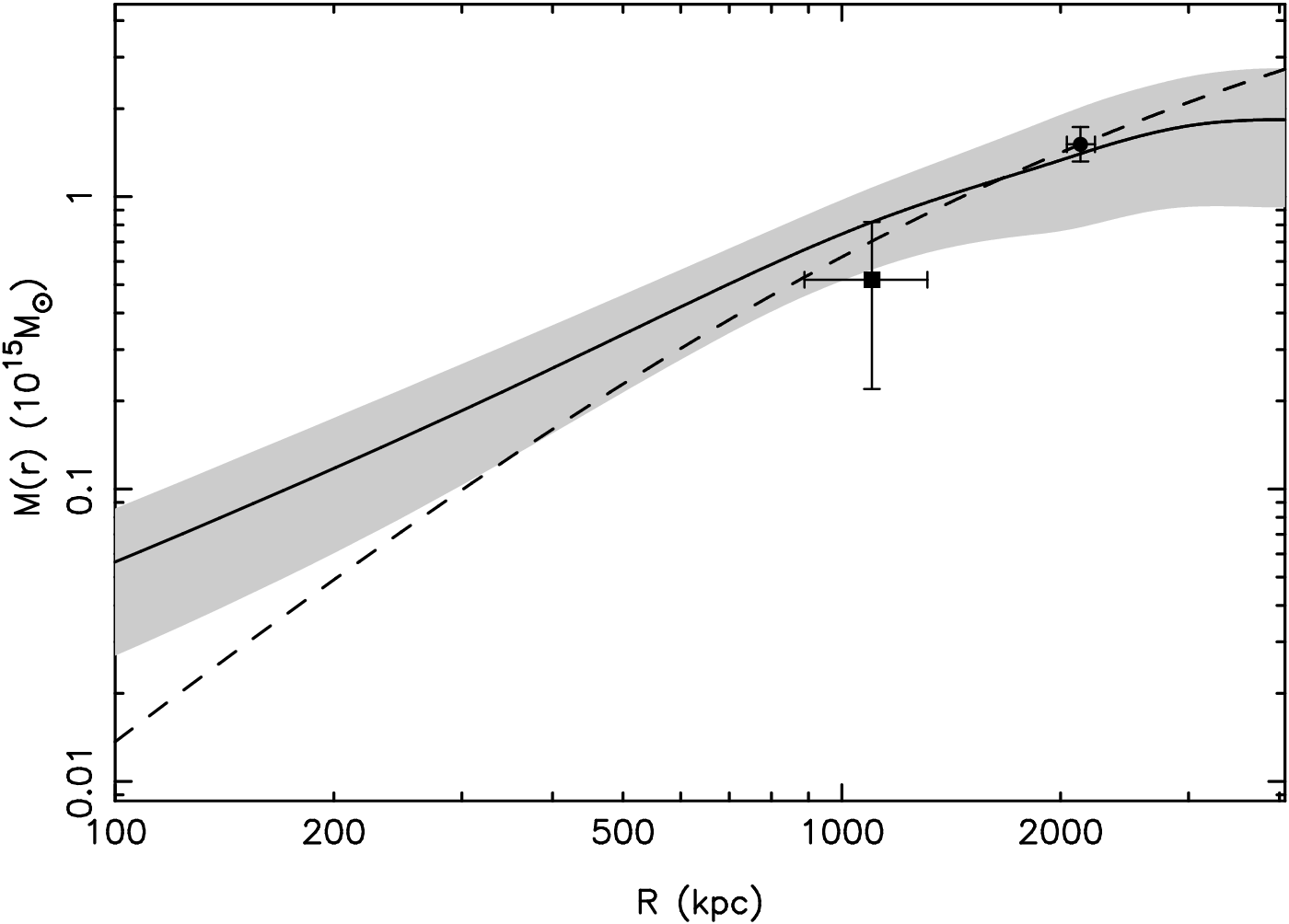}\\
\includegraphics[width=0.23\textwidth, angle=-90]{RXCJ1206_Mr.eps}\hspace{0.02\textwidth}%
\includegraphics[width=0.23\textwidth, angle=-90]{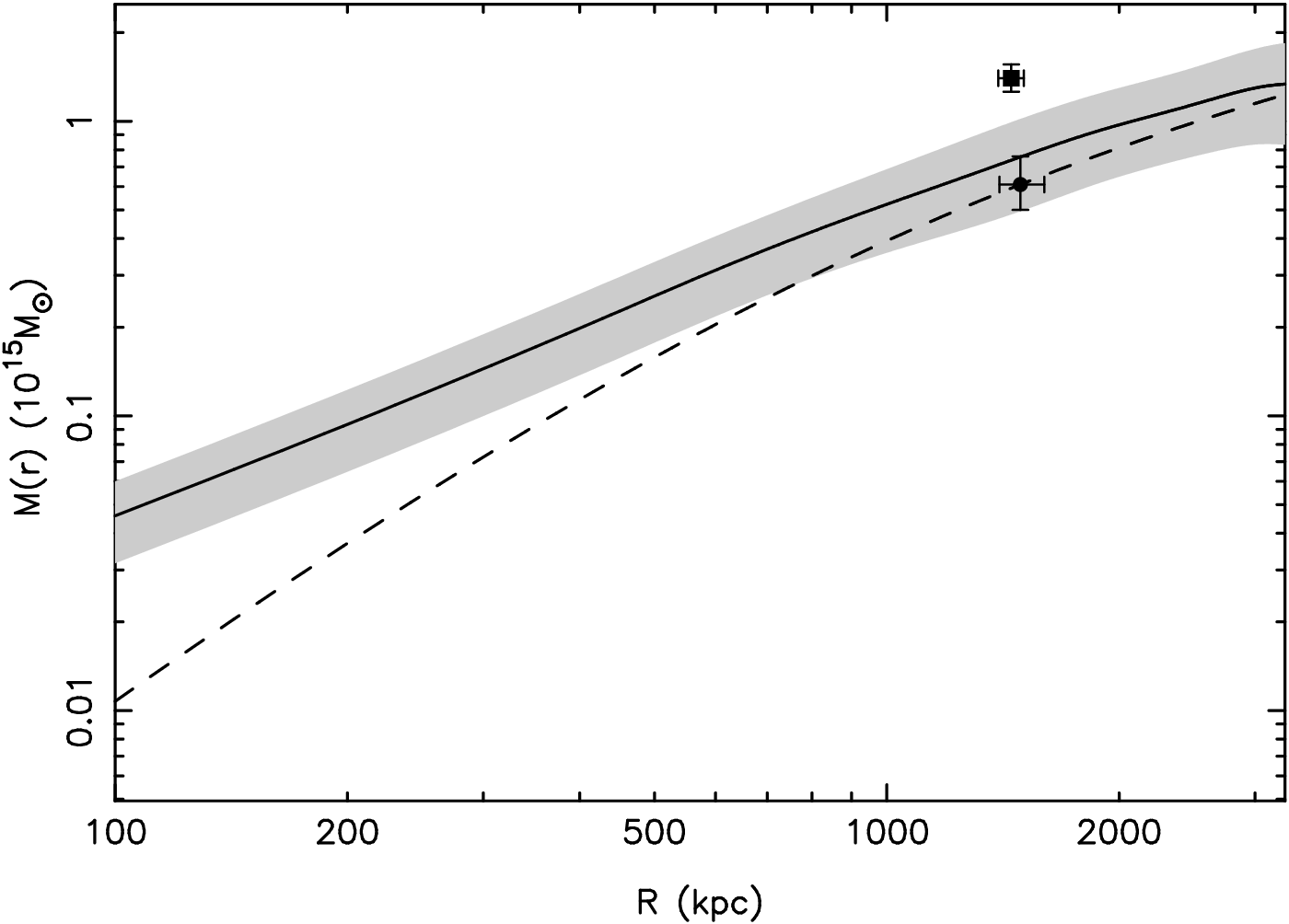}\hspace{0.02\textwidth}%
\includegraphics[width=0.23\textwidth, angle=-90]{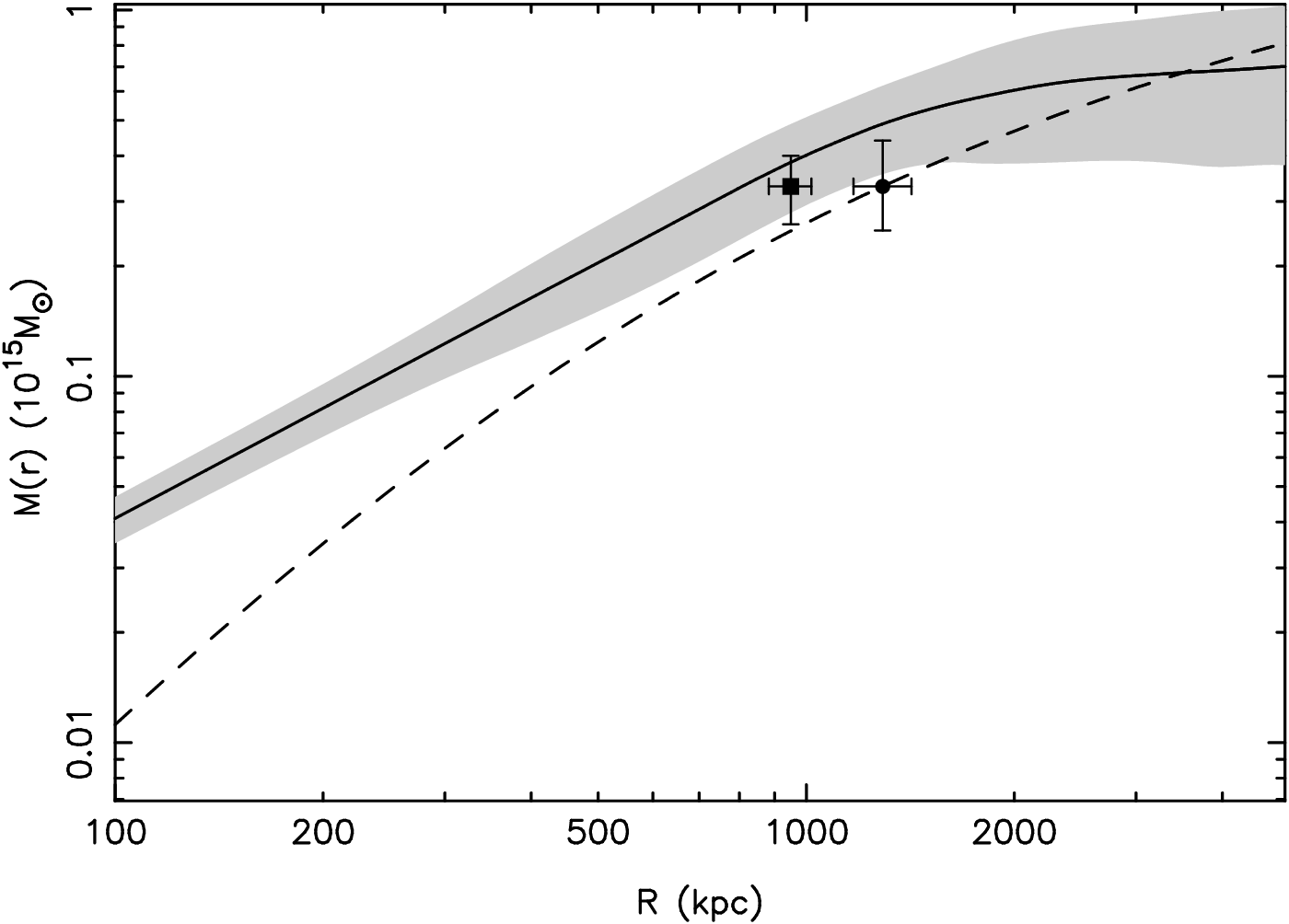}\\
\includegraphics[width=0.23\textwidth, angle=-90]{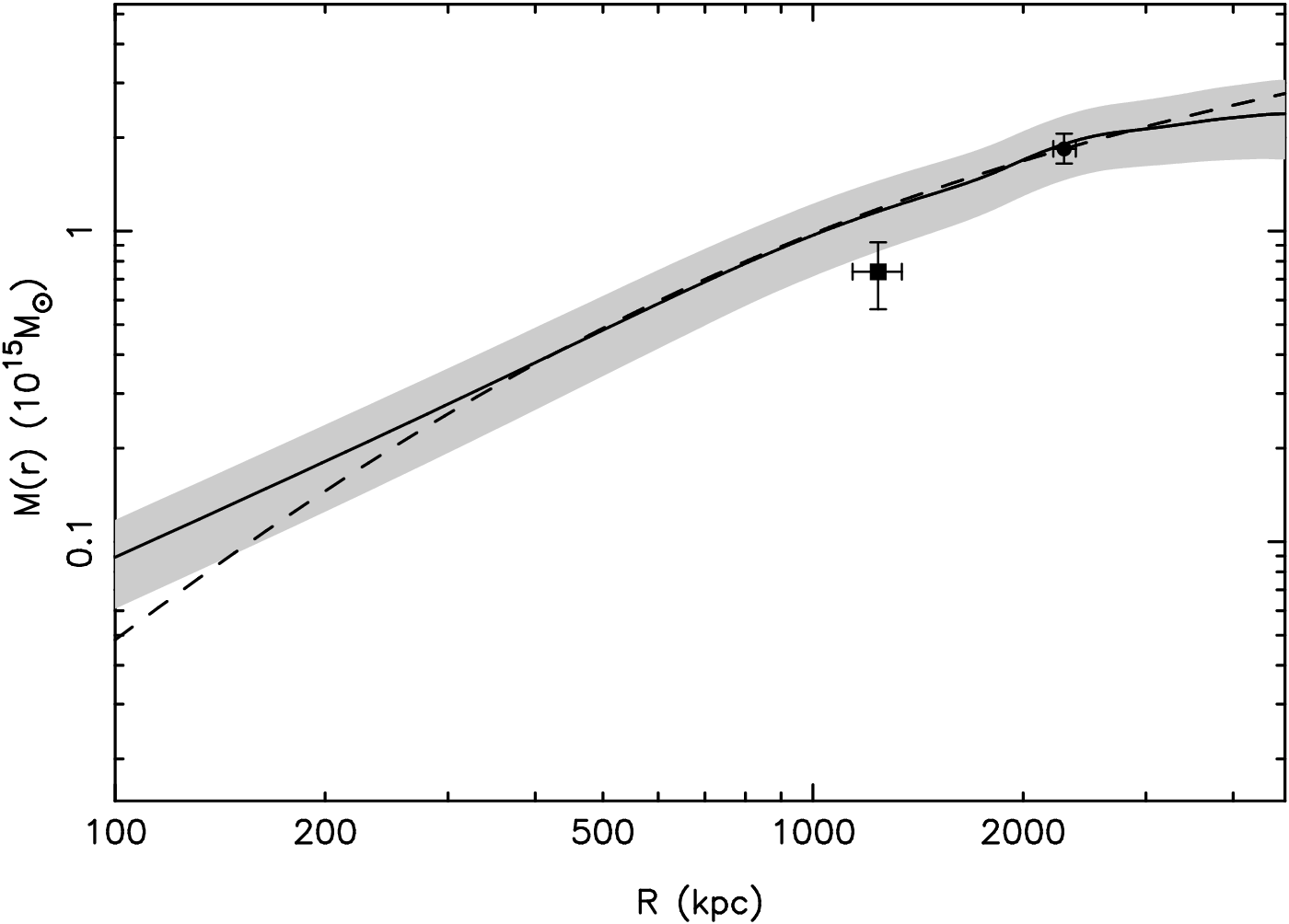}
\caption{Caustic mass profile (solid curve) and its $1\sigma$ confidence interval (shaded area). The dashed curve corresponds to the NFW mass profile derived from the Jeans analysis; it passes through the point $(R_{200},M_{200})$. The second point shows the couple $(R_{500},M_{500})$ derived from the X-ray analysis. From left to right, top to bottom: RXCJ0014, RXCJ0225, RXCJ0516, RXCJ0528, RXCJ0658, RXCJ1131, RXCJ1206, RXCJ1347, RXCJ2011, and RXCJ2308.}\label{fig:Mr_all}
\end{figure*}

\begin{figure*}
\includegraphics[width=0.43\textwidth, angle=-90]{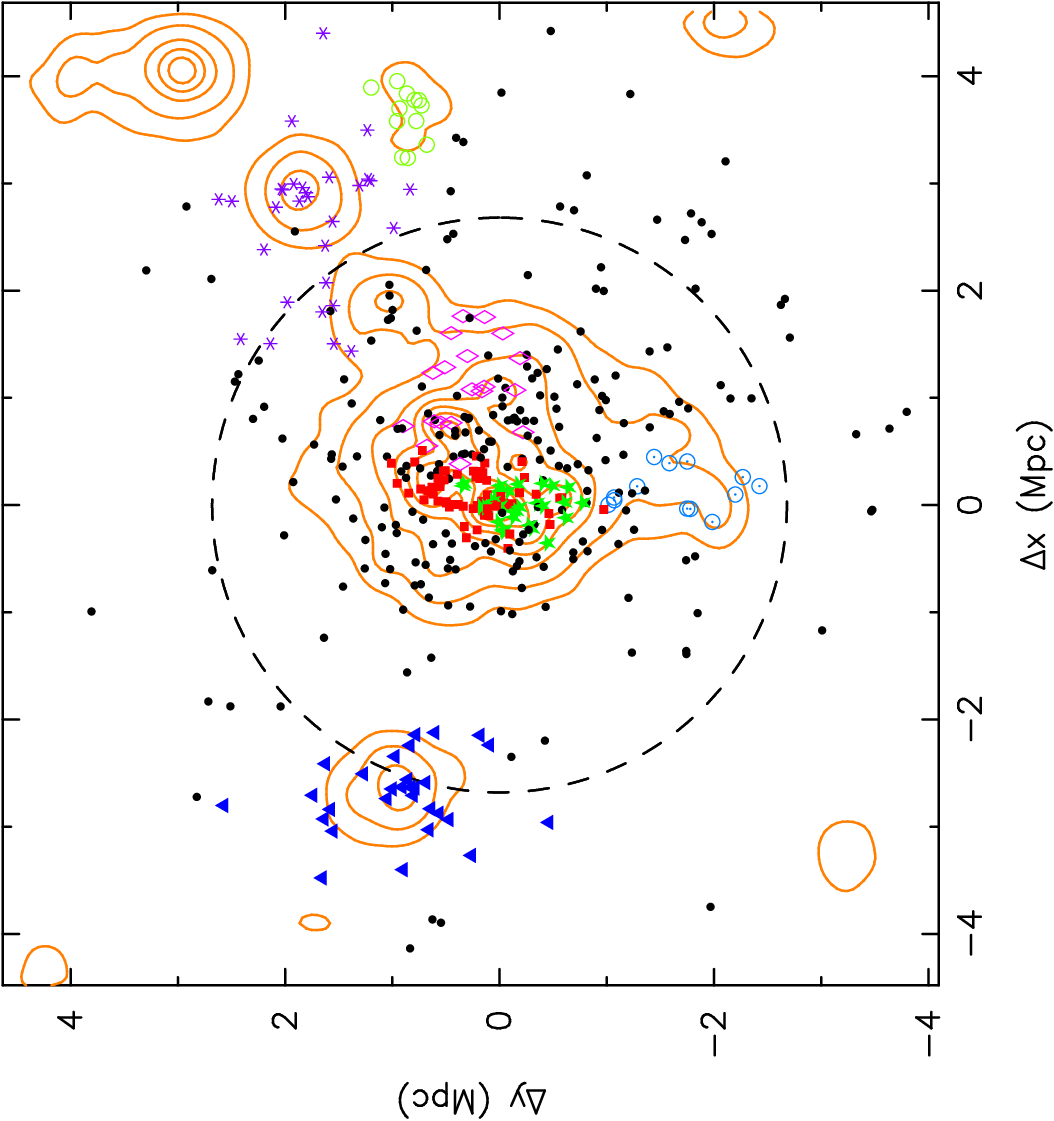}\hspace{0.02\textwidth}%
\includegraphics[width=0.43\textwidth, angle=-90]{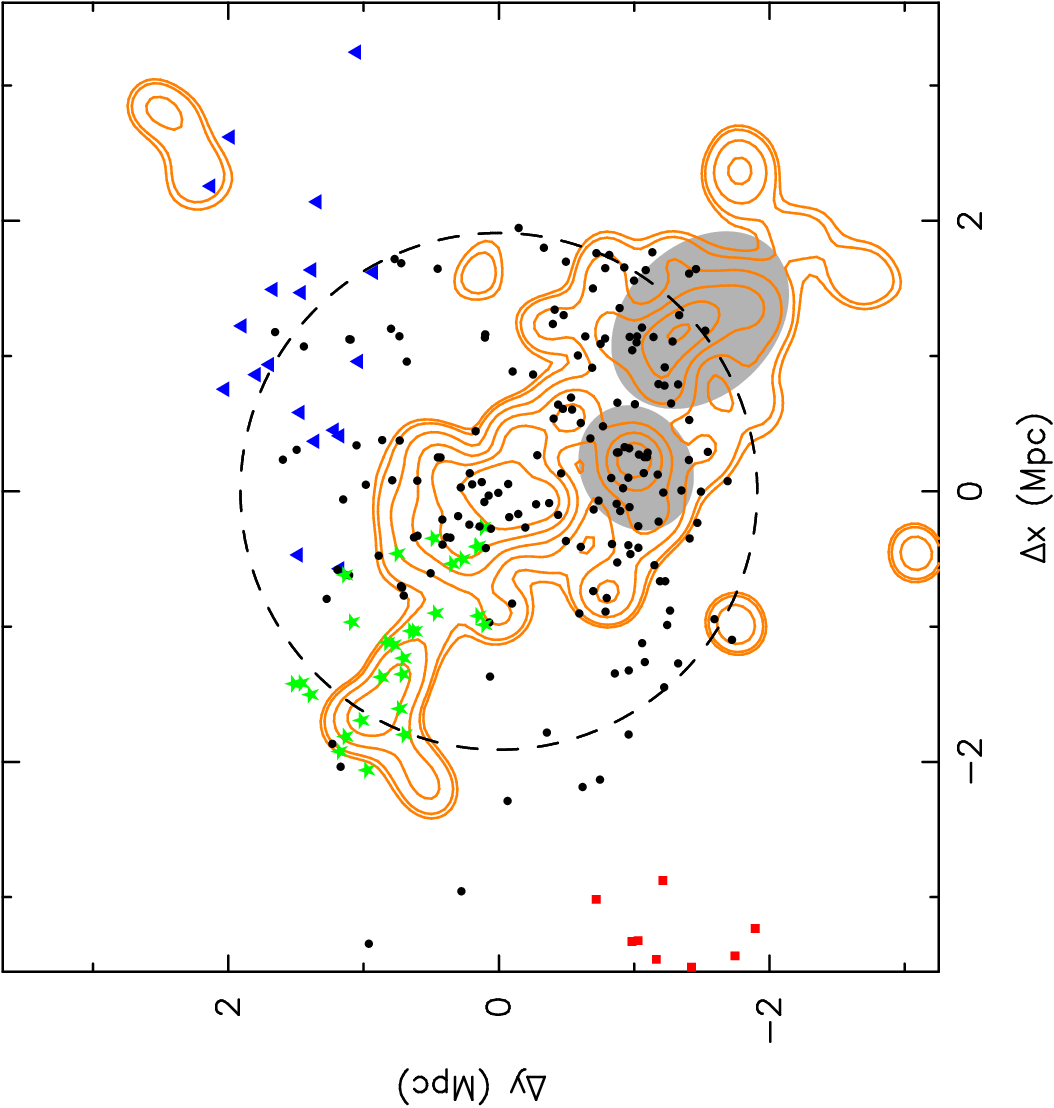}\\
\includegraphics[width=0.43\textwidth, angle=-90]{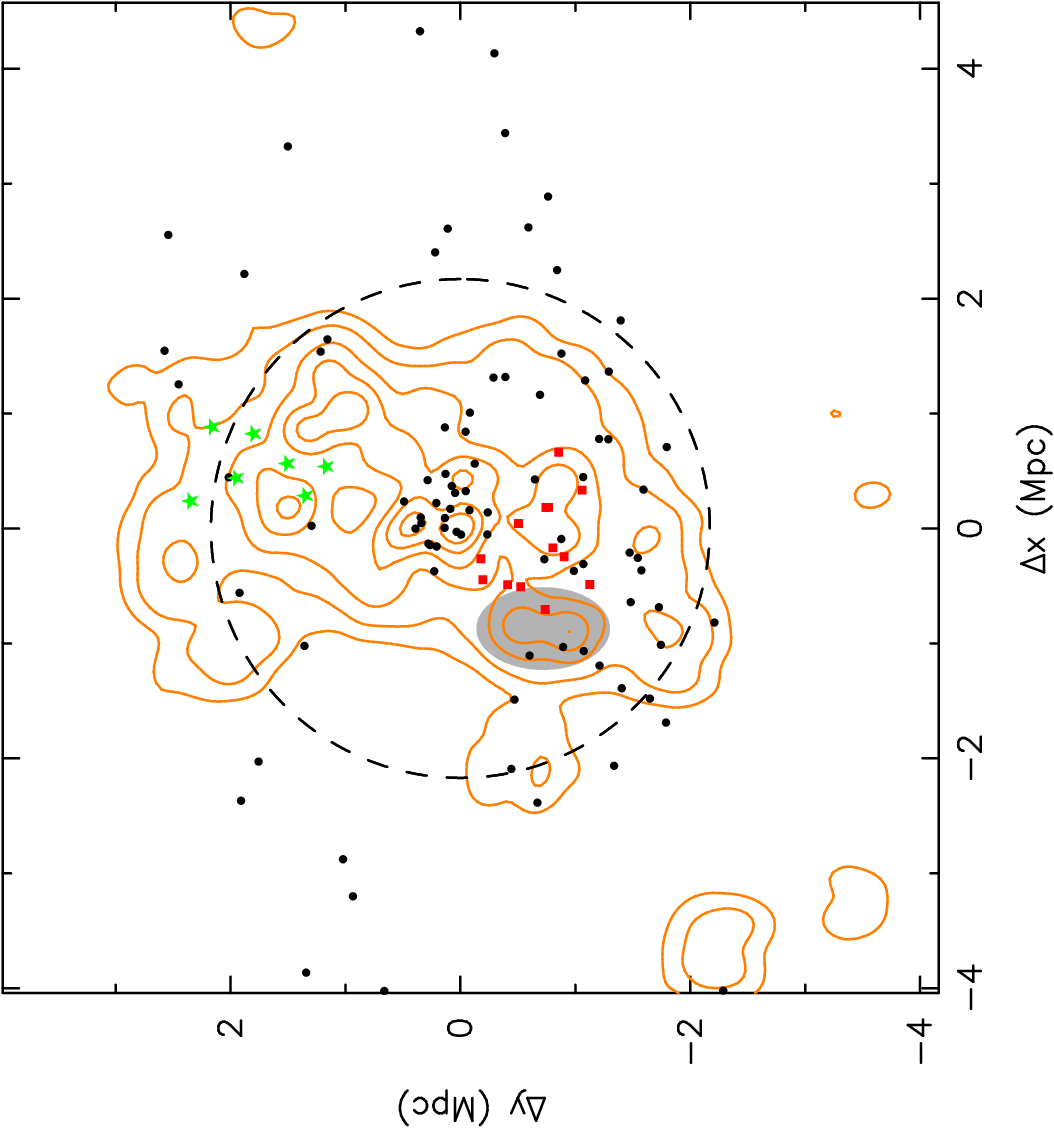}\hspace{0.02\textwidth}%
\includegraphics[width=0.43\textwidth, angle=-90]{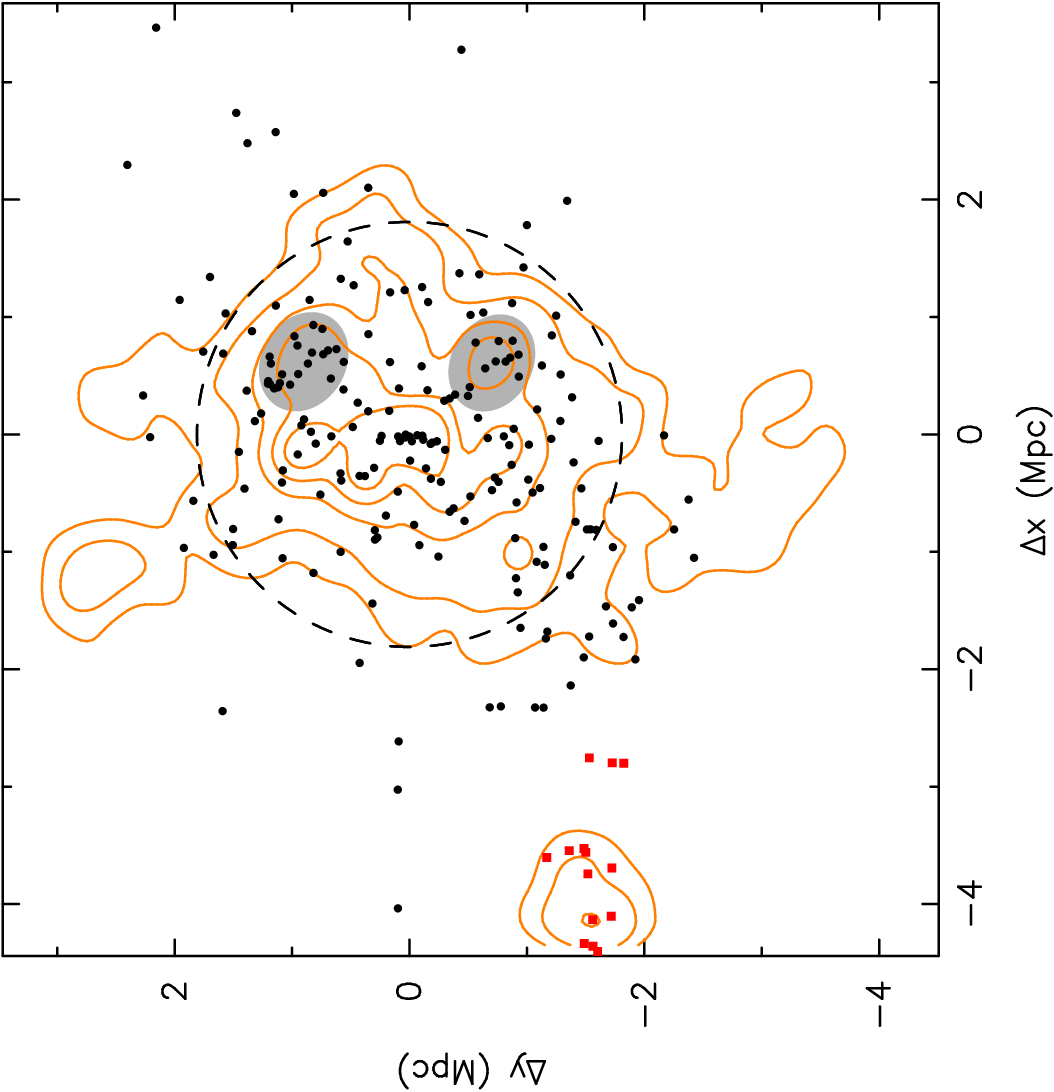}
\caption{Structure of RXCJ0014 (top left), RXCJ0225 (top right), RXCJ0516 (bottom left), and RXCJ0528 (bottom right). The different symbols show the KMM partitions (the black dots are for the main body of the cluster). The orange contours trace the surface density of the red-sequence galaxies; they start at $5\sigma$ above the mean background density. The dashed circle has a radius $R_{200}$. The light-grey ellipses trace the additional substructures not detected by the DS test; the galaxies inside them were also excluded for the updated dynamical analysis}\label{fig:KMM_1}
\end{figure*}

\begin{figure*}
\includegraphics[width=0.43\textwidth, angle=-90]{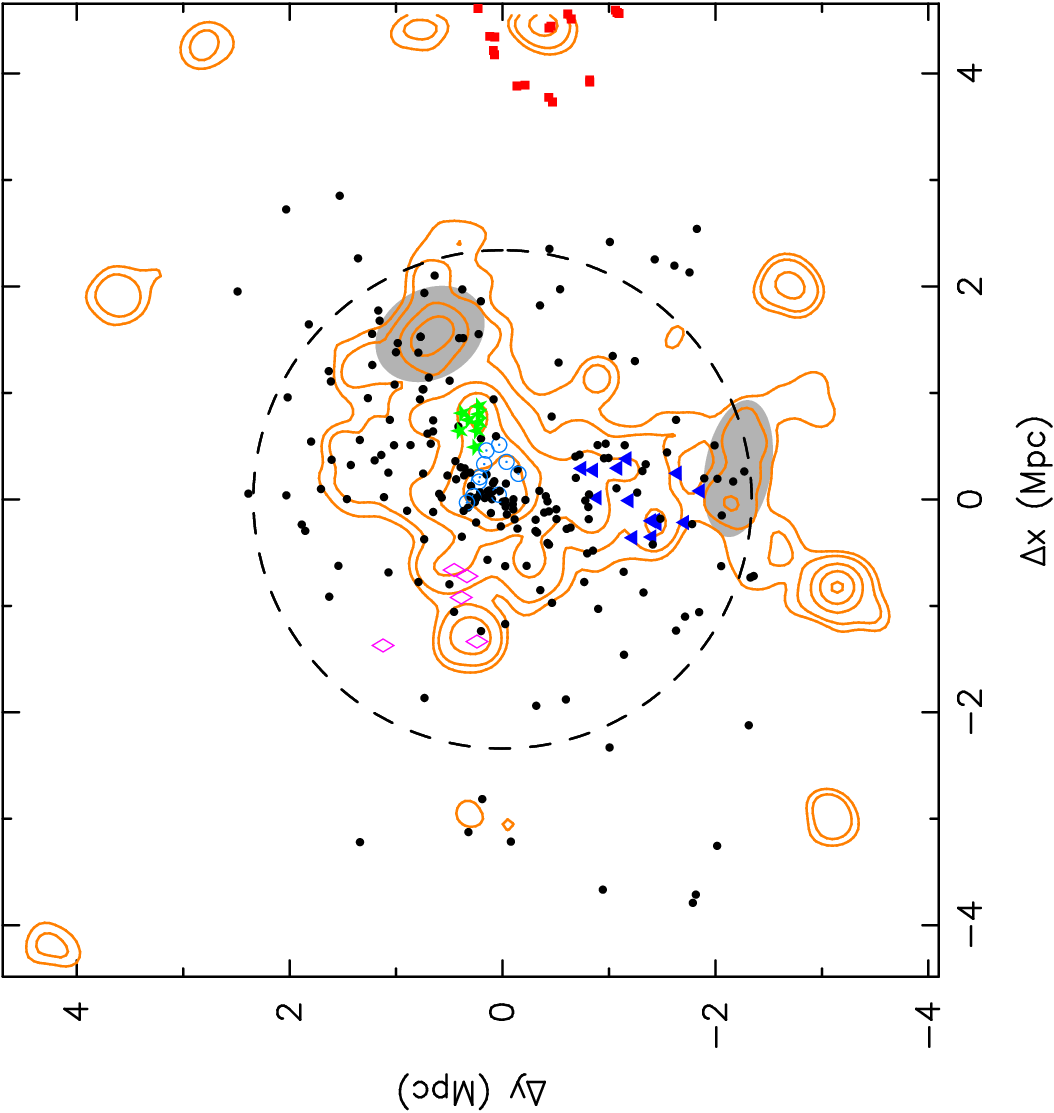}\hspace{0.02\textwidth}%
\includegraphics[width=0.43\textwidth, angle=-90]{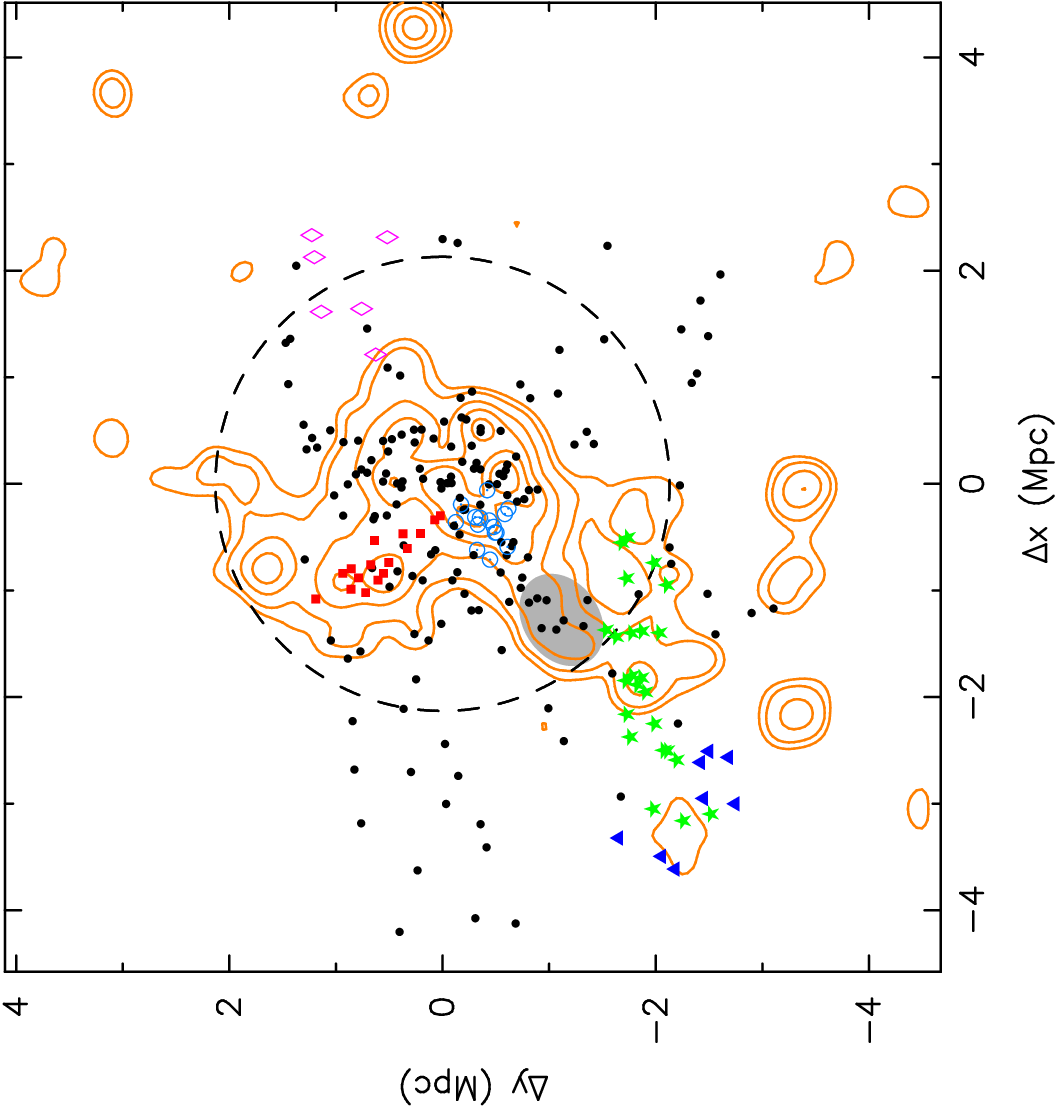}\\
\includegraphics[width=0.43\textwidth, angle=-90]{RXCJ1206_KMM.eps}\hspace{0.02\textwidth}%
\includegraphics[width=0.43\textwidth, angle=-90]{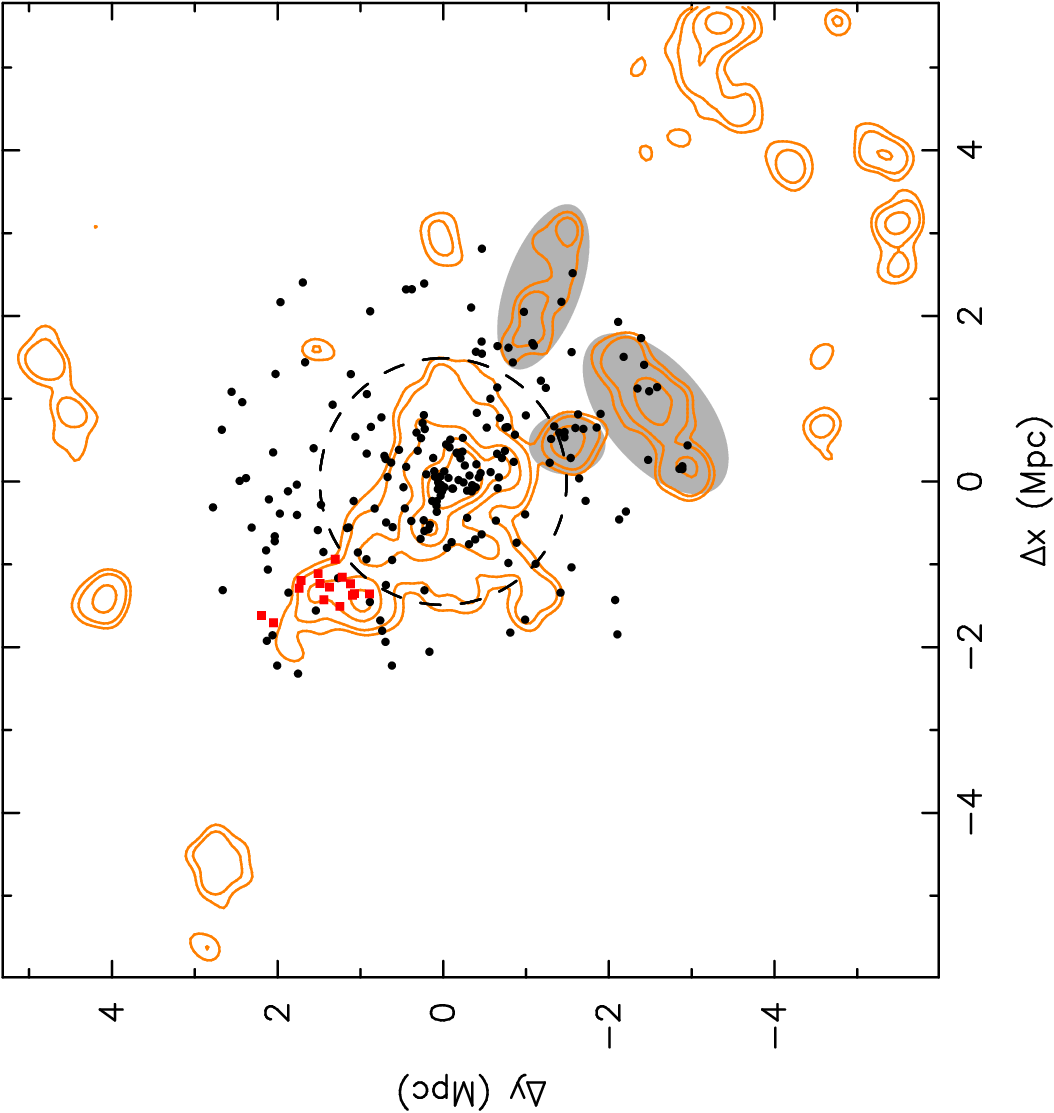}
\caption{Same as Fig. \ref{fig:KMM_1}, but for RXCJ0658 (top left), RXCJ1131 (top right), RXCJ1206 (bottom left), and RXCJ1347 (bottom right).}\label{fig:KMM_2}
\end{figure*}

\begin{figure*}
\includegraphics[width=0.43\textwidth, angle=-90]{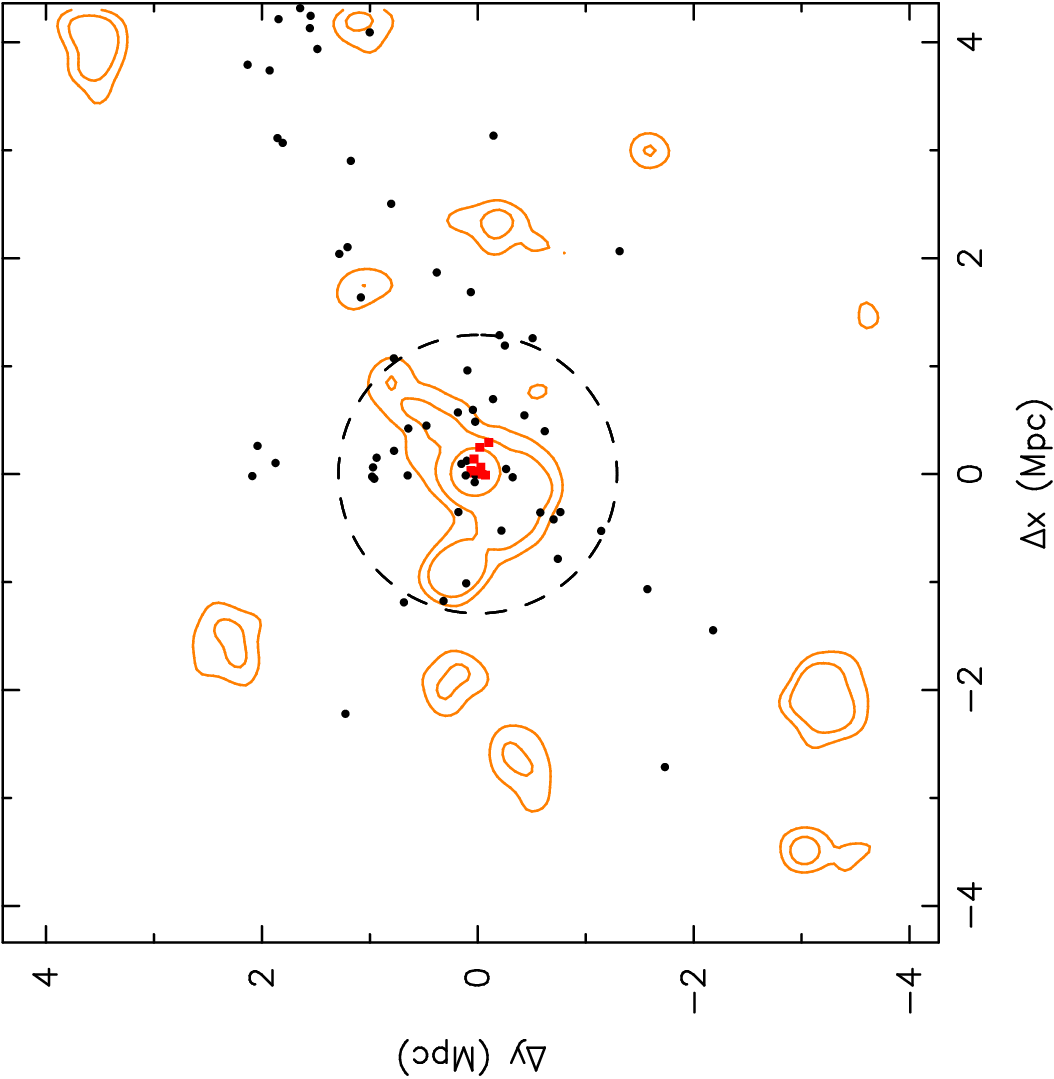}\hspace{0.02\textwidth}%
\includegraphics[width=0.43\textwidth, angle=-90]{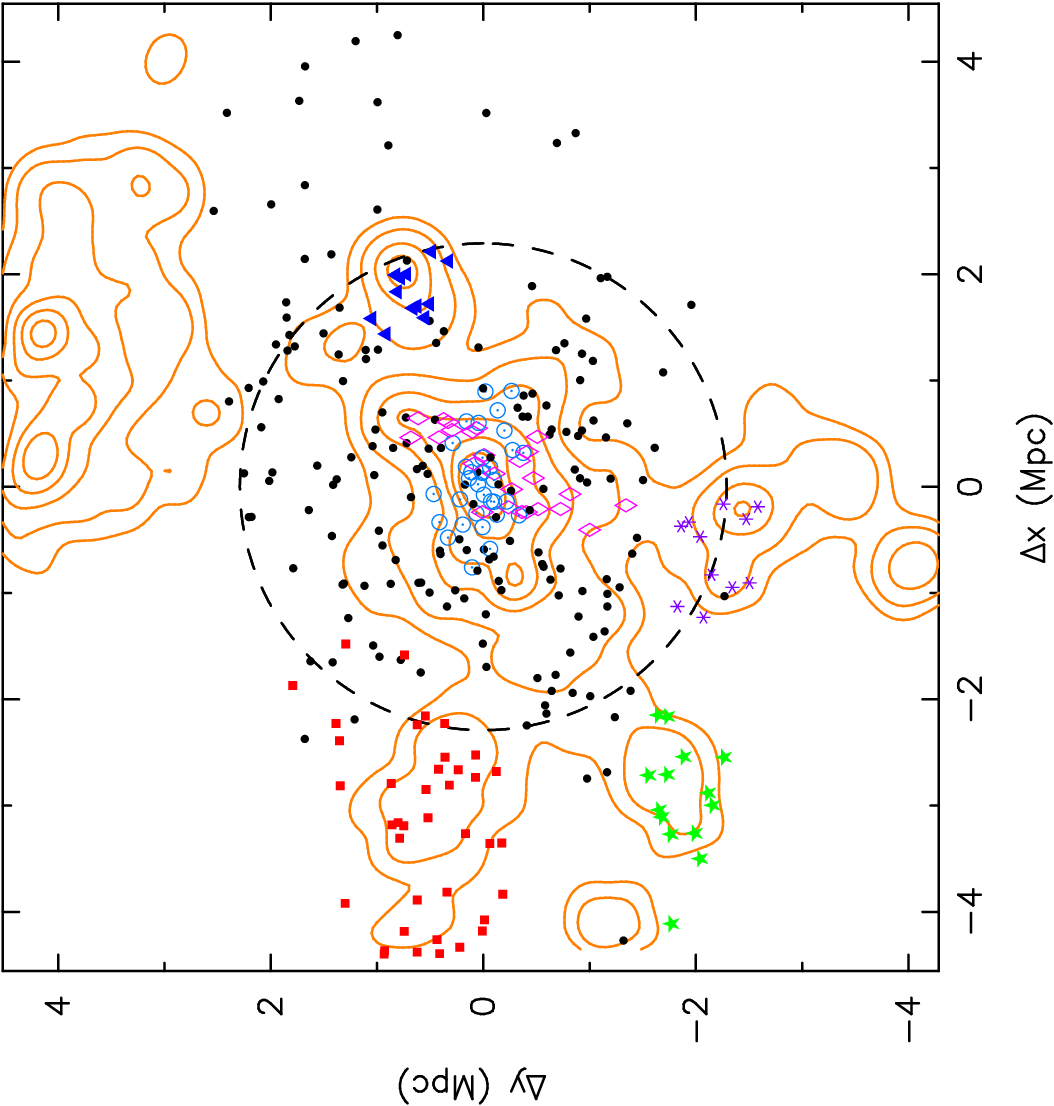}
\caption{Same as Fig. \ref{fig:KMM_1}, but for RXCJ2011 (left panel) and RXCJ2308 (right panel).}\label{fig:KMM_3}
\end{figure*}

\section{Redshifts}
Table \ref{table:app} contains the list of the photometric and spectroscopic cluster members used for this work. The full list is available at the CDS.

\setcounter{table}{0}
\renewcommand{\thetable}{D\arabic{table}}

\begin{table*}
\centering
\begin{threeparttable}
\caption{Sky position and redshift of the cluster members.}
\label{table:app}
\begin{tabular}{l c c c c c c c}
\hline\hline\noalign{\smallskip}
Cluster & RA & DEC & $z_{s}$ & F$_{\mathrm{EZ}}$ & $z_{p}$ & $\delta z_{p}$ & F$_{\mathrm{RS}}$\\
& (J2000) & (J2000) & & & & \\
\noalign{\smallskip}\hline\noalign{\smallskip}
RXCJ0516&05:16:18.8&-54:38:17.9&0.2888&4&-&-&1\\
RXCJ0516&05:16:36.1&-54:38:17.3&-&-&0.29&0.01&1\\
RXCJ0516&05:17:21.4&-54:38:16.4&0.2957&4&-&-&1\\
RXCJ0516&05:18:16.3&-54:38:13.1&-&-&0.34&0.08&1\\
RXCJ0516&05:16:43.2&-54:38:14.1&-&-&0.27&0.06&1\\
RXCJ0516&05:18:00.5&-54:38:11.4&-&-&0.30&0.03&0\\
RXCJ0516&05:17:00.7&-54:38:11.9&-&-&0.29&0.04&0\\
RXCJ0516&05:17:14.6&-54:38:10.5&-&-&0.29&0.01&1\\
RXCJ0516&05:16:35.4&-54:38:07.6&-&-&0.32&0.04&0\\
RXCJ0516&05:17:03.7&-54:38:06.4&0.2930&4&-&-&0\\
\noalign{\smallskip}\hline
\end{tabular}
    \begin{tablenotes}
      \small
      \item Columns: (1) Cluster host. (2,3) Equatorial coordinates of the galaxy. (4) VIMOS spectroscopic redshift. (5) EZ flag of the spectroscopic redshift estimate. (6) WFI photometric redshift, equal to the spectroscopic value when available. (7) Uncertainty of the photometric redshift. (8) Red-sequence membership flag. F$_{\mathrm{RS}}=1$ for the red-sequence galaxies, F$_{\mathrm{RS}}=0$ otherwise.
    \end{tablenotes}
  \end{threeparttable}
\end{table*}

\end{document}